\documentclass[reprint,superscriptaddress,preprintnumbers,amsmath,amssymb,aps,prd]{revtex4-2}
\usepackage{float}
\usepackage{graphicx}
\usepackage{dcolumn}
\usepackage{bm}
\usepackage{bbold}
\usepackage{braket}
\usepackage{amssymb,amsmath}
\usepackage{bbm}
\usepackage{slashed}

\usepackage{color}
\usepackage{amsfonts}
\usepackage{subfigure}
\usepackage{array}
\usepackage{enumerate}

\newcommand{\Tr}{\ensuremath{\operatorname{Tr}}}
\newcommand{\tr}{\ensuremath{\operatorname{tr}}}

\newcolumntype{L}{>{\centering\arraybackslash}m{3cm}}
\newcommand{\imag}{\text{i}}

\newcommand{\LEGO}{LEGO\textsuperscript{\textregistered}}

\definecolor{blue}{rgb}{0,0,1}

\definecolor{green}{rgb}{0,1,0}

\definecolor{red}{rgb}{1,0,0}

\definecolor{gray}{rgb}{.5,.5,.5}

\definecolor{darkgreen}{rgb}{.0,.5,.0}

\usepackage{xspace}
\usepackage{xfrac}
\usepackage{hyperref}
\usepackage[nameinlink]{cleveref}
\usepackage{xifthen}
\usepackage{xcolor}
\hypersetup{
	colorlinks,
	linkcolor={red!75!black},
	citecolor={blue!75!black},
	urlcolor={blue!75!black}
}

\def\lA0{{\langle A_0 \rangle}}
\def\bA0{{\bar{A}_0}}

\def\0#1#2{\frac{#1}{#2}}

\graphicspath{{./figures/}{./}}

\begin{document}
	
\title{Four-quark scatterings in QCD III}

\newcommand{\getHeidelbergAffiliation}{\affiliation{Institut f{\"u}r Theoretische Physik, Universit{\"a}t Heidelberg, Philosophenweg 16, 69120 Heidelberg, Germany}}

\newcommand{\getEMMIAffiliation}{\affiliation{ExtreMe Matter Institute EMMI, GSI, Planckstr. 1, D-64291 Darmstadt, Germany}}

\newcommand{\getDalianAFfiliation}{\affiliation{School of Physics, Dalian University of Technology, Dalian, 116024, P.R. China}}
	
\author{Wei-jie Fu}\getDalianAFfiliation

\author{Chuang Huang}\email{huang@thphys.uni-heidelberg.de}\getHeidelbergAffiliation

\author{Jan M. Pawlowski}\getHeidelbergAffiliation\getEMMIAffiliation

\author{Yang-yang Tan}\getDalianAFfiliation

\author{Li-jun Zhou}\getDalianAFfiliation

\begin{abstract}

We study the full infrared dynamics of 2+1 flavour QCD with the functional renormalisation group approach. We resolve  self-consistently the glue dynamics as well as the dynamics of chiral symmetry breaking. The computation hosts no phenomenological parameter or external input. The only ultraviolet input parameters are the physical ones in QCD: the light and strange quark masses. They are adjusted to the physical ratios of the pion and kaon masses, divided by the pion decay constant. The results for other observables of current first-principles computations are in quantitative agreement with the physical ones. This work completes the series of papers, initiated and furthered in \cite{Fu:2022uow, Fu:2024ysj}, on dynamical chiral symmetry breaking and the emergence of mesonic bound states within the functional renormalisation group. As a first application we discuss the formation of light mesonic bound states. Amongst other applications such as the phase structure of QCD, the current work paves the way for studying QCD parton distribution functions within the functional renormalisation group approach to first-principles QCD. \\[15ex]

\end{abstract}

\maketitle 

\tableofcontents

\newpage
\section{Introduction}
\label{sec:introduction}

Infrared QCD with its strongly correlated dynamics hosts a few of the most intriguing phenomena or problems fundamental physics has to offer. Open challenges range from the 
quantitative description and understanding of quark scattering and the formation of hadron resonances to the phase structure of QCD and the non-equilibrium physics of QCD in heavy ion collisions or in the early universe. The resolution of these phenomena requires, either in parts or fully, non-perturbative realtime approaches with small systematic errors and a high computational accuracy. 

First-principles functional approaches to QCD are tailor-made for these tasks: Typically they are used for the computation of spacelike (Euclidean) or equilibrium QCD correlation functions.  However, as diagrammatic approaches, unlike lattice simulations, they allow for a straightforward continuation to timelike and non-equilibrium QCD correlation functions. For recent reviews on QCD applications see \cite{Eichmann:2016yit, Fischer:2018sdj, Dupuis:2020fhh, Fu:2022gou}. These unique advantages in the timelike regime of QCD come at the price of rather intricate computer-algebraic and numerical challenges, that are posed by the hierarchy of coupled integral and/or integral-differential set of diagrammatic relations between QCD correlation functions. The respective tasks can again be divided into two interrelated but separate ones: the quest for quantitative precision including a small systematic error in a full first-principles approach, and the quest for a systematic continuation of functional approaches to QCD into the timelike domain.       

In the present work we concentrate on the former task and put forward a functional fRG approach to first-principles 2+1 flavour QCD. We also provide QCD results on the latter task with the first analysis of bound state properties of light mesons. The present work completes a series of three works: The first two, \cite{Fu:2022uow, Fu:2024ysj}, were concerned with the matter dynamics of low energy QCD. In these works only the quark dynamics was included, specifically concentrating on the quantitative resolution of the four-quark scattering sector of QCD. In the present work the matter dynamics is augmented with the gluonic one. In combination this gives us access to the full dynamics of QCD from the perturbative ultraviolet regime to the strongly correlated infrared regime.  

Special emphasis is given to the interface between the pure glue sector of the theory and the matter sector. This dynamic is encoded in the quark-gluon vertex, as well as in higher quark-gluon scatterings. This interface has been studied thoroughly in \cite{Mitter:2014wpa, Williams:2014iea, Williams:2015cvx, Cyrol:2017ewj, Gao:2021wun, Aguilar:2024ciu}, and its quantitative resolution is key to a small systematic error in functional QCD approaches. Together with the quantitative resolution of the pure matter sector put forward \cite{Fu:2022uow,Fu:2024ysj}, we obtain quantitative first-principles results in 2+1 flavour QCD. With the current approach, to our best knowledge, this is achieved for the first time in 2+1 flavour QCD without the need of external input such as gluon correlation functions, adding to the respective two-flavour work in \cite{Cyrol:2017ewj}. Our fRG approach is then used for a first application to the bound state properties of light mesons. 

This work is organised as follows: In \Cref{sec:QCD-EffAction}, we discuss the functional renormalisation group approach to 2+1 flavour QCD. In particular we describe and evaluate the approximations for the full effective action of QCD. This is done separately for the pure glue, glue-matter interface, and pure matter sectors. In \Cref{sec:QCDCorrelations} we discuss the determination of the fundamental parameters of 2+1 flavour QCD, that is the light and strange current quark masses. We also present results for momentum-dependent QCD correlation functions, and in particular for all quark, ghost and gluon two-point functions, the dressings of the ghost-gluon, three-gluon vertex, and quark-gluon and four-quark vertices. In \Cref{sec:meson-part}, we focus on the bound state properties of the light mesons. 
In \Cref{sec:conclusion} we briefly summarise the progress made in the present work.

\section{Functional 2+1 flavour QCD}
\label{sec:QCD-EffAction}

The functional renormalisation group approach to QCD is governed by the flow equation for the one-particle irreducible effective action, the Wetterich equation \cite{Wetterich:1992yh}. This flow equation describes the differential change of the QCD effective action. The current work builds on \cite{Fu:2022uow,Fu:2024ysj} and the QCD works in the fQCD collaboration in the past decade, \cite{Mitter:2014wpa, Braun:2014ata, Rennecke:2015eba, Cyrol:2016tym, Cyrol:2017ewj, Corell:2018yil, Fu:2019hdw, Gao:2020fbl, Gao:2020qsj, Gao:2021wun, Lu:2023mkn, Ihssen:2024miv}, or rather the respective systematic error analysis based on the \LEGO principle \cite{Ihssen:2024miv}. It also draws from countless fRG works in low-energy effective theories for QCD. For these and further QCD works we refer to the recent QCD reviews \cite{Dupuis:2020fhh, Fu:2022gou}. 

The classical gauge-fixed action and some notational details are discussed in \Cref{app:SQCD}. The flow for the effective action and that for (inverse) propagators and vertices are discussed in \Cref{app:Flow}. The respective QCD effective action can be divided into the pure glue sector and the matter sector, 
\begin{subequations} 
	\label{eq:QCDEffAction}
\begin{align}
    \Gamma_{k}[\Phi]=\Gamma_{\mathrm{glue},k}[A,\bar{c},c]+\Gamma_{\textrm{mat},k}[A,\bar{q},q]\,, 
\label{eq:QCDEffActionSplit}
\end{align}
with 
\begin{align}
\Phi=(A,c, \bar{c},q,\bar{q})\,.
\label{eq:Phi}
\end{align}
In \labelcref{eq:QCDEffAction}, the subscript ${}_k$ indicates the infrared cutoff scale in the functional renormalisation group. 
The pure glue sector in \labelcref{eq:QCDEffAction} carries the purely gluonic interactions, also including the gauge fixing part with the auxiliary ghost fields. Its approximation is provided and discussed in detail in \Cref{sec:GammaGlue}. 

The matter sector of the effective action splits into the glue-matter interface between the glue and quark fields and a pure matter sector, 
\begin{align}
	\Gamma_{\textrm{mat},k}[\Phi]= \Gamma_{\textrm{inter},k}[A,\bar{q},q]+\Gamma_{\textrm{quark},k}[\bar{q},q]\,.  
	\label{eq:QCDmat}
\end{align}
\end{subequations}
In \labelcref{eq:QCDmat}, the part $\Gamma_{\textrm{quark},k}$ contains the quark--anti-quark scattering terms to all order and $\Gamma_{\textrm{inter},k}[A,\bar{q},q]$ comprises the interface between $\Gamma_{\textrm{quark},k}$ and the pure glue sector. 

It is well-known that the matter sector is sourced and dominated by quark-gluon interactions for perturbative and semi-perturbative momentum scales $p\gtrsim 1$\,GeV, see \cite{Braun:2014ata, Rennecke:2015eba, Cyrol:2017ewj, Fu:2019hdw, Ihssen:2024miv}. 
This regime is followed by a small interface regime with mixed quark-gluon and four-quark dynamics at $p \approx 0.5$\,GeV, before the system finally settles in a regime dominated by pion dynamics for $p \lesssim 0.5$\,GeV, and hence the pure matter sector. 
For a recent detailed study and resolution of this change of the relevant degrees of freedom as well as their diagrammatic resolution see \cite{Ihssen:2024miv}. Note also that the infrared regime is the validity regime of chiral perturbation theory $\chi$PT, and hence, amongst other insights, functional QCD flows provide us with an interpolation between the perturbative high energy regime of QCD and $\chi$PT. 
Indeed, the present approach allows us to determine the low energy effective constants in $\chi$PT. 

In combination, these previous investigations suggest the following quantitative approximation for the two parts of the matter sector in 2+1 flavour QCD in \labelcref{eq:QCDmat}, 
\begin{align}\nonumber 
 \Gamma_{\textrm{inter},k}[A,q,\bar{q}]\approx &\, \Gamma_{d,k}[A,q,\bar q]\,,\\[2ex] \Gamma_{\textrm{quark},k}[q,\bar{q}]\approx&\,\Gamma_{4q,k}[q,\bar{q}]\,. 
\label{eq:QCDmatApprox}
\end{align}
The interface action is expanded in terms of gauge-invariant operators, which has been shown to work quantitatively in \cite{Mitter:2014wpa, Cyrol:2017ewj} in two-flavour QCD: there the quark-gluon vertex and higher order gluon scatterings with two-quarks have been expanded in the operators $\bar q \slashed{D}^n q$ with $n=1,2,3$ and $\bar q [\slashed{D},\slashed D] q$. Within such an expansion, the dressings of vertices with different powers of gauge fields are related on the symmetric point, subject to the irrelevance of further operators.  These relations have indeed been confirmed numerically in \cite{Mitter:2014wpa, Cyrol:2017ewj}, corroborating such an expansion. The part of the action built from these operators can be seen as a generalised Dirac part, hence the name $\Gamma_{d,k}$. It hosts the interface terms between the pure glue sector and the matter sector. Its approximation is discussed and evaluated in \Cref{sec:GammaInterface}. 

The pure matter sector in the matter action hosts the quark scatterings to all orders. We use a four-quark approximation, $\Gamma_{4q,k}$, whose dynamics and quantitative nature were the central focus of the first two papers of the present series \cite{Fu:2022uow, Fu:2024ysj}. This analysis did not require dynamical gluons as they decouple below the gluon mass gap of approximate $1$\,GeV. The approximation used here is that put forward and evaluated in \cite{Fu:2024ysj}. Its main properties are recapitulated in \Cref{sec:Gamma4q}. In combination, the analyses in \Cref{sec:GammaGlue,sec:GammaInterface,sec:Gamma4q} lead us to the quantitative approximation \labelcref{eq:GammaGlue,eq:GammaInter,eq:Gamma4q}. Its convergence properties and the respective systematic error estimate are discussed separately in \Cref{sec:Wrapup+SystematicError}.

\subsection{Pure glue sector}
\label{sec:GammaGlue}

The approximation for the pure glue sector of QCD draws from the quantitative study of the dynamics of Yang-Mills theory in \cite{Cyrol:2016tym, Corell:2018yil} with the functional renormalisation group, and similar ones with Dyson-Schwinger equations, \cite{Huber:2018ned, Huber:2020keu,Eichmann:2021zuv}. These studies, and references therein, also include a careful evaluation of the importance of higher-order tensor structures for the quantitative dynamics of this system. In combination this suggests the approximation detailed in \Cref{app:truncation-gauge} and the full glue effective action is provided there in \labelcref{eq:GammaGlueDetailed}. In the following we discuss the important features of the approximation at the example of selected terms, 
\begin{widetext} 
\begin{align}\nonumber 
    \Gamma_{\textrm{glue},k}=&\int_{p}\Bigg\{
 A_\mu(-p) \, p^2\left[Z_A(p)\, \Pi^\perp_{\mu\nu}(p)\,+ \frac{1}{\xi} \Pi^\parallel_{\mu\nu}(p) \right]\,A_\mu +\bar c(-p) \,Z_c(p)\,p^2\, c(p)\Biggr\} \\[2ex] 
  & +\int_{p,q}\, Z^{1/2}_c(q) Z^{1/2}_c(-p-q) Z^{1/2}_A(p)\, \lambda_{A \bar c c}(p,q)\,
\bar c(q) {\cal T}^{\textrm{(cl)}}_{A \bar c c}(p,q) A_\mu(p)\,c(-p-q)+ \textrm{glue interactions} \,,
\label{eq:GammaGlue}
\end{align}
\end{widetext} 
with the transversal and longitudinal projection operators $\Pi^\bot\,,\,\Pi^\parallel$ defined in \labelcref{eq:ProjectionOps} in \Cref{app:truncation-gauge}, and 
\begin{align}
    \int_{p}\equiv\int \frac{d^{4}p}{(2\pi)^{4}}\,.
\end{align}
\Cref{eq:GammaGlue} contains the full momentum-dependent kinetic terms of the ghost and gluon fields in the first line. We do not consider the longitudinal dressing of the gluon, which is unity for $k=0$ due to the respective Slavnov-Taylor identity (STI). At $k\neq 0$ a non-trivial longitudinal dressing is enforced by the modified STIs (mSTIs). All explicit computations here are done in the Landau gauge 
\begin{align}
    \xi = 0\,.
    \label{eq:Landaugauge}
\end{align}
and hence the longitudinal dressing drops out of the gluon propagator which is transverse. The choice  \labelcref{eq:Landaugauge} conveniently optimises both: computational simplicity (both algebraically and numerically), and the smallest systematic error in a given approximation, for a respective discussion see \cite{Fischer:2008uz, Cyrol:2016tym, Dupuis:2020fhh}.  The transverse propagator provides an important benchmark test for our approximation: 
Its momentum dependence including the absolute scales is computed within the present work and is then compared to the respective lattice results, see \Cref{fig:YM-twopoint} in \Cref{sec:QCDCorrelations} and the discussions therein. 
We achieve quantitative agreement including the absolute scales. 

The purely gluonic vertices are detailed in \Cref{app:truncation-gauge} and are abbreviated with `glue interactions' in the second line in \labelcref{eq:GammaGlue}. 
In short, we only use the classical tensor structures for the three- and four-gluon interactions and approximate their dressings with the symmetric point momentum configuration. 

This approximation is illustrated with the example of the ghost-gluon vertex in the second line of \labelcref{eq:GammaGlue}: 
This vertex contains only one transverse tensor structure, the transverse projection of the classical one, $[{\cal T}^{\textrm{(cl)}}_{A \bar cc }]^{abc}_{\mu}(p,q) = \imag q_\mu f^{abc}$ with the gluon momentum $p$ and the anti-ghost momentum $q$, see \labelcref{eq:YM-ccA}. Hence, the ghost-gluon vertex in \labelcref{eq:GammaGlue} is simply a parametrisation of the full vertex without any approximation. The vertex dressing is split in a factor, that carries the wave functions of the fields attached to it, and hence its renormalisation group (RG) scaling. This parametrisation is applied to all vertices and we parametrise  
\begin{align}
    \Gamma_{\Phi_{i_1}\cdots \Phi_{i_n}}^{(n)}(\boldsymbol{p}) = \left[\prod_{j=1}^n Z^{1/2}_{\Phi_{i_j}}(p_j) \right]\, \bar \Gamma_{\Phi_{i_1}\cdots \Phi_{i_n}}^{(n)}(\boldsymbol{p})\,,  
    \label{eq:GR-InvariantGn}
\end{align}
with $\boldsymbol{p} =(p_1,...,p_n)$. In \labelcref{eq:GR-InvariantGn}, we have introduced the RG invariant vertices $\bar \Gamma^{(n)}(\boldsymbol{p})$. While \labelcref{eq:GR-InvariantGn} constitutes a mere reparametrisation of the vertex, this split becomes important if further approximations are applied to either $\Gamma^{(n)}$ or $\bar\Gamma^{(n)}$, for a detailed discussion see \cite{Ihssen:2024miv}: In general, the RG-invariant vertices $\bar \Gamma^{(n)}(\boldsymbol{p})$ have a smoother momentum dependences, for a non-trivial evaluation and application in quantum gravity see \cite{Denz:2016qks}. 

In view of these considerations we discuss approximations for the full momentum dependence of the dressing $\lambda_{A\bar c c}(p,q)$, that is the RG-invariant dressing of the vertex $\bar \Gamma_{A\bar c c}(p,q, -p-q)$. In the present work we use an approximation that is informed from the investigation of the full vertex in \cite{Mitter:2014wpa, Cyrol:2017ewj}. We evaluate the vertex dressing within the symmetric point approximation: Instead of the two momenta $p,q$ (two radial variables and one angle) we consider only an average momentum dependence, 
\begin{align} 
\lambda_{A\bar c c}(p,q) \to \lambda_{A\bar c c}(\bar p)\,, 
\label{eq:lambdaAbarccbarp}
\end{align}
with 
\begin{align}
    \bar p^2 = \frac{p^2+q^2+(p+q)^2}{3}\,. 
    \label{eq:SymPoint3Average}
\end{align}
This average momentum dependence of $\lambda_{A\bar c c}(\bar p)$ is extracted from the flow equation by an evaluation of $\partial_t \lambda_{A\bar c c}(p,q)$ on the maximally symmetric simplex configuration. This procedure is also applied to the gluonic three- and four-point functions as well as the quark-gluon vertex, and their momentum dependence is reduced to that on a single scalar parameter. At this simplex configuration with $n=3,4$ the momenta satisfy 
\begin{align}
    p_i\cdot  p_j  = \frac{ n\delta_{ij} -1}{n-1} \bar p^2\,,\qquad \bar p^2 = \frac{1}{n}\sum_{i=1}^n p_i^2\,. 
    \label{eq:SymPoint-npoint}
\end{align}
In \labelcref{eq:SymPoint-npoint}, all momenta are counted as incoming and momentum conservation at the vertices is guaranteed by 
\begin{align} 
p_n=-(p_1+\cdots+p_{n-1})\,. 
\label{eq:MomentumConservation}
\end{align}
In the present ghost-gluon example we have the incoming momenta $p_1=p$ (gluon), $p_2=q$ (anti-ghost) and $p_3=-p-q$ (ghost). 

The symmetric point approximation discussed for the ghost-gluon vertex is also used for the other vertices in the pure glue sector and in the glue-matter interface evaluated in \Cref{sec:GammaInterface}. This approximation is also discussed in \Cref{app:truncation-gauge} and all symmetric point dressings are given by 
\begin{align}
    \lambda_{i}(p_{1},p_{2},\cdots,p_{n_i})\approx \lambda_{i}(\bar{p})\,,
    \label{eq:SymmetricPointDressings}
\end{align}
with $i=A^3\,, A^4\,,\,A\bar c c\,,\,A\bar l l,A \bar s s$ and $n_i = n-1$ for $n$-point functions. 

The ghost-gluon sector allows us to define two avatars of the running strong coupling,   
\begin{subequations}
\label{eq:StrongCouplingGlueSector}
\begin{align}
\alpha_{A\bar{c}c}(p)=\frac{\lambda_{A\bar{c}c}^{2}(p)}{4\pi} \,,\quad  \alpha^{\ }_T(p)=\frac{1}{ 4 \pi} \frac{g_s^2}{Z_A(p) Z_c(p)^2} \,.
\label{eq:GhostGluonAvatars}
\end{align}
The Taylor coupling $\alpha_T^{\ }(p)$ in \labelcref{eq:GhostGluonAvatars} uses the non-renormalisation of the ghost-gluon vertex in the Landau gauge, and is related to the process-independent coupling, see \cite{vonSmekal:1997ohs,Boucaud:2008gn,vonSmekal:2009ae,Blossier:2012ef,Zafeiropoulos:2019flq,Gao:2024gdj}. Note that, while they carry the same RG-scaling at two-loop, they have a different momentum-dependence already at this order. 

The symmetric-point dressings of the three- and four-gluon vertices discussed in \Cref{app:truncation-gauge} give rise to two further avatars of the strong coupling, 
\begin{align}
    \alpha_{A^{3}}(p)=\frac{\lambda_{A^{3}}^{2}(p)}{4\pi},\qquad \alpha_{A^{4}}(p)=\frac{\lambda_{A^{4}}(p)}{4\pi}\,. 
    \label{eq:GluonAvatars}
\end{align}
\end{subequations} 
Seemingly this leaves us with three different initial conditions for the strong coupling at the initial cutoff scale $k=\Lambda$. The fourth one, the Taylor coupling is a derived quantity and does not show up in the flow. However, at vanishing cutoff scales, $k=0$, all avatars agree with each other at asymptotically large momenta due to the respective Slavnov-Taylor identities (STIs), leading to a single coupling parameter to be fixed, as it should be. At $k\neq 0$, the three avatars of the couplings do not agree, but we encounter minor deviations due to the regulator-terms in the modified STI (mSTI). In \cite{Cyrol:2016tym,Cyrol:2017ewj} these differences have been resolved by tuning the initial couplings such that the STIs at $k=0$ are satisfied for perturbative momenta. 

As these differences are small and are even reduced within the current symmetric point approximation, we resort to using the approximate identity 
\begin{align}
\alpha_{i,\Lambda}(p) = \alpha_{s,\Lambda}\,,\qquad i= A^3\,,\, A^4\,,\, A\bar c c\,. 
\label{eq:Initialalphaglue}
\end{align}
We also emphasise that \labelcref{eq:Initialalphaglue} is not a free coupling parameter in QCD: Using a specific (small) value of $\alpha_{s,\Lambda}$ merely defines the physical value of the scale $\Lambda$. Finally, based on the results in \cite{Cyrol:2016tym,Cyrol:2017ewj} we identify the four-gluon avatar by the three-gluon one, see also \labelcref{eq:lambdaA34} in \Cref{app:truncation-gauge}, 
\begin{align} 
\alpha_{A^4,k}(p) = \alpha_{A^3,k}(p)\,,
\end{align}
for all $k$ and $p$. This approximation is quantitative for momenta $p \gtrsim 1.5$\,GeV, and the increasing difference below this scale is irrelevant due to the decoupling of the glue dynamics. 

This concludes our discussion of the pure glue sector of the effective action, for a detailed discussion of the vertex approximation and further details see \Cref{app:truncation-gauge}.

\subsection{Glue-matter interface}
\label{sec:GammaInterface}

The glue-matter interface of the effective action contains the kinetic term of the quark with its full momentum dependence, as well as the quark-gluon interactions within the RG-invariant expansion scheme with the symmetric point approximation explained above around  \labelcref{eq:GR-InvariantGn,eq:SymPoint3Average,eq:lambdaAbarccbarp,eq:SymPoint-npoint}. Similarly to the ghost-gluon vertex we shall exemplify this approximation of the full interface effective action with the quark--anti-quark--gluon interaction with the classical tensor structure 
\begin{align}
\Bigl[{\cal T}^{(1)}_{A\bar q q}(p,q)\Bigr]^a_\mu = {\cal T}^{(1)}_{A\bar q q}=\imag \gamma_\mu T_c\,,
\label{eq:QuarkGluonClassical}
\end{align}
with the gauge group generator $T_c$ in the fundamental representation. In contradistinction to the ghost-gluon term, the quark-gluon term in \labelcref{eq:GammaInter} is pivotal for the infrared dynamics of QCD as it is by far the most relevant quark-gluon interaction. Moreover, as in the pure glue sector the classical tensor structure is but one basis element of a complete tensor basis of the quark--anti-quark--gluon interaction interactions whose transverse part hosts eight basis elements, see \labelcref{eq:qqA-tensor} in \Cref{app:truncation-qqA}. The full glue-matter interface of the effective action includes further quark-gluon three-point terms as well as higher order vertices. The complete form is deferred to \Cref{app:truncation-qqA}, see \labelcref{eq:GammaInterApp}. 

With this preparation the vertex expansion of the glue-matter interface reads 
\begin{widetext}
\begin{align}
    \Gamma_{d,k}= &\int\limits_{p}\bar{q}(-p)\, Z_{q}(p)\,\Big[\mathrm{i}\slashed{p}+ M_{q}(p)\Big]\,q(p)+\int\limits_{p,q}\bar q(q)\, Z_{A}^{\frac{1}{2}}(p) Z_{q}^{\frac{1}{2}}(q)Z_{q}^{\frac{1}{2}}(r)\lambda_{A\bar{q}q}^{(1)}(p,q)\,\Bigl[{\cal T}^{(1)}_{A\bar q q}(p,q)\Bigr]^a_\mu\, A^a_\mu(p) q(r)+ \cdots\,,
    \label{eq:GammaInter}
\end{align}
\end{widetext}
with $r=-p-q$. The right side in \labelcref{eq:GammaInter} contains the full momentum-dependent kinetic terms of the light and strange quarks, and we have only displayed one of the eight quark-gluon interaction terms. Its  RG-invariant  dressing $\lambda^{(1)}_{A\bar q q}(p,q)$ is that of the respective tensor structure of the RG-invariant vertex $\bar\Gamma^{(3)}_{A\bar q q}$ defined in \labelcref{eq:GR-InvariantGn}. 

We also note that the sum of the two terms in \labelcref{eq:GammaInter} can be viewed as a fully momentum-dependent generalisation of the Dirac term in the classical 2+1 flavour QCD action \labelcref{eq:SA+Sq} in \Cref{app:SQCD}. This action has an isospin symmetry: $m_u=m_d=m_l$ with the light quark field $l=(u,d)$, and $m_s> m_l$, see \labelcref{eq:mqall}. We shall not consider flavour-mixing terms in our approximation and hence the quark wave and mass functions also carry the symmetry, 
\begin{align}\nonumber 
    Z_{q}(p)=&\,\mathrm{diag}(Z_{l}(p),Z_l(p)\,,\,Z_s(p))\,,\\[2ex] 
    M_q(p)=&\,\mathrm{diag}(M_{l}(p)\,,\, M_l(p),M_s(p))\,.
    \label{eq:Zq-Mq}
\end{align}
Note that the mass function $M_q$ carries the only two physical parameters of the theory: we shall tune  
\begin{align} 
M_{q,k=\Lambda}(p) \approx (m_l,m_l,m_s)\,,
\label{eq:CurrentQuarkMassesLambda}
\end{align}
such that the physical values of the ratios of the pion and kaon pole masses to the pion decay constants are obtained at $k=0$. 

We proceed with a brief discussion of the additional terms taken into account here (indicated by the `$\cdots$') and displayed in \Cref{app:truncation-qqA}, before we close this Section with an evaluation of the symmetric point approximation: 

The second term in \labelcref{eq:GammaInter} is but one of eight transverse of twelve altogether) quark-gluon terms. A complete basis 
\begin{align}
\bigl\{{\cal T}^{(1)}_{A\bar q q}(p,q)\,,\,...\,,\,{\cal T}^{(8)}_{A\bar q q}(p,q)\bigr\}\,,
\label{eq:QuarkGluonBasis1}
\end{align}
whose transverse projections span the transversal tensors, is provided in \Cref{app:truncation-qqA}. Only three of these eight tensors are relevant and will be considered in our approximation: two chirally symmetric ones, ${\cal T}_{A\bar q q}^{(1,7)}$, and one chiral symmetry breaking one, ${\cal T}^{(4)}_{A\bar q q}$. 
Here, ${\cal T}^{(1)}_{A\bar q q}=\imag \gamma_\mu T_c$ is the classical tensor structure. 
This relevance ordering has been thoroughly tested in functional approaches, see e.g.~\cite{Mitter:2014wpa, Williams:2014iea, Williams:2015cvx, Cyrol:2017ewj, Gao:2021wun, Aguilar:2024ciu}. We have also included two-gluon--quark-antiquark scattering terms that are related by STIs to the quark-gluon terms with ${\cal T}_{A\bar q q}^{(1,4,7)}$, see \cite{Mitter:2014wpa, Cyrol:2017ewj}. The discussion of these terms is deferred to \Cref{app:truncation-qqA}. All vertex dressings are treated self-consistently: their flow is computed and fed back into the diagrammatic part of the flow equations. 

So far, no approximation has been applied to the quark-gluon vertex, apart from dropping the tensor structures that do not contribute to the dynamics. Now we apply the symmetric point approximation already used in the pure glue sector. We exemplify it with the quark-gluon dressings $\lambda^{(i)}_{A\bar q q}(p,q)$, where we reduce the full momentum dependence on gluon and anti-quark momenta $p,q$ to that at the symmetric point with \labelcref{eq:SymPoint-npoint,eq:MomentumConservation}. Then, the approximation of the quark-gluon vertex is summarised with 
\begin{align}
    \lambda_{A\bar q q}^{(1,4,7)}(p,q) \to \lambda_{A\bar q q}^{(1,4,7)}(\bar p)\,, \qquad \lambda_{A\bar q q}^{(2,3,5,6,8)}(p,q) \approx 0\,.
    \label{eq:QuarkGluonApproximation}
\end{align}
The same approximation is applied to the two-quark-two-gluon vertices considered here, more details can be found in \Cref{app:truncation-qqA}. 

The light and strange quark dressings $\lambda^{(1)}_{A\bar q q}$ of the quark-gluon vertex provide us with two avatars of the strong coupling, 
\begin{align}
    \alpha_{A\bar{l}l}(p)=\frac{1}{4\pi}\left[\lambda_{A\bar{l}l}^{(1)}\right]^{2}(p)\,,\quad \alpha_{A\bar{s}s}=\frac{1}{4\pi}\left[\lambda_{A\bar{s}s}^{(1)}\right]^{2}(p)\,. 
    \label{eq:StrongCouplingqqA}
\end{align}
These avatars complement that obtained in the pure glue sector, see \labelcref{eq:Initialalphaglue}. 

Functional QCD admits a highly modular structure, the modules being the pure glue, the glue-matter interface and the pure matter sectors. These modules are only connected by a few diagrammatic interfaces, and this structure, allows for very reliable systematic error estimates and the apparent convergence of the results. This is called the \LEGO principle, for a detailed analysis see \cite{Ihssen:2024miv}. The quark-gluon vertex takes a special rôle among these interfaces and we close this Section with a thorough analysis of this interface and its stability. To begin with, 
the feedback of the quark-gluon and quark fluctuations in the glue-matter interface and the pure matter sectors into the pure glue sector are transmitted by closed quark loops with external gluon lines. The respective interface vertices are the quark-gluon vertex and higher two-quark--$n$-gluon vertices. It has been already checked in \cite{Mitter:2014wpa, Cyrol:2017ewj, Fu:2019hdw, Ihssen:2024miv} that this feedback is rather insensitive to the infrared details for $p\lesssim 5$\,GeV of the approximation of the interface action. The current analysis corroborates these findings. In turn, quark correlation functions and in particular the quark mass function are rather sensitive to changes in the quark-gluon vertex for $p\lesssim 5$\,GeV. 

Moreover, while the inclusion of $\lambda_{A\bar q q}^{(4,7)}$ is relevant for the gap equation, their inclusion into the flows of the quark-gluon vertex dressings $\lambda_{A\bar q q}^{(1,4,7)}$ is not. Consequently, it is solely the dressing $\lambda^{(1)}_{A\bar q q}$ of the classical tensor structure, which governs dynamical chiral symmetry breaking and  the respective systematic error analysis. 

In the present work we use the STI-procedure introduced in \cite{Cyrol:2017ewj} for achieving quantitative accuracy and gauge consistency: for cutoff scales $k\gtrsim 5$\,GeV, 
\begin{align}
    \alpha_{A\bar{l}l,k}(p)=\alpha_{A\bar{s}s,k}(p)=\alpha_{A\bar{c}c,k}(p)\,, \qquad k\geq \Lambda_\textrm{STI}\,.
    \label{eq:STI-constraint}
\end{align}
In \labelcref{eq:STI-constraint}, the lower bound for the STI construction is varied in the regime 
\begin{align}
    3\,\textrm{Gev} \leq \Lambda_\textrm{STI} \leq 7\,\textrm{GeV}\,.  
    \label{eq:LambdaSTI}
\end{align}
For smaller cutoff scales, $k<\Lambda_\textrm{STI}$ we use the flow equation for the dressings. This flow captures the decoupling of the quark-gluon dynamics due to the gluon mass gap and the quark constituent masses. This procedure quantitatively captures both the gauge-consistent running in the perturbative and semi-perturbative regimes as well as the dynamic decoupling of the quark-gluon dynamics in the infrared. The choice of a very large regime for $\Lambda_\textrm{STI}$ also allows for a conservative systematic error estimate. 

Finally we remark, that the above STI-procedure can be dropped if implementing flows that accommodate the full mSTIs for the vertex dressings and in particular the quark wave function $Z_q(p)$, for a recent discussion see \cite{Ihssen:2024miv}: the regulator insertions lead to a sizable flow of $Z_{q,k}(p)$ for $k\gtrsim 5$\,GeV while the wave function $Z_{q,k=0}(p)\approx 1$. The respective flows and initial conditions are intertwined, leaving us with a logarithmic fine-tuning problem whose resolution will be considered elsewhere. 

In summary, the STI-procedure implemented in the current work, together with the conservative systematic error estimate, captures the quark-gluon dynamics well, and gives us a quantitative grip on dynamical chiral symmetry breaking. With this Section and \Cref{sec:GammaGlue} we also have collected all relevant parameters in QCD: the strong coupling $g_s$, \labelcref{eq:Initialalphaglue,eq:STI-constraint},  and the current quark asses $m_l,m_s$, \labelcref{eq:CurrentQuarkMassesLambda}. For the sake of convenience we collect them here, 
\begin{align}
\alpha_{i,\Lambda}(p) = \alpha_{s,\Lambda}\,,\qquad  M_{q,\Lambda}(p)=(m_l,m_l,m_s)\,,
\label{eq:InitialConditions}
\end{align}
with $i= A^3\,,\, A^4\,,\, A\bar c c\,,\,A\bar l l,A \bar s s$. As already discussed below \labelcref{eq:Initialalphaglue}, the initial condition $\alpha_{s,\Lambda}$ for the strong coupling merely defines $\Lambda$ in terms of physical scales, and it is the current quark masses that are tuned for the physical ratios of the pion and kaon mass with the pion decay constant. This is explicitly done in \Cref{sec:InitialConditions}.

\subsection{Pure matter sector}
\label{sec:Gamma4q}

The third sector of the effective action \labelcref{eq:QCDEffAction} is the pure matter sector. It contains all multi-quark scatterings, the lowest order being the four-quark interactions. 
This sector has been thoroughly analysed in \cite{Fu:2022uow, Fu:2024ysj} for 2+1 flavour QCD, and we shall simply use the results obtained there. 
For a comprehensive analysis we refer to these works and only briefly summarise the important findings, concentrating on the investigation of the relevance order of different terms. 
This analysis also took into account previous analyses, and in particular those in the fRG approach to QCD with emergent composites in \cite{Braun:2014ata, Mitter:2014wpa, Rennecke:2015eba, Cyrol:2017ewj, Fu:2019hdw, Ihssen:2024miv}. 
In these works the relevance of higher order scatterings in the scalar-pseudoscalar channel was evaluated. The structure of these terms suggests a rapid drop in their relevance for physical pion masses and larger ones. 
The numerical findings confirmed impressively these structural arguments, for a comprehensive analysis see \cite{Ihssen:2024miv}. We emphasise that the higher order scatterings in the scalar-pseudoscalar channel still lead to subleading contributions and based on the analysis in \cite{Ihssen:2024miv} this informs our total systematic error estimate of 10$\%$ discussed in \Cref{sec:Wrapup+SystematicError}.  Moreover, a by-product of these findings is the complete irrelevance of four-quark channels with sufficiently massive lowest lying resonances with $m_\textrm{res} \gtrsim 500$\,MeV. One can show readily that even the contributions of the four-quark scattering vertices in these channels to low-order correlation functions (two-, three-, four-point function) are irrelevant, and their feedback to higher order ones is even more suppressed.    

In summary this leads us to the four-quark approximation $\Gamma_{4q,k}$ for the complete pure matter sector of the QCD effective action. As already mentioned above, this approximation has been discussed judiciously in \cite{Fu:2022uow, Fu:2024ysj}, in particular in view of dynamical chiral symmetry breaking and the emergence of the bound states in low energy regime. We use  
\begin{align}\nonumber 
 \Gamma_{4q,k}[q,\bar q] =&-\int\prod_{i=1}^4\left[ \frac{d^4 p_i}{(2 \pi)^4}Z_{q}^{\frac{1}{2}}(p_{i})\right] \, (2 \pi)^4\delta\left(\sum_{i=1}^4 p_i\right)\\[2ex]
 & \hspace{-1cm}\times\lambda_{\alpha}(\boldsymbol{p}) \,
 {\mathcal{T}}^{(\alpha)}_{ijkm}(\boldsymbol{p})\,\bar q_i(p_1) q_j(p_2) \bar q_k(p_3) q_m(p_4)\,, 
 \label{eq:Gamma4q}
\end{align}
with 
\begin{align}
    \boldsymbol{p}=(p_1,p_2,p_3,p_4)\,.
    \label{eq:boldp}
\end{align}
where $\lambda_{\alpha}(\boldsymbol{p})$ is the respective dressing of the four-quark vertex. At the lowest momentum-independent order, the set of Fierz-complete basis $\mathcal{T}^{(\alpha)}$ includes 10 tensors in $N_{f}=2$ QCD. In the case of $N_{f}=2+1$, because of the flavor symmetry breaking between the light and strange quarks, the number of the basis tensors is increased to 26, see the details in \Cref{app:truncation-matter}. The complete basis with momentum-dependent tensors is far bigger, for the two flavour case see \cite{Eichmann:2011vu}. 
Moreover, the tensors have symmetric and anti-symmetric parts with anti-symmetric and symmetric dressings, that is $\lambda^\mp_\alpha(\boldsymbol{p}) \,\mathcal{T}^{(\alpha\pm)}$, see \labelcref{eq:Gamma4q-final} and the respective discussion in \Cref{app:truncation-matter}.   
Consequently, we consider 52 terms, related to the 26 Fierz-complete momentum-independent tensors, thus generalising the computation in \cite{Fu:2024ysj} for 2 flavours to 2+1 flavours. 

While it is possible to solve the system within this approximation, it can be further reduced without the loss of quantitative precision. Here we draw in particular from the results in \cite{Fu:2024ysj}, where it has been shown in the two-flavour setting that the scalar and pseudoscalar channels are the only relevant ones, and the contributions of all other channels add up to less than 1\% for the strength of chiral symmetry breaking, measured in observables such as the chiral condensate.  The suppression of the contribution of the other channels relates to their large 'resonance' masses with $m_\textrm{res} > 500$\,MeV. This leads to a very efficient suppression of these channels. In the present work in 2+1 flavour QCD we use this analysis and only consider the channels with $m_\textrm{res} \lesssim 500$\,MeV. Accordingly, we include the $(\sigma,\pi)$-channels and the  $(\kappa, K)$ channels, the latter being the scalar and pseudoscalar channels of the light-strange quark interaction respectively. While encoding off-shell scattering processes, the pole position of the dressings ($t$-channels) provide the pole masses of the mesons with the respective quantum numbers. 

%
\begin{figure}[t]
	\includegraphics[width=0.3\textwidth]{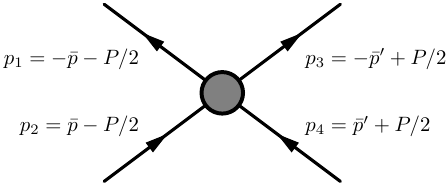}
	\caption{Diagrammatic representation of the quark four-point function, where all momenta are counted as incoming.}
	\label{fig:G4quark-Mom}
\end{figure}
%

In \Cref{app:truncation-matter} this has been discussed in more detail.  The tensor structures of the $(\sigma,\pi,\kappa, K)$-channels taken into account here are combination of tensors in the complete basis in \Cref{app:truncation-matter}, see \labelcref{eq:4qBasis-sigmaPiKkappa}. This reduces the size of the contributions from the other tensor structures.  \Cref{eq:4qBasis-sigmaPiKkappa} take the explicit form 
\begin{align}\nonumber 
 \mathcal{T}^{\sigma}_{ijkm}{\bar{q}}_iq_j{\bar{q}}_kq_m=&(\bar{q}\,T^0 q)^2\,,\\[2ex]\nonumber 
 \mathcal{T}^{\pi}_{ijkm}{\bar{q}}_iq_j{\bar{q}}_kq_m=&-(\bar{q}\,\gamma_5 T^{(1-3)} q)^2\,,\\[2ex]\nonumber 
 \mathcal{T}^{\kappa}_{ijkm}{\bar{q}}_iq_j{\bar{q}}_kq_m=&(\bar{q}\,T^{(4-7)} q)^2\,,\\[2ex] 
 \mathcal{T}^{K}_{ijkm}{\bar{q}}_iq_j{\bar{q}}_kq_m=&-(\bar{q}\,\gamma_5 T^{(4-7)} q)^2\,, 
 \label{eq:4Tensors2+1}
\end{align}
where the generators $T^{(i)}$ are the respective Gell-Mann matrices, and the quark field $q$ is given by $q=(l,s)$ with the light quark field $l=(u,d)$. 

Finally we reduce the full momentum dependence of the dressings $\lambda^{(\alpha)}(\boldsymbol{p})$, while still keeping the quantitative nature of the approximation. This task has been evaluated in  \cite{Fu:2024ysj} and informs the present approximation. We briefly recall the arguments and analysis done there: Specifically we want to keep the quantitative access to hadron resonances which are encoded in the resonant behaviour of the respective momentum channel of a given four-quark tensor that carries the same quantum numbers as the resonance under investigation. Here, momentum channel refers to the Mandelstam momenta $s,t,u$ of the four-quark vertex depicted in \Cref{fig:G4quark-Mom}. We define 
\begin{align}
    s=(\bar{p}+\bar{p}')^2,\quad t=P^{2},\quad u=(\bar{p}-\bar{p}')^2\,,
    \label{eq:stu-genenral}
\end{align}
with 
\begin{align}
    P=-(p_{1}+p_{2}),\quad\bar{p}=\frac{p_{2}-p_{1}}{2},\quad\bar{p}'=\frac{p_{4}-p_{3}}{2}\,.
    \label{eq:Pp-p1234}
\end{align}
Then, the full momentum dependence of dressings of the four-quark vertex $\lambda_{\alpha}(p_{1},p_{2},p_{3},p_{4})$ is reduced to a $s,t,u$-dependence, i.e.,
\begin{align}
    \lambda_{\alpha}(p_{1},p_{2},p_{3},p_{4})=&\lambda_{\alpha}(s,t,u)+\Delta\lambda_{\alpha}(p_{1},p_{2},p_{3},p_{4})\nonumber\\[2ex]
    \approx & \lambda_{\alpha}(s,t,u)\,,
\end{align}
where the contribution of $\Delta\lambda_{\alpha}$ is found to be smaller than 1.5$\%$ \cite{Fu:2024ysj}. We extract the $s,t,u$-channel momentum dependences from the configuration 
\begin{align}
    P_\mu=&\sqrt{P^{2}}\,\Big(1,\,0,\,0,\,0 \Big)\,,\nonumber\\[2ex]
    \bar p_\mu=&\sqrt{p^{2}}\, \Big(1,\, 0,\,0,\,0 \Big)\,,\nonumber\\[2ex]
    \bar p^\prime_\mu=&\sqrt{p^{2}}\, \Big(\cos \theta,\, \sin \theta,\,0,\,0 \Big)\,,
    \label{eq:4quark-mom}
\end{align}
This choice is not unique and the ambiguity from choosing different momentum configurations has been studied in \cite{Fu:2024ysj}: it turns out that this ambiguity amounts to a less than 1\textperthousand variation of the result. Accordingly, the respective error is negligible.  

This concludes the discussion of the pure matter sector of the effective action.

\section{Systematic error control and apparent convergence}
\label{sec:Wrapup+SystematicError}

In this Section we present an analysis of the apparent convergence of the present approximation introduced in \Cref{sec:QCD-EffAction}, based on an assessment of the systematic error estimate. While chiefly important for the assessment of the quantitative reliability of the current functional QCD approach, it requires some basic technical understanding of functional approaches. The technically less proficient or interested reader may skip this Section in a first reading.  

\begin{figure}[t]
\includegraphics[width=0.48\textwidth]{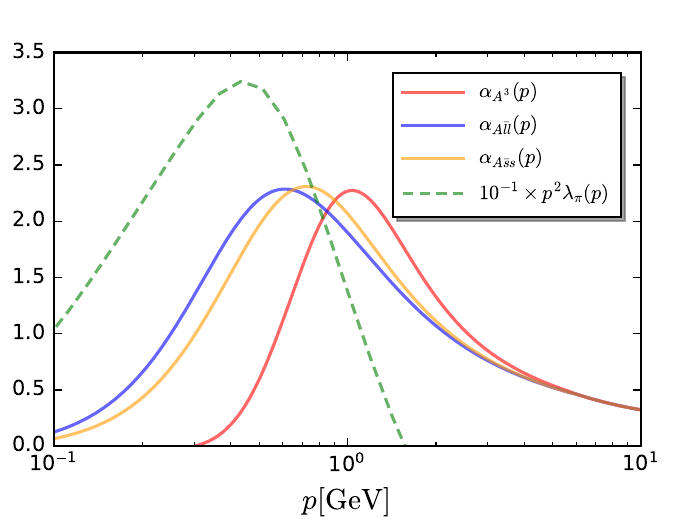}
\caption{Relevant interface couplings: Glue and quark-gluon avatars $\alpha_i$ of the strong coupling, where $i=A^3,A\bar l l,A\bar s s$, see also \Cref{fig:alpha} and $10^{-1} p^2 \lambda_\pi(p)$, see also \Cref{fig:lambda4q}. All avatars of the strong couplings are evaluated at the symmetric point. The pion four-quark coupling $\lambda_\phi(p)=\lambda_\phi(0,t=p^2,0)$ is evaluated in  the $t$-momentum channel. For better visibility we have multiplied it with $10^{-1}$. The avatars of the strong couplings peak in the regime $p_\textrm{peak} \approx 1$\,GeV and tend towards zero for $p\lesssim 0.5$\,GeV. The dimensionless pseudoscalar four-quark coupling $p^2\,\lambda_\pi(p)$ peaks at about $p_\textrm{peak} \approx 0.5$\,GeV and tends towards zero for $p\lesssim m_\pi$. It also vanishes rapidly in the ultraviolet for $p\gtrsim 1$\,GeV proportional to $\alpha_s^2(p)$. }
\label{fig:InterfaceCouplings}
\end{figure}

In \Cref{sec:GammaGlue,sec:GammaInterface,sec:Gamma4q} we have introduced and discussed the approximations of the different parts of the effective action, see \labelcref{eq:GammaGlue,eq:GammaInter,eq:Gamma4q} and the respective \Cref{app:truncation-gauge,app:truncation-qqA,app:truncation-matter}. 
These three parts constitute well-defined modules of QCD in terms of general expansion schemes in functional approaches as discussed in \cite{Ihssen:2024miv, Lu:2023mkn}. The three modules are connected via small diagrammatic interfaces which enable us to discuss the convergence of the approximations of the modules separately. The most important interface coupling is summarised in \Cref{fig:InterfaceCouplings}.
In particular this allows us to take advantages of convergence and stability analyses done within these sub-systems in the literature.  
Moreover, the stability of the whole system can be assessed by evaluating the stability of the different modules under changes of the interfaces. 
This is called the \LEGO principle and is discussed in detail in \cite{Ihssen:2024miv}. 

Following the \LEGO principle, we assess the stability of the pure glue, glue-matter interface and pure matter sectors of the full system in  \Cref{sec:StabilityPureGlue,sec:StabilityInterface,sec:Stability4q}, leading to the combined error estimate in \Cref{sec:StabilityAll}. We close this Section with \Cref{sec:DiagrammticDescription}, which comprises a structural comparison of the diagrammatic resummations included here with that in standard approximations to bound state computations with DSE, BSE and Faddeev equations. 

%
%

\subsection{Stability of the pure glue sector}
\label{sec:StabilityPureGlue}

The vertex expansion scheme and different approximations of the pure glue sector have been studied in detail in \cite{Cyrol:2016tym}. The best approximation studied there surpasses the one used here in terms of full momentum dependences of vertices. We have resorted to the symmetric point approximation for the gluonic and ghots-gluon vertices in the present work as the detailed analysis in \cite{Cyrol:2016tym} has shown their quantitative reliability. For the details of this analysis we refer the reader to this work. 

It is left to discuss the impact of the interface of the pure glue sector to the matter sector. To that end we evaluate the impact of \textit{large} changes of this interface on the pure glue correlation functions. Here, 'large' accommodates very conservatively potential deficiencies of the interface and the matter sector. This analysis is simplified by the fact that the pure glue sector of the action \labelcref{eq:GammaGlue,eq:GammaGlueDetailed} is connected to the pure matter sector only via the interface action discussed in \Cref{sec:GammaInterface} with its quark-gluon vertices. In short, fluctuations of the matter sector enter the flows for the ghost, gluon and ghost-gluon correlation functions only via closed quark loops with external gluon and ghost lines. We have completely dropped the latter interface couplings as quarks and ghosts only interact via diagrams with internal gluons, quarks and ghosts: These diagrams are highly subdominant in the perturbative regime as they require high orders in the strong coupling, and decouple quickly in the infrared due to the efficient decoupling of the gluon dynamics. In contradistinction to the resonant chiral dynamics they are also suppressed in the momentum regime with momentum scales $p,k\approx 1$\,GeV. 

The pure quark loops with external gluons effectively only change the momentum scale running of the gluonic correlation functions from that in Yang-Mills theory to that in QCD. It has been checked already in \cite{Mitter:2014wpa, Cyrol:2017ewj} that even large changes in the quark-gluon dressings in the semi-perturbative and infrared regime only have a subdominant effect on the gluonic correlation functions. 

In short, the pure glue sector of the action is well converged and is very stable under potential even large changes of the interface as well as the matter sector.

\subsection{Stability of the glue-matter interface}
\label{sec:StabilityInterface} 

The stability and convergence of the glue-matter interface have been evaluated in great detail in \cite{Mitter:2014wpa, Cyrol:2017ewj, Ihssen:2024miv} in the fRG approach and in \cite{Williams:2014iea, Williams:2015cvx, Gao:2021wun, Gao:2020qsj, Lu:2023mkn, Aguilar:2024ciu} in the DSE approach. 
There it has been found that only three of the eight transversal tensor structure of the quark-gluon vertex, namely $\lambda^{(1,4,7)}_{A\bar q q}$, are relevant for quantitative accuracy, and full quantitative precision requires taking into account a few higher order scatterings. 
Still, the by far dominant contribution is hosted by the dressing of the classical tensor structure of the quark-gluon vertex, and the results for the correlation functions quickly converge. The latter fact is also related to the fact that the pure glue correlation functions are rather insensitive to the respective changes as well as to the presence or absence of chiral symmetry breaking. 

The above entails that the largest systematic error in the glue-matter interface originates in the treatment of the flow of the dressing $\lambda^{(1)}_{A\bar q q}(p)$ of the classical tensor structure of the quark-gluon vertex. While it is well controlled by the mSTI in the perturbative and semi-perturbative regime, its (integrated) flow in the onset regime of confinement has a relatively large systematic error, see the discussion in \Cref{sec:GammaInterface} around \labelcref{eq:LambdaSTI}. This error is accounted for with a very conservative evaluation of this interface within the regime $3\,\textrm{GeV} \lesssim k\lesssim 7$\,GeV, \labelcref{eq:LambdaSTI}. In summary this leads to a systematic error of approximately 5\% in the most sensitive correlation function, the quark mass functions, while further correlation functions are less sensitive to these variations.

\subsection{Stability of the pure matter sector}
\label{sec:Stability4q}

This leaves us with an evaluation of the systematic error for the pure matter sector of the effective action. This sector has been studied thoroughly in the first two works in the series \cite{Fu:2022uow, Fu:2024ysj}. The results there informed the approximation of the four-quark sector of the effective action \labelcref{eq:Gamma4q}, and a detailed analysis is found there. It is left to evaluate the potential impact of higher order quark vertices, such as $(\bar q q)^n$ and $(\bar q \gamma_5 T^i q)^n$ with $i=1,2,3$ and $n\geq 3$. As in the case of the four-quark interactions, there is a clear hierarchy with the pseudoscalar ones being by far the dominant ones. Their impact has been thoroughly studied in the fRG approach to QCD with emergent composites in \cite{Braun:2014ata, Rennecke:2015eba, Cyrol:2017ewj, Fu:2019hdw} and in particular in the recent work \cite{Ihssen:2024miv}. In short, their relevance drops quickly with $n$, the exception being the chiral limit with pion masses $m_\pi \lesssim 1$\, MeV, see \cite{Braun:2023qak}. Still, $n=3,4$ leads to sizeable though subdominant corrections below $5\%$, for a discussion see the works above. In a forthcoming work we shall augment the current approximation with emergent composites in the scalar-pseudoscalar sector. This eliminates the biggest source of our systematic error as well as gives us more direct and simpler access to hadronic bound state properties and timelike correlation functions.

\subsection{Combined systematic error estimate}
\label{sec:StabilityAll}

We conclude that the by far dominating sources for the systematic error are the lack of full control of the flow of $\lambda^{(1)}_{A\bar q q}$ discussed in \Cref{sec:StabilityInterface} and the missing higher order quark--anti-quark scattering terms discussed in \Cref{sec:Stability4q}. Both of them are of the order $5\%$. Moreover, they are partially accommodated in the value of the current quark masses leading to the physical ratio of the pion decay constant and the pion/kaon pole masses. In summary we are led to a conservative systematic error estimate of $10\%$ for our observables.

\subsection{Dynamics and higher order scatterings}
\label{sec:DiagrammticDescription}

We close the discussion of the functional 2+1 flavour QCD setup in \Cref{sec:QCD-EffAction} and its systematic error estimate in \Cref{sec:StabilityPureGlue,sec:StabilityInterface,sec:Stability4q,sec:StabilityAll} with a brief evaluation of the dynamics included in the present approximation. For further analyses of the present and related approximations we refer the reader to \cite{Cyrol:2017ewj, Fu:2019hdw, Ihssen:2024miv} and the reviews \cite{Dupuis:2020fhh, Fu:2022gou}. 

Here we concentrate on a discussion in the light of another application areas of the present approach: 
the hadron resonance spectrum, the formation of hadrons, parton distributions functions and related observables. 
While part of this evaluation can also be found in \cite{Fu:2022uow, Fu:2024ysj}, we aim at a self-contained discussion here. 
To begin with, the most important part of the coupled set of flow equations is the four-quark flow and its feedback into the quark-gluon vertex and the quark propagator. 
In terms of (off-shell) resonances this includes $\sigma$-mode, pion, $K$ and $\kappa$ exchanges in the flow of the quark-gluon vertex. 

\begin{figure}[t]
\includegraphics[width=0.45\textwidth]{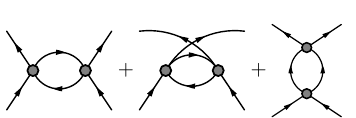}
\caption{Flow diagrams of four-quark vertices in the pure matter sector for the $t$-, $u$-, and $s$-momentum channels, respectively}
\label{fig:stu-equ}
\end{figure}

We proceed with a discussion of the off-shell scattering processes included in the present approximation of the matter sector: the four-quark vertex is sourced by all topologies of the quark-gluon boxes. 
Importantly, the resummation in the flow does not only generate ladder diagrams (gluon exchanges between a quark and an anti-quark), but also generic topologies including e.g. also the annihilation of a quark--anti-quark pair into two gluons, followed by the re-creation of such a pair from two-gluons, not necessarily in the same tensor channel. 
The general resummation is obtained from the flow: in the next flow step at a given scale $k$, all the topologies and diagrams generated before are fed back into the flow. Apart from successively dressing the vertices and propagators in the quark-gluon boxes, this iterative procedure generates triangle and fish diagrams as well as the tadpole diagrams with six-point vertices. Again it is important for the evaluation of the resummation scheme that all topologies of the fish and triangle diagrams are considered, see \Cref{fig:stu-equ}. Moreover, all four-quark vertices considered depend on all Mandelstam variables $s,t,u$. Diagrammatically this implies that all different momentum and tensor channels are fed back into each other. This rather complete, self-consistent, resummation of the four-quark vertex originates in the fact, that flow diagrams only contain fully dressed vertices in contradistinction to DSEs with one classical one and BSEs with a given BSE kernel. Note that the latter statement only concerns the standard formulation of the DSE-BSE framework and elaborated schemes have been set up that circumvent this diagrammatics, see e.g.~\cite{Huber:2018ned, Huber:2020keu} and the review \cite{Eichmann:2016yit}. For a recent example with the skeleton expansion, leading to diagrammatics similar to that in the fRG flows, see \cite{Gao:2024gdj}. 

%
\begin{figure}[t]
\includegraphics[width=0.48\textwidth]{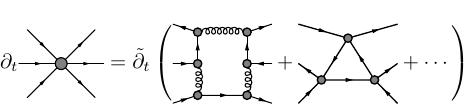}
\caption{Diagrammatic depiction of the flow equations of the six-quark vertex. For the sake of brevity we only depict one diagram of the whole class of quark-gluon and four-quark diagrams. The operator $\tilde \partial_t$ only hits the $k$-dependence in the explicit regulator terms in the propagators. }
\label{fig:QuarkGluon6Point}
\end{figure}

%
In conclusion, in terms of all orders of perturbative diagrams, such a self-consistent flow generates all nested diagrams of quark-gluon diagrams except those that are sourced by a quark-gluon six-point function at one-loop level. Its flow is schematically depicted in \Cref{fig:QuarkGluon6Point}, where we have left out diagrams with six-point and eight-point functions as well as mixed diagrams with quark four-point and quark-gluon vertices. 

We emphasise that this includes but goes far beyond rainbow-ladder diagrams. While a full analysis is beyond the scope of the present work, we collect some of the scattering processes that can be derived from the above diagrammatic dissection. First of all, all two-to-two four-quark scattering processes are included. Due to the iteration, this also includes two-to-many quark processes. Note that almost all of these processes are negligible as their resonance mass scale is very large $\gtrsim 1$\,GeV and these resonances emerge or get dynamical at about $k\approx 500$\,MeV, the scale of dynamic chiral symmetry breaking. Finally, in terms of two-to-two quark and higher order resonant interactions, the dropped tadpole diagrams comprise part of these interactions that, however, are phase-space suppressed. Moreover, at the physical point and for smaller pion masses these scatterings are dominated by multi-pion scatterings. These contributions, while present, have been shown to be subleading, see in particular \cite{Braun:2014ata, Mitter:2014wpa, Rennecke:2015eba, Cyrol:2017ewj, Fu:2019hdw} and in particular the recent comprehensive systematic error analysis in \cite{Ihssen:2024miv}. In summary, the error introduced by their omission is part of our 10\% error estimate and the multi-meson scatterings will be added in a forthcoming work that combines the approximations of \cite{Cyrol:2017ewj, Ihssen:2024miv} with the present one. 
%
%

We close this Section with a brief diagrammatic translation of this fRG setup with the Dyson-Schwinger--Bethe-Salpeter equation setup commonly used for bound state analyses. To begin with, the inclusion of the four-quark vertices in the fRG framework and the respective approximation in their flow corresponds to the inclusion of the BSE and the choice of the Bethe-Salpeter kernel. This relation is even more evident in the fRG approach with emergent composites. Then, the pion--quark--anti-quark (Yukawa) vertex simply reduces to the pion BSE wave function on the pion pole, see \cite{Mitter:2014wpa, Cyrol:2017ewj}. For further direct relations between the two setups we refer to works on $n$PI flows, see \cite{Pawlowski:2005xe, Blaizot:2010zx, Fu:2013sku, Carrington:2012ea, Carrington:2014lba, Carrington:2017lry, Carrington:2019fwp, Blaizot:2021ikl}. In both frameworks, a self-consistent treatment of the chiral dynamics is of utmost importance and is achieved by deriving the loop relations for the correlation functions from the functional relation of a suitably chosen effective action. 

In the  DSE-BSE setup self-consistent relations between the correlation functions can be derived from $n$-particle irreducible effective actions: then, the choice of the BSE-kernel also determines the approximation of the quark-gluon vertex in the gap equation. These choices are dictated by the $n$PI expansion scheme underlying the analysis. Such a treatment leads to a consistent implementation of chiral symmetry breaking and in particular ensures the consistency of the dynamics of the pseudo-Goldstone bosons. 

In the fRG approach, the flows for the correlation functions are derived from the functional flow equation of the effective action, see \labelcref{eq:FunFlow} in \Cref{app:Flow}. Accordingly, the consistency of the chiral dynamics is hard-wired in a \textit{self-consistent} solution of the connected set of flow equations for all vertices. Here, \textit{self-consistent} entails that all vertices considered in the approximation of the effective action, are fed-back into the diagrammatic flows on the right hand side. Roughly speaking, the BSE-kernel is encoded in the approximation to propagators and vertices, specifically the quark-gluon vertices. Hence, phrased in the DSE-BSE language, it is guaranteed that the same BSE-kernel is used throughout the system. The hard-wired self-consistency allows us to choose the approximations for all vertices and flows separately. Accordingly, we can implement the best approximation of each subsystem within the full system. This brief analysis concludes the discussion of the approximation of the effective action and the flow, including its relation to other functional approaches.

\section{QCD correlation functions}
\label{sec:QCDCorrelations}

In this Section we report on our results for the quark-gluon correlation functions in 2+1 flavour QCD. To our knowledge this is the first self-consistent computation of QCD correlation functions in 2+1 flavour QCD, i.e.~not resorting to any input for part of the correlation functions such as the ghost and gluon propagators as commonly done. These correlation functions will then be used for computing bound state properties of mesons such as the pion Bethe-Salpeter amplitude in \Cref{sec:meson-part}. 

In \Cref{sec:InitialConditions} we discuss how to adjust the UV initial conditions in our fRG approach to QCD. 
This is tantamount to fixing the renormalisation group point and conditions in Dyson-Schwinger equations (DSEs) in the MOM${}^2$ scheme underlying the fRG approach. For a discussion and comparison with the DSE approach see \cite{Gao:2021wun}. 
In 
\Cref{sec:CorrelationsGlue} we discuss the results for gluon and ghost propagators, and in \Cref{sec:CorrelationsInterface} we discuss the results for the quark dressings and the avatars of the strong coupling. In \Cref{sec:4qCorrelations} we present our results on the four-quark dressings, from which the meson masses, Bethe-Salpeter amplitude and the decay constants are derived. The respective results are discussed in \Cref{sec:meson-part}.

\subsection{Initial conditions for functional QCD}
\label{sec:InitialConditions}

In this Section we describe the tuning of ultraviolet parameters \labelcref{eq:InitialConditions} in the fRG approach to QCD. This starts with choosing a small initial strong coupling at the initial cutoff scale $k=\Lambda$, 
\begin{align}
    \alpha_{s,\Lambda}= 0.179\,. 
    \label{eq:alphasLambda}
\end{align}
Implicitly this provides the value of the initial cutoff scale, when the physical scales are fixed by tuning appropriate mass ratios in the theory. However, in the absence of any physical scale all scales are measured in the initial scale. Hence, apart from \labelcref{eq:alphasLambda}, the theory is determined by the two current quark masses, measured in the initial scale. We have scanned the parameter range   
\begin{align}
0\leq \frac{m_l}{\Lambda} \leq 9.1\times 10^{-5}\,, \qquad 0 \leq \frac{m_s}{\Lambda}\leq 2.4\times 10^{-3} \,.
\label{eq:mCurrentLambdaRange} 
\end{align}
The three parameters in \labelcref{eq:alphasLambda,eq:mCurrentLambdaRange} are the fundamental parameters in 2+1 flavour QCD. We emphasise that these are the only parameters at our disposal in the current first-principles fRG approach to QCD. No external input or phenomenological parameter is used. Moreover, as elucidated above, the strong coupling is not tuned, its initial value simply sets the physical scale. Hence we fix the masses \labelcref{eq:mCurrentLambdaRange} with the two ratios $m_\pi/f_\pi$ and $m_K/f_\pi$, and all further observables and correlation functions are genuine predictions. 

\begin{figure}[b]
	\includegraphics[width=0.48\textwidth]{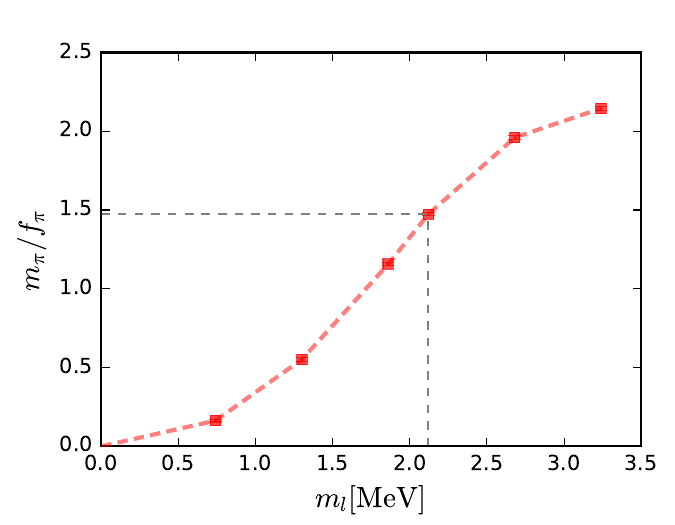}
	\caption{$m_{\pi}(m_l)/f_{\pi}(m_l)$ as a function of the light current quark mass $m_{l}$. The strange current quark mass is fixed to its physical value $m_s=55.9$\,MeV, see \Cref{tab:Parameters}. The gray dashed lines indicate the physical parameter values in \Cref{tab:Parameters}, leading to the physical value $m_{\pi}/f_{\pi}=137/93$.}
	\label{fig:ml-mpi/fpi}
\end{figure}

\begin{table}[t]
  \begin{center}
  \begin{tabular}{|c|c||c|c|}
    \hline & & & \\[-2ex]
    Observables  & Value & Parameters & Value  \\[1ex]
    \hline & & &   \\[-2ex]
    $m_{\pi}/f_{\pi}$ & \, 137.0/93.0  \, & $m_{l}$ & \, 2.1 MeV \, \\[1ex]
    \hline & & & \\[-2ex] 
    $m_{K}/f_{\pi}$ & 494.0/93.0 & $m_{s}$ & 55.9 MeV \\[1ex]
    \hline & & & \\[-2ex] 
    \hline & & & \\[-2ex] 
    $M_{l}$ & 344.5 MeV &  &  \\[1ex]
    \hline & & & \\[-2ex] 
    $M_{s}$ & 487.3 MeV &  &  \\[1ex]
    \hline & & & \\[-2ex] 
    $m_{\sigma}$ & 515.2 MeV &  &  \\[1ex]
    \hline & & & \\[-2ex] 
    $f_{K}$ & 114.1 MeV &  &  \\[1ex]
    \hline
  \end{tabular}
  \caption{Current quark masses and observables in 2+1 flavour QCD: we choose a small strong coupling $\alpha_{s,\Lambda}=0.179$ at the initial cutoff scale $\Lambda$, see \labelcref{eq:alphasLambda}. This choice is safely in the perturbative regime, and the respective ultraviolet (UV) cutoff turns out to be $\Lambda=35.7\,\mathrm{GeV}$, see \labelcref{eq:LambdaPhys}. The light and strange current quark masses $m_l,m_s$ are the only fundamental parameters and are determined by the two ratios of meson pole masses and the pion decay constant, $m_\pi/f_{\pi}$ and $m_K/f_{\pi}$. Both, the decay constant and the pole masses are determined from the QCD correlation functions computed in \Cref{sec:QCDCorrelations}. The decay constants are determined with \labelcref{eq:def-fpifK}. The pole masses $m_{\pi}$ and $m_{K}$ are reconstructed from the four-quark dressings within an Pad\'e approximation, see \Cref{fig:InverseLambdaPoleMasses}. We also show some first predictions on the constituent quark masses $M_q=M_q(p=0)$, see \Cref{fig:Quark}, the pole mass of the $\sigma$-mode, see \Cref{fig:lambda4q} and the kaon decay constant \labelcref{eq:def-fpifK}.}
  \label{tab:Parameters}
  \end{center}\vspace{-0.5cm}
\end{table}
%
These ratios are fixed as follows. The QCD correlation functions computed in \Cref{sec:CorrelationsGlue,sec:CorrelationsInterface} can be used to compute meson masses, see \Cref{fig:InverseLambdaPoleMasses} in \Cref{sec:4qCorrelations} and the discussions therein, and the decay constants, see \labelcref{eq:Pi+K-DecayConstants} in \Cref{sec:meson-part}. 
This allows us to determine the physical choice for the initial condition as that, leading to the physical ratios $m_\pi/f_\pi$ and $m_K/f_\pi$, see \Cref{tab:Parameters}. In \Cref{fig:ml-mpi/fpi} we illustrate the respective tuning of the current quark masses with that of tuning $m_{l,\Lambda}=m_l$ to its physical value, while keeping $m_{s,\Lambda}$ already fixed at its physical value, $m_{s,\Lambda}=55.9$\,MeV: the ratio $m_\pi(m_l)/f_\pi(m_l)$ rises monotonously from its vanishing value in the chiral limit with $m_l=0$ and $m_\pi=0$ to its physical one. Note that the full procedure amounts to the tuning of the pair $m_l,m_s$, and this two-dimensional search is simplified by the mass hierarchy and the monotonicity in both masses. 

%
%
%
\begin{figure*}[t]
	\includegraphics[width=0.48\textwidth]{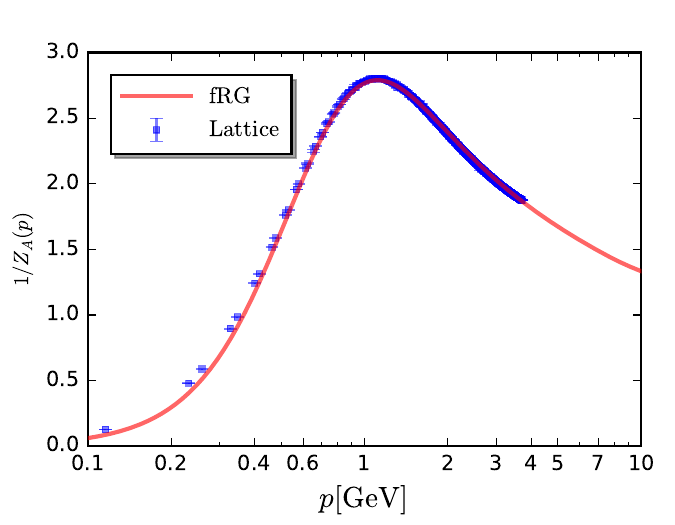}\hspace{0.5cm}
	\includegraphics[width=0.48\textwidth]{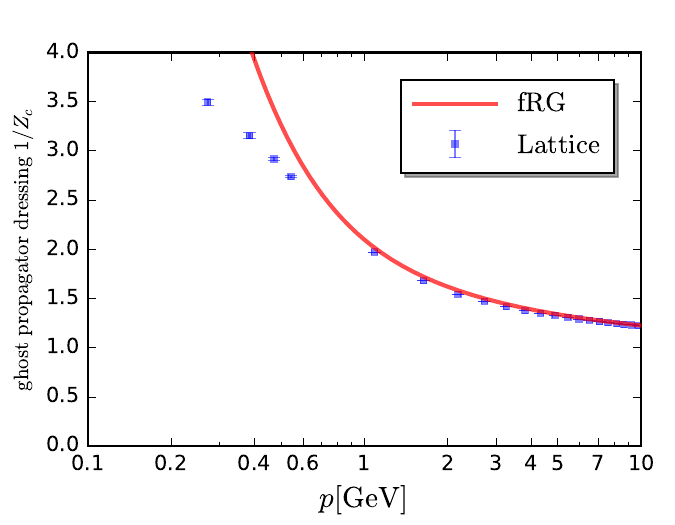}
	\caption{Left panel: Gluon propagator dressing $1/Z_{A}(p)$ in the Landau gauge in comparison to the unquenched lattice QCD results for $N_{f}=2+1$ flavours \cite{Boucaud:2018xup}. Right panel: Ghost propagator dressing $1/Z_{c}(p)$ in the Landau gauge in comparison to the lattice results for $N_{f}=2+1$ flavours \cite{Boucaud:2018xup}. }
	\label{fig:YM-twopoint}
\end{figure*}
%

In terms of the initial cutoff scale, it is given by  
\begin{align}
\frac{m_l}{\Lambda} \approx  5.88 \times 10^{-5} \,, \qquad \frac{m_s}{\Lambda} \approx  1.57 \times  10^{-3} \,. 
\label{eq:mCurrentLambda} 
\end{align}
which implies the ratio of the current quark masses,  
\begin{align}
\frac{m_s}{m_l} \approx 26.6\,,   
\label{eq:ratiomCurrent}
\end{align}
close to that used in lattice computations. Moreover, for \labelcref{eq:mCurrentLambda} the pion mass is the physical one, $m_\pi = 137$\,MeV, which finally sets our physics scales. Measured in these physical units, the initial cutoff scale is given by 
\begin{align} 
\Lambda= 35.7\,\textrm{GeV}\,. 
\label{eq:LambdaPhys}
\end{align}
The physical parameters as well as some first predictions of fundamental observables and correlation functions are also summarised in 
\Cref{tab:Parameters}.

\subsection{Correlation functions in the glue sector}
\label{sec:CorrelationsGlue}

We begin our discussion of the numerical results with the pure glue sector. As discussed in \Cref{sec:StabilityPureGlue}, it is the most insensitive of the modules when it comes to its stability under changes of correlation functions in the other modules including the interface, the quark-gluon scatterings. In terms of its relevance for the other modules, the interface and the matter sectors, the gluon propagator, or rather its dressing, is the most relevant correlation function. The gluon, ghost and quark propagators or dressings are also is the only correlation function, where precision data from the lattice exist and provide benchmark tests within the 10\% systematic error estimate discussed in \Cref{sec:StabilityAll}. 

It has been discussed at length in the literature, see \cite{Braun:2007bx, Fischer:2008uz, Fister:2013bh, Mitter:2014wpa, Cyrol:2016tym, Cyrol:2017ewj}, that the feedback of the gluon propagator into other correlation functions is via its dressing $1/Z_A(p)$ due to the momentum integral in the flows. This dressing in 2+1 flavour QCD is depicted in the left panel of \Cref{fig:YM-twopoint} together with the 2+1 flavour lattice data from \cite{Boucaud:2018xup}. Our result is in excellent agreement with the lattice data, well within the 10\% error estimate. The increasing deviation in the infrared comes from the tuning of the scaling solution in the glue sector instead of the decoupling solution found on the lattice, for a detailed discussion see \cite{Fischer:2008uz,Cyrol:2016tym}. It has been argued there that this difference may originate in an infrared gauge fixing ambiguity in the Landau gauge. Then, this difference does not show in observables. It has been shown in \cite{Braun:2007bx, Fister:2013bh, Cyrol:2017ewj} at the example of the quark mass function and the confinement-deconfinement temperature that this is indeed the case. 

This infrared difference to the lattice is also shown in the ghost correlation function shown in the right panel of \Cref{fig:YM-twopoint} together with the 2+1 flavour lattice data from \cite{Boucaud:2018xup}. This difference does not feed back into the rest of the system as it is transmitted via ghost-gluon diagrams, all of which contain at least one gluon propagator.  The scaling of ghost and gluon dressings (and hence of the vertices) are related and in combination they cancel out very efficiently. 
This is also evident in the quark-gluon avatars of the strong coupling, which is plotted together with the purely gluonic avatars in \labelcref{fig:alpha}. As they also involve correlation functions in the interface module of the effective action, we defer the respective discussion to \Cref{sec:CorrelationsInterface}.

In the present fRG approach to first principle QCD, the overall momentum scale is provided by $m_\pi=137$\,MeV for the physical ratio $m_\pi/f_\pi$. All other scales, such as the initial cutoff scale $\Lambda$, \labelcref{eq:LambdaPhys}, and the momentum scale in \Cref{fig:YM-twopoint} are measured in $m_\pi$. Accordingly, the excellent agreement of the ghost and gluon dressings with the lattice data for all momenta $p\gtrsim 1$\,GeV is an impressive test of the overall scale setting and the quantitative nature of the present approximation across the whole system: the scale setting is done in the pure matter sector of the system and the momentum scales of the ghost and gluon dressings is an 'observable' in the pure glue sector. We note that in particular the peak position of the gluon propagator, $p\sim 0.955$ GeV, which indicates the confinement mass gap, is in quantitative agreement with the peak position of the lattice data.

\subsection{Correlation functions in the quark-gluon interface}
\label{sec:CorrelationsInterface}

\begin{figure}[t]
\includegraphics[width=0.48\textwidth]{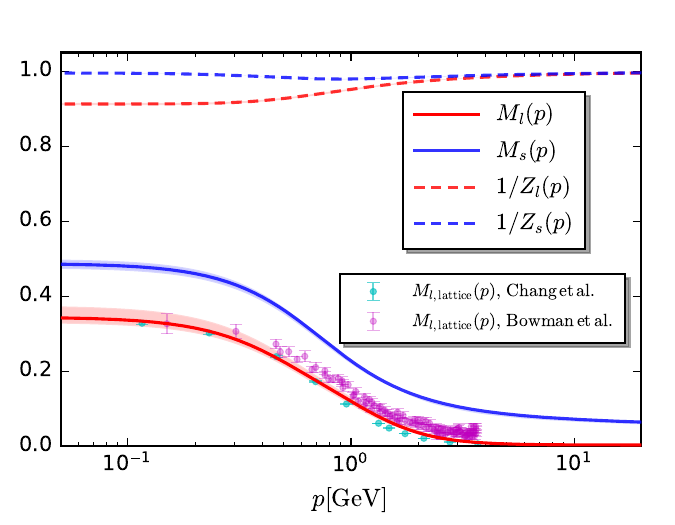}
\caption{Quark mass function $M_{q}(p)$ (solid lines, in unit of GeV) and inverse quark wave functions $1/Z_{q}(p)$ (dashed lines) for light (red line) and strange (blue line) quarks with $\Lambda_{\mathrm{STI}}=5\,\mathrm{GeV}$. The error bands stand for the results by varying $\Lambda_{\mathrm{STI}}$ from 3 to 7 GeV. The light quark mass function is also compared with the lattice results in \cite{Chang:2021vvx} (cyan points) and \cite{Bowman:2005vx} (purple points).}
\label{fig:Quark}
\end{figure}

The quark propagators, or more precisely their dressings, as well as that of the quark-gluon vertex are the most relevant building blocks in the interface module. The propagator, at $k=0$ has the form 
\begin{align}
    G_q(p) = \frac{1}{Z_q(p)} \frac{1}{\imag \slashed{p} +M_q(p)} \,,
    \label{eq:QuarkPropagator}
\end{align}
which is just the inverse of the kinetic term in \labelcref{eq:GammaInter}. It hosts the RG-invariant mass functions $M_q(p)=\textrm{diag}(M_l(p)\,,\,M_l(p)\,, \,M_s(p) )$ and the quark wave functions $Z_q(p)=\textrm{diag}(Z_l(p)\,,\,Z_l(p)\,,\, Z_s(p) )$. In \Cref{fig:Quark} we show both, $M_q(p)$ and $1/Z_q(p)$ for the physical point together with lattice data for the mass function from \cite{Chang:2021vvx}.  Note that the dressing $Z_q(p)$ is not RG-invariant and drops out of all our observables as both the vertices and the propagator have global $Z_q$-factors which cancel out in the diagrams. 

The RG-invariant mass functions agree well with the respective lattice results and the deviations are well within the 10\% systematic error of the current computation. In \Cref{fig:Quark} we also indicate the systematic error from the STI procedure applied to the quark gluon vertex. Even though we have taken a very large regime $3\,\textrm{GeV} \lesssim p \lesssim 7$\,GeV for the potential onset regime of confinement, the error is relatively small. This corroborates the quantitative nature of the current approximation.

%
\Cref{fig:Quark} also displays the inverse wave functions $1/Z_q$ for the light and strange quarks. The strange quark wave function is consistent with unity within the systematic errors of our calculations, while $1/Z_q$ for the light quark is significantly smaller than one, that is consistent with previous results of functional QCD, see \cite{Mitter:2014wpa, Williams:2014iea, Williams:2015cvx,  Aguilar:2016lbe, Cyrol:2017ewj, Gao:2021wun, Ihssen:2024miv}. Note that there are still considerable systematic errors in lattice results of quark wave functions, see \cite{Chang:2021vvx} for a discussion.

We close the discussion of the mass function with a dissection of its flow. The flow consists of the sum of diagrams in three diagrammatic classes, see \Cref{fig:twopoint-feyn}: the first class is the quark-gluon diagram with two quark-gluon vertices which dominates the flow in the UV. Due to the isospin symmetry of the propagator and vertices the internal quark propagator in these diagrams is that of the external quark. The second diagram is a quark-gluon tadpole whose contribution is negligible. The third diagram class are the quark tadpole whose contribution is as relevant for chiral symmetry breaking as the quark-gluon diagram. Due to the mixing of quark flavours in the four-quark vertices all quarks contribute to the two-point function flow of a given quark. 

In general, we expect the quark-gluon diagram to dominate the ultraviolet flow for $p\gtrsim 1\sim 5$\,GeV. While this also holds true here, all contributions approach zero quickly, as the flow of the mass function is proportional to itself: in the UV it is the only scale of chiral symmetry breaking, as the size of the chiral symmetry breaking vertices is negligible. Accordingly, due to the relatively small current quark masses the flows are small. In the infrared both contributions are rising and peak at about the cutoff scale of dynamical chiral symmetry breaking $k_\textrm{peak} \approx k_\chi$. 

%
\begin{figure}[t]
\includegraphics[width=0.48\textwidth]{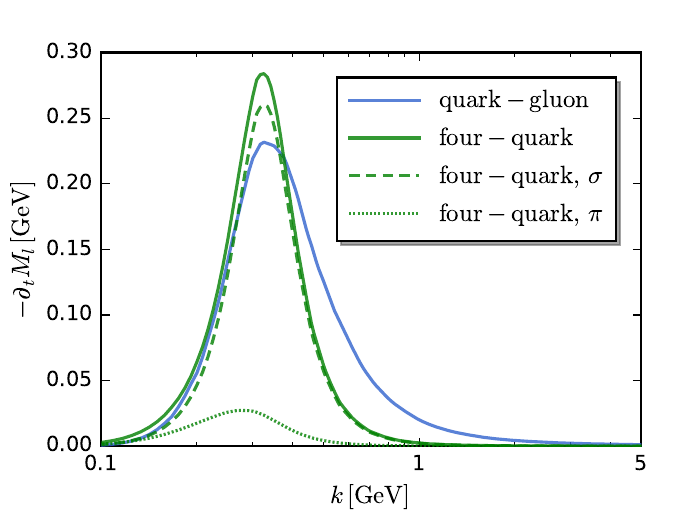}
\caption{Different diagrammatic contributions to the flow of the light quark mass at $p=0$.}\label{fig:dtMq-k}
\end{figure}
%
We also find that the quark tadpole is dominated by the contribution from the scalar $\sigma$-channel of the four-quark vertex, see \Cref{fig:dtMq-k}. A more detailed analysis elucidates the decomposition of the full QCD dynamics in both, the fRG approach to QCD with emergent composites and the present approach. A closer look at the $\sigma$-channel reveals that it hosts two contributions in contrast to one in the pion channel: both contain one diagram, where the external colour and flavour runs through the internal quark line. These two diagrams agree up to a relative minus sign for the pions and the multiplicity three. In short, this part is dominated by the pions. These contributions agree with the meson diagrams in the fRG approach to QCD with emergent composites. 

However, in the $\sigma$-channel there is an additional one proportional to $\tr T^0$, where colour and flavour of the internal propagator are not connected to the external colour and flavour. It is this diagram which dominates the contribution from the quark contribution. The respective diagram is absent in the pion channel as it is proportional to $\tr \gamma_5\, T^i=0$ and $\tr \gamma_5\, \slashed q\, T^i=0$ with the loop momentum $q$ and $i=1,2,3$. In the fRG approach to QCD with emergent composites this contribution is generated by the flow of the expectation value of the $\sigma$-mode, and the equation of motion of the $\sigma$-mode contains the respective closed quark-loop.  

We also emphasise, that the peak position $k_\textrm{peak}$ of the four-quark contributions stays at $k_\textrm{peak}\approx k_\chi$ also in the chiral limit. Finally, we note that the four-quark contribution is crucial for the size of the quark mass function and the inclusion of the four-quark scatterings, or rather the scalar and pseudoscalar part of it, is essential for the quantitative nature of the approximation. This concludes our detailed analysis of the different contributions in the flow of the quark mass function, and hence that of dynamical chiral symmetry breaking.

%
\begin{figure}[t]
\includegraphics[width=0.48\textwidth]{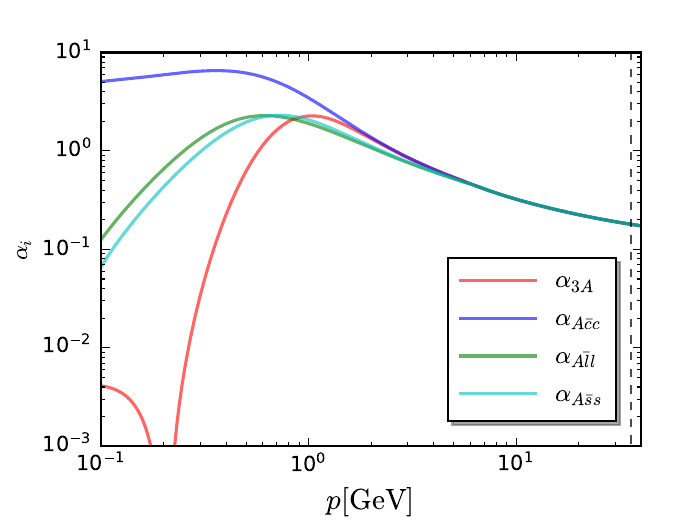}
\caption{Avatars of the strong coupling, $\alpha_{i}$ for $i=A\bar c c, A^3, A\bar l l, A\bar s s$ defined from the dressings of the ghost-gluon vertex, three-gluon vertex and quark-gluon vertices in \labelcref{eq:GhostGluonAvatars,eq:GluonAvatars,eq:StrongCouplingqqA}. 
All dressings converge towards each other for momenta $p\gtrsim 5$\,GeV due to the STI and start deviating below, see also \Cref{fig:InterfaceCouplings}. The dashed vertical line on the right indicates $p=\Lambda=35.7$\,GeV, see \labelcref{eq:LambdaPhys}. This momentum scale can be understood as a (soft) UV-momentum cutoff. } 
\label{fig:alpha}
\end{figure}
%

The avatars of the strong coupling from the ghost gluon, three gluon and quark-gluon vertices, \labelcref{eq:StrongCouplingGlueSector,eq:StrongCouplingqqA}, are shown in \Cref{fig:alpha}. 
For perturbative and semi-perturbative momenta, $p\gtrsim 3$\,GeV, all avatars of the strong coupling converge towards each other. 
This convergence is consistent with the respective STIs, which constrain the longitudinal parts of the respective classical tensor structures. 
For smaller momenta, $p\lesssim 3$\,GeV one has to consider two effects: 
First of all, with the growth of the strong couplings or rather the growing strength of the respective vertices, the scattering kernels in the STIs get increasingly important and the equivalence $\alpha_i(p)\approx \alpha_s(p)$ is successively lost. 
Moreover, confinement necessitates non-uniform vertices as otherwise the confining mass gap cannot be generated, see e.g.~\cite{Aguilar:2011xe, Cyrol:2016tym} and the recent review \cite{Ferreira:2025anh}. 
Here, non-uniform refers to the fact that longitudinal dressings (related to the STI) and transverse ones are different. 
Obviously, no conclusion can be inferred for the non-uniform parts of the transversal vertices from the STIs. 
While the non-trivial scattering kernels can be accommodated, the non-uniform nature of the vertices is a dynamical effect and is one of the reasons for using the flows of the dressing $\lambda^{(1)}_{A q \bar q}$ of the quark-gluon vertex in the infrared.  

In summary, for momenta $p\lesssim 3 $\,GeV all avatars of the strong coupling differ sizably, and these differences are dynamical and not a truncation artifact. 
Moreover, while the deviations are visible for $p\lesssim 3$\,GeV, they are already building up dynamically before in terms of the mass gap of the gluon propagator. 
This has been investigated thoroughly in pure Yang-Mills theory in \cite{Cyrol:2016tym} within the fRG approach and is well-understood. 
The underlying dynamical mechanism and its momentum dependence have been investigated judiciously in the literature, see \cite{Ferreira:2025anh}. 
These investigations inform our lower boundary of $\Lambda_\textrm{STI}$ in \labelcref{eq:LambdaSTI}, at which both, the scattering kernels in the STIs and the confining dynamics are already present.  

In the infrared for $p\lesssim 1$\,GeV, all avatars either level off (ghost-gluon) or decay towards zero. The latter is clearly seen for the avatars $\alpha_{A \bar l l}, \alpha_{A \bar s s}$ of the quark-gluon coupling. 
The strange avatar drops earlier due to the larger strange quark mass. 
For momenta at about 1 GeV the strange quark avatar is slightly larger than the light quark one, which is related to the respective size or order of the quark dressings $1/Z_l(p),1/Z_s(p)$, \Cref{fig:Quark}, which implicitly enter the definitions. 

The sign of the three-gluon dressing $\lambda_{A^3}$ changes at about 200\,MeV and approaches zero from below, see \Cref{fig:alpha}.  This behaviour is triggered by the ghost loop in the fRG or DSE for the three-gluon vertex which is dominating the deep infrared and leads to the negative infrared dressing. The location of the zero crossing depends on the chosen solution (scaling or decoupling), for further discussions and results  see e.g.~\cite{Pelaez:2013cpa, Aguilar:2013vaa, Eichmann:2014xya, Blum:2014gna, Cyrol:2016tym, Athenodorou:2016oyh, Barrios:2022hzr, Ferreira:2023fva, Huber:2018ned, Huber:2020keu} (Yang-Mills) and \cite{Cyrol:2017ewj, Ihssen:2024miv} (QCD). Accordingly, the respective strong coupling $\alpha_{A^3}$ in \labelcref{eq:StrongCouplingqqA} drops to zero at about 200\,MeV before recovering and approaching zero from below.  

\begin{figure}[t]
\includegraphics[width=0.48\textwidth]{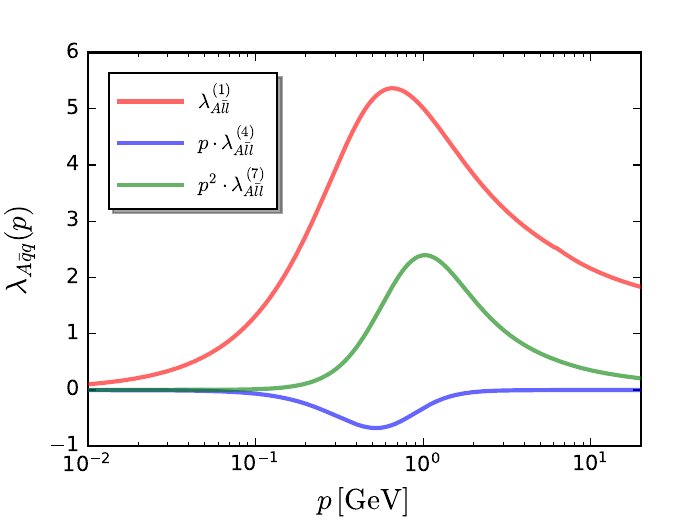}
\caption{Dimensionless dressings $\lambda_{A\bar l l}^{(1,4,7)}(p)$ of the quark-gluon vertex for the tensor structures ${\cal T}^{(1,4,7)}$ in \labelcref{eq:T147}, as functions of the symmetric point momentum $p$.}
\label{fig:qqA}
\end{figure}

We close this Section with the discussion of $\lambda_{A \bar l l}^{(4,7)}$, the dressings of the other two relevant tensor structures of the light-quark--gluon vertex. 
They are shown in \Cref{fig:qqA} together with that of the classical tensor structure, $\lambda^{(1)}_{A\bar l l}$. 
The full basis is presented in \labelcref{eq:qqA-tensor} in \Cref{app:truncation-qqA}, and for the benefit of the reader we recall the three relevant tensor structures, 
\begin{align}\nonumber 
   \left[\mathcal{T}^{(1)}_{A\bar{q}q}(p,q)\right]_{\mu}
    =&\mathrm{i}\gamma_{\mu}\,,\qquad 
  \left[\mathcal{T}^{(4)}_{A\bar{q}q}(p,q)\right]_{\mu}
  =\mathrm{i}\sigma_{\mu\alpha} p_{\alpha}\,,\\[2ex]
   \left[\mathcal{T}^{(7)}_{A\bar{q}q}(p,q)\right]_{\mu}
   =&\frac{{\textrm{1}}}{3}\,\sigma_{\alpha\beta}\gamma_\mu \left({p+q}\right)_\alpha (p-q)_\beta +\textrm{permut.}\,. 
\label{eq:T147}
\end{align}
Here, $p$ is the gluon momentum and $q$ is the anti-quark momentum, and all ${\cal T}^{(i)}_{A\bar q q}$ are 
multiplied by $\Pi^\perp(p) \, T^a_c$. 

%
%
The dimensionless dressings $\lambda_{A \bar l l}^{(4,7)}(p)$ peak in the regime $0.5\,\textrm{GeV} \lesssim p\lesssim 1$\,GeV. 
While it is suggestive to measure the importance of the dimensionless dressings for, e.g.~the flow of the constituent quark mass function $M_l(p)$ by their size, this size can be deceiving for two reasons. 
First of all, it is the size of the respective diagrammatic contributions that enters $\partial_t M_l(p)$ and this depends on the contraction of the tensor structures: 
while ${\cal T}^{(7)}_{A \bar l l}$ is chirally symmetric as is the classical tensor structure, ${\cal T}^{(4)}_{A \bar l l}$ breaks chiral symmetry, see \labelcref{eq:T147}. 
The latter fact explains already why the respective dressing switches on at smaller momenta. 
Second of all, the flows are non-linear. 
The impact of the different dressings in the flow or DSE of the quark mass function has been studied in detail in \cite{Cyrol:2017ewj} and in particular in \cite{Gao:2021wun}. 
In these works the functional equation for the quark mass function has been analysed as follows: 
we use the full solution for the vertex dressings and the gluon propagator and solve the functional relation for the quark propagator by dropping either of the dressings $\lambda^{(4,7)}_{A \bar l l}$. 
There it has been found that dropping $\lambda_{A \bar l l}^{(4)}$ leads to a more than 50\% decrease of the constituent quark mass, while dropping $\lambda_{A \bar l l}^{(7)}(p)$ increases the light quark mass function by roughly 10\%.

\subsection{Correlation functions in the matter sector}
\label{sec:4qCorrelations}

\begin{figure}[t]
\includegraphics[width=0.48\textwidth]{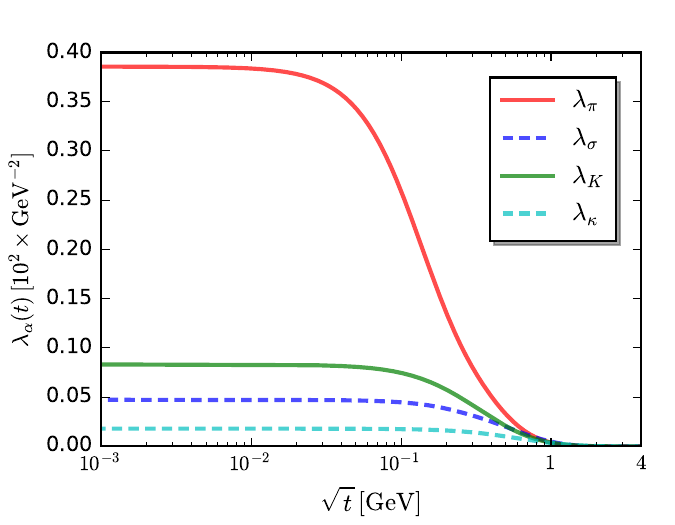}
\caption{Four-quark dressings of the $\pi$, $\sigma$, $K$, $\kappa$ meson channels as functions of $t$-channel momentum with $s=u=0$.}\label{fig:lambda4q}
\end{figure}

In this Section we present and discuss our results for the dressings of the four-quark vertex. A full tensor basis with 26 elements can be found in 
\labelcref{eq:4qBasisI,eq:4qBasisII,eq:4qBasisIII,eq:4qBasisIV,eq:4qBasisV,eq:25Channels} in 
\Cref{app:truncation-matter}. 
We have not taken into account all of them in the flows as most decouple very efficiently in all flows including their own ones. 
This will be discussed below. 
In short, we have only considered the four channels \labelcref{eq:4qBasis-sigmaPiKkappa}: the scalar and pseudoscalar channels with the dressings $\lambda_\sigma,\lambda_\pi$ and the $K,\kappa$-meson channels with the dressings $\lambda_K,\lambda_\kappa$. 
These dressings are shown in \Cref{fig:lambda4q} in $10^{2}\, \textrm{GeV}^{-2}$ units as functions of the $t$-channel momentum at $u=s=0$, 
\begin{align}
    \lambda_\alpha(t) = \lambda_\alpha(s=0,t,u=0)\,.
\label{eq:lambdat}
\end{align}
This channel contains the pion resonance and other mesonic resonances. 
This gives rise to the large value of $\lambda_\pi(t=0) \propto 1/m_\pi^2$. 
This also explains the ratio of $\lambda_\pi/\lambda_\sigma \propto \,m_\sigma^2/m_\pi^2$. 

%
%
%
\begin{figure}[t]
\includegraphics[width=0.48\textwidth]{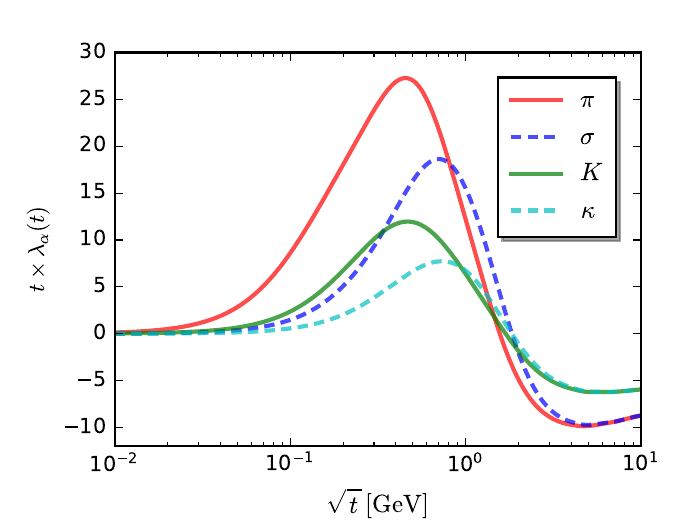}
\caption{Dimensionless four-quark dressings $t\times\lambda_\alpha$ of the $\pi$, $\sigma$, $K$, $\kappa$ meson channels as functions of $t$-channel momentum with $s=u=0$. }
\label{fig:lambda4qDimless}
\end{figure}
%

It is important to realise, that, while the size of the dressings exhibits the mass hierarchy of the resonances, the importance of the different tensor structures of the four-quark vertex has to be measured in terms of the size of the diagrammatic contributions in a given flow. 
The relative size of these contributions depends not only on the dressings, but also on the internal propagators of the diagrams and the contractions of all the tensors in the diagrams, for a detailed discussion see \cite{Ihssen:2024miv}. 
Hence, such an analysis has to be done separately for each flow of a correlation function, and the size of the dressing can be deceiving. 

Moreover, the momentum-dependence of the dressings is integrated over in the diagrams and it is rather their dimensionless form $t\,\lambda_\alpha(t)$ that enters the overall analysis. 
This combination is shown in \Cref{fig:lambda4qDimless}. 
Clearly, the hierarchy seen in \Cref{fig:lambda4q} of the dressings is less pronounced in its dimensionless form. Also, $t\,\lambda_\sigma(t)$ is significantly larger than $t\,\lambda_K(t)$, leave aside $t\,\lambda_\kappa(t)$. 
This order reflects better on the actual importance of the contributions, but as we have discussed above, such an importance analysis has to be done with the diagrammatic contributions.  

Far more relevant than such an in detailed analysis of the four-quark sector is that of the relative relevance of the quark-gluon (box diagrams), quark-gluon--four-quark (triangle diagrams) and four-quark (fish diagrams) contributions in the flow of the dressings of the four-quark vertex, see \Cref{fig:fourpoint-feyn} in \Cref{app:Flow}. Its relevance comes from its importance for the systematic error analysis in \Cref{sec:Wrapup+SystematicError}. As discussed in \Cref{sec:Gamma4q,sec:Stability4q}, the flows $\partial_t \lambda_\alpha$ are dominated by the quark-gluon box, proportional to $\alpha_{A\bar q q}^2$, for momenta $p\gtrsim 1$\,GeV. 

\begin{figure}[t]
	\includegraphics[width=0.48\textwidth]{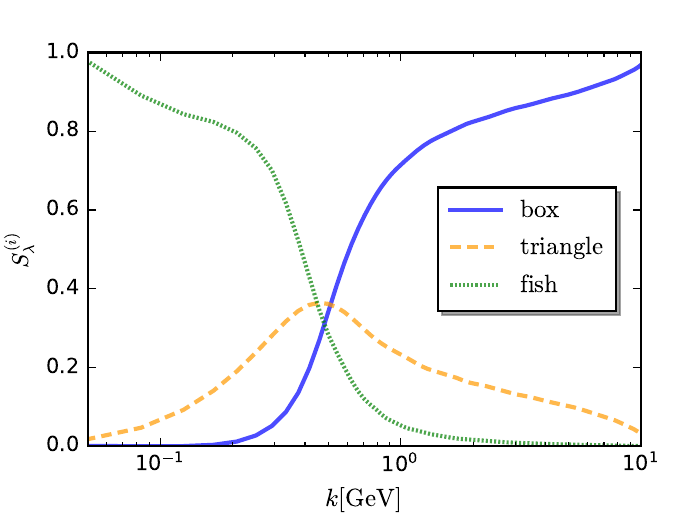}
	\caption{Relative strength of the dimensionless flow of four-quark dressing of the $\pi$ channel as functions of scale $k$ at $s=t=u=0$.}
	\label{fig:dtlambda_RelativeStrength}
\end{figure}

In the following we concentrate on the dressing $\lambda_\pi$ of the pion channel. For a discussion of the relevance order we define the relative strength of a given diagrammatic contribution (box, triangle, fish) with 
\begin{align}
	{\cal S}_\lambda^{(i)} = \frac{\left |\textrm{Flow}^{(i)}_{\lambda}\right|}{\sqrt{ \sum_j \left(\textrm{Flow}^{(j)}_{\lambda}\right)^2}}\,,\quad i,j= \textrm{box, triangle, fish}\,. 
	\label{eq:Strength}
\end{align} 
The different $S_i$ are depicted in \Cref{fig:dtlambda_RelativeStrength}. 
One clearly sees the dominance of the quark-gluon box for large cutoff scales $k\gtrsim 1$\,GeV. 
This UV regime is followed by one that is dominated by the triangle diagrams, proportional to $\alpha_{A \bar q q} \lambda_\pi$. 
In the infrared with $k\lesssim 0.5$\,GeV, the fish diagrams dominate. More precisely, this contribution is dominated by the diagram proportional to $\lambda_\pi^2$. This successive order of dominance is sourced by the two mass scales at play: the gluon mass gap $m_\textrm{gap} \approx 0.5 - 1$\,GeV and the pion pole mass $m_\pi = 137$\,MeV present in the $t$-channel $\lambda_\pi(s=0, t, u=0)$. 

%
%
We note that the peak position of the four-quark contribution of the flow of $\lambda_\pi$ is proportional to $m_\pi$, in contradistinction to that of the flow of the quark mass function which peaks at the chiral symmetry breaking scale, see the respective discussion in \Cref{sec:CorrelationsInterface}. This proportionality originates in the pseudo-scalar channel, whose $t$-channel is proportional to $1/(t +m_\pi^2)$, the pion propagator. In particular, in the chiral limit this triggers a logarithmic infrared divergence of the fish diagram at vanishing external momentum due to the two internal pion propagator. In contrast, the tadpole diagram in $\partial_t M_q$ is infrared finite. 

%
\begin{figure}[t]
\includegraphics[width=0.48\textwidth]{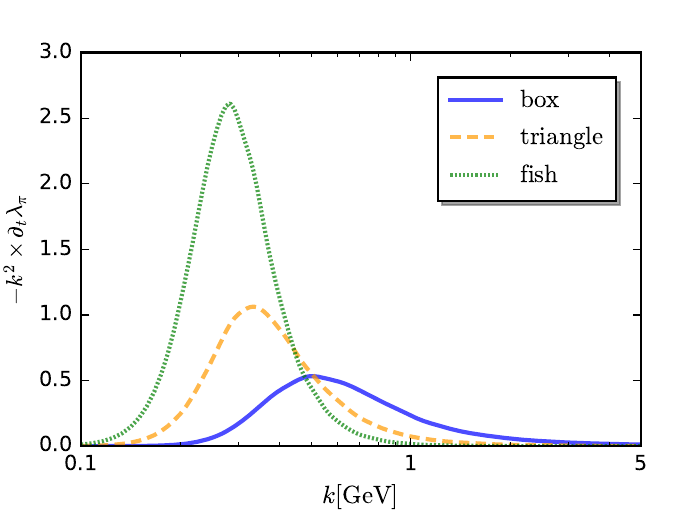}
\caption{Different diagrammatic contributions to the dimensionless flow of four-quark dressing of the $\pi$ channel as functions of scale $k$ at $s=t=u=0$.}
\label{fig:dtlambda4q-pi}
\end{figure}
%

%
\begin{figure}[b]
\includegraphics[width=0.48\textwidth]{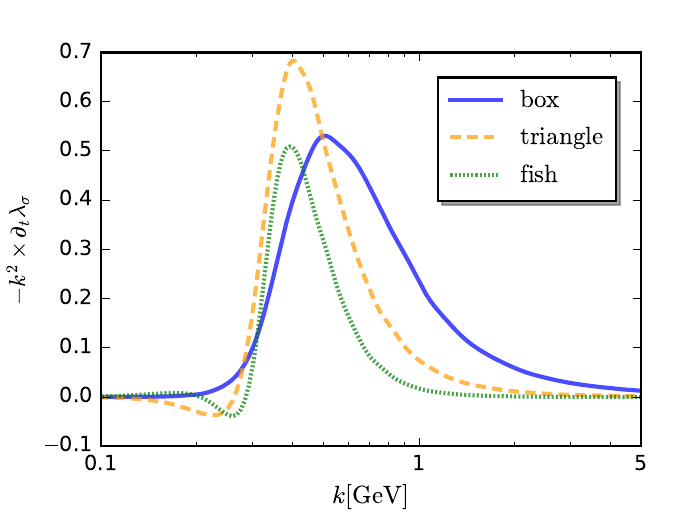}
\caption{Different diagrammatic contributions to the dimensionless flow of four-quark dressing of the $\sigma$ channel as functions of scale $k$ at $s=t=u=0$.}
\label{fig:dtlambda4q-sigma}
\end{figure}
%

%
\begin{figure}[t]
\includegraphics[width=0.48\textwidth]{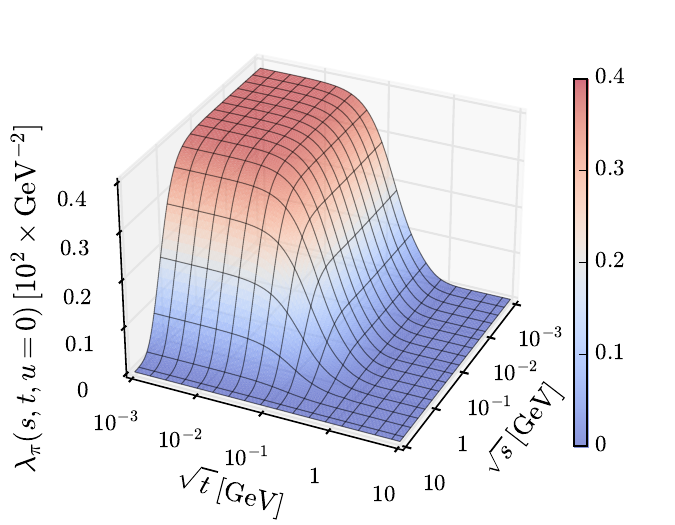}
\caption{3D plot of the four-quark coupling in the $\pi$ channel as a function of $\sqrt{t}$ and $\sqrt{s}$ with $u=0$.}
\label{fig:lambdaTS-3D}
\end{figure}
%
Note also that this successive dominance order is more pronounced in the fRG approach with emergent composites: there, also the pion channel is augmented with a regulator mass, effectively leading to $m^2_\pi\to m^2_\pi+k^2$.  
This shifts the interface regime and the pion dominance regime towards the infrared, leading to a larger separation of the three regimes. For a detailed discussion see \cite{Ihssen:2024miv}. The direct comparison with the results there is further complicated by the different basis choice there. 
Roughly speaking, the pion channel defined in \cite{Ihssen:2024miv} is an average of the pion and $\sigma$-channel in the present work. Evidently, in the cutoff regime with dynamical chiral symmetry breaking with $k\lesssim 500$\,MeV, the pion contributions are dominating the flows of the $\lambda_\alpha$, if these contributions come with a sufficiently large prefactor or overlap. In a general basis this overlap is uniquely determined by the residue of the pion pole obtained for $t\to -m_\pi^2$. A full comparison also involves powers of the relative mass scales of the respective channels, for example powers of $m_\pi^2/m_\textrm{res}^2$ for a given meson resonance. This comparison also extends to gluon exchanges where  $m_\textrm{res}$ is substituted by the gluon mass gap. Accordingly, if the overlap with the pion channel is missing, it is the channel with the lowest resonance mass (and a sufficient overlap), that sets the scales. 

We illustrate this structure within a discussion of the different diagrammatic contributions to the flow of $\lambda_\pi$ and $\sigma$, depicted in \Cref{fig:dtlambda4q-pi,fig:dtlambda4q-sigma}. In the flow of $\lambda_\pi$, the overlap of the pion exchange diagrams (fish and triangle) is sufficiently large and we have successive dominance regimes of the box (UV), triangle (small interface) and fish (IR) diagrams, see \Cref{fig:dtlambda_RelativeStrength} and the respective discussion around \Cref{eq:Strength}. In contrast, no such order is found in the flow of  $\lambda_\sigma$, see \Cref{fig:dtlambda4q-sigma}. Specifically, the fish diagram is not infrared dominant, which indicates a rather small contribution of the pion resonance. 

%
\begin{figure}[b]
	\includegraphics[width=0.48\textwidth]{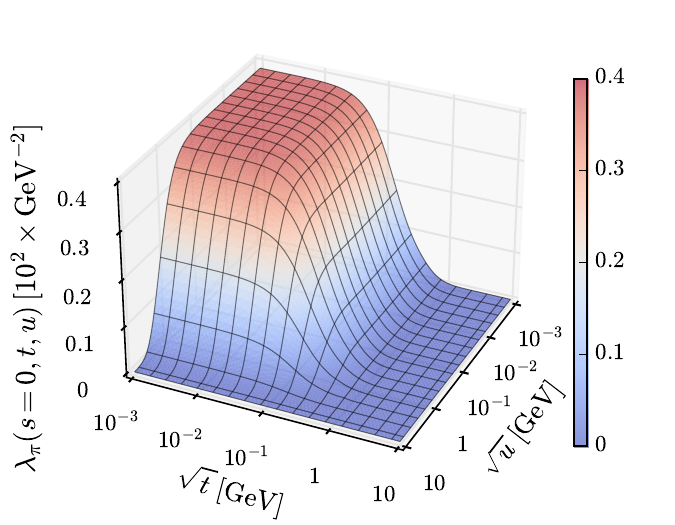}
	\caption{3D plot of the four-quark coupling in the $\pi$ channel as a function of $\sqrt{t}$ and $\sqrt{s}$ with $u=0$. }
	\label{fig:lambdaTU-3D}
\end{figure}
\begin{figure}[t]
\includegraphics[width=0.48\textwidth]{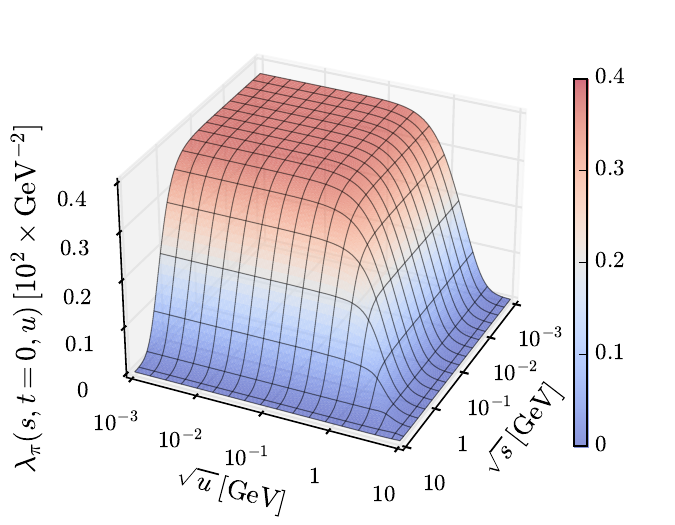}
\caption{3D plot of the four-quark coupling in the $\pi$ channel as a function of $\sqrt{s}$ and $\sqrt{u}$ with $t=0$.} \label{fig:lambdaSU-3D}
\end{figure}

%
This concludes our discussion of the $t$-channel dressings and their relevance order. 
In the present work we have not only taken into account the $t$-channel dressing, but dressings that depend on all momentum channels. 
Apart from its more quantitative nature it also allows us to evaluate one-momentum channel approximation, for a more detailed discussion see \cite{Fu:2022uow, Fu:2024ysj}. 
For this discussion we concentrate on the most important pion channel with the dressing $\lambda_\pi(s,t,u)$. Its evaluation is based on an analysis of our results for two-dimensional momentum-channel planes in \Cref{fig:lambdaTS-3D} ($s-t$ plane), \Cref{fig:lambdaTU-3D} ($t-u$ plane) and \Cref{fig:lambdaSU-3D} ($s-u$ plane). 

We start with the results for the two $t$-channel momentum planes depicted in \Cref{fig:lambdaTS-3D,fig:lambdaTU-3D}. We find that in the $s,u$ directions the dressing is constant for momenta $p\lesssim 1$\,GeV, while it drops earlier in the $t$ direction. This indicates that in the relevant regime with $s,u\lesssim 1$\,GeV, the dressing shows no dependence on these momenta. This suggests that a $t$-momentum channel approximation is already quantitative at the physical point. Given, that the flows themselves provide an additional momentum average due to the presence of the infrared cutoff scale $k$, the approximation $\lambda_\pi(s,t,u)\to \lambda(t)$ is a quantitative one. We emphasise that this analysis does not hold true in the chiral limit, where $\lambda_\pi$ is divergent and all momentum and angular dependences are amplified. Finally we note that the two plots in \Cref{fig:lambdaTS-3D,fig:lambdaTU-3D} suggest an approximate symmetry between the $s$ and $u$ momentum-channel dependence. This is confirmed in \Cref{fig:lambdaSU-3D}, where we show $\lambda_\pi$ in the $s,u$- plane. 

%
This concludes our analysis of the four-quark correlation functions. In short, the analysis in the present Section as well as that in \Cref{sec:CorrelationsGlue,sec:CorrelationsInterface} entail, that the present approximation is a quantitative one for vacuum QCD, and the 10\% systematic error estimate in \Cref{sec:Wrapup+SystematicError} is indeed a very conservative one.

\section{Meson properties and Bethe-Salpeter Amplitudes}
\label{sec:meson-part}

In combination, the first-principles approach to 2+1 flavour QCD, set up in this paper, and that with emergent composites, allow for many applications, that include the QCD phase structure, parton distribution functions and other timelike observables. Respective applications go beyond the scope of the present work and we concentrate on the light spectrum. This serves as a proof of principle as well as providing us with the observables we have used for determining the physical point with $m_\pi/f_\pi= 137/93$: we compute the pole masses of the pion and the $\sigma$ resonance within a Padé approximation and extract the pion and kaon decay constants from their Bethe-Salpeter amplitudes. 

The direct computation of pole masses requires a functional approach for timelike momenta. Such an approach has been set up in the past years based on the Källén-Lehmann representation of correlation functions, see \cite{Horak:2020eng, Horak:2021pfr, Horak:2022myj, Horak:2022aza, Horak:2023hkp, Braun:2023qak, Fukushima:2023wnl}, for Keldysh contour setups see \cite{Pawlowski:2015mia,Tan:2021zid}. This complements the standard approach used in bound state analyses, see e.g.~the review \cite{Eichmann:2016yit}. 

%
\begin{figure}[t]
	\includegraphics[width=0.48\textwidth]{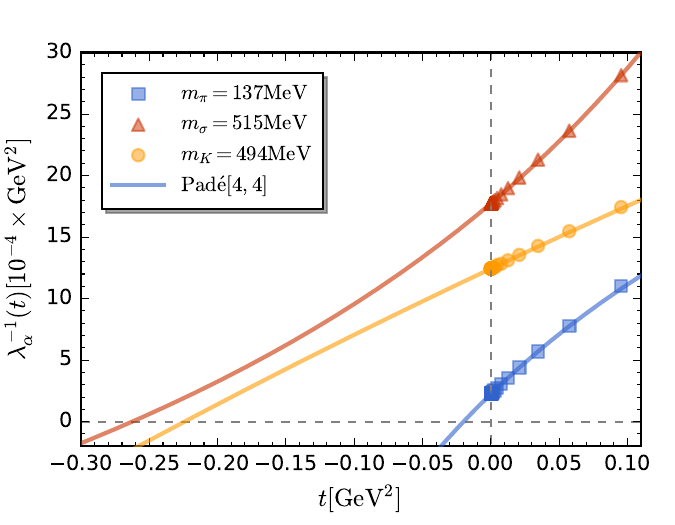}
	\caption{Inverse four-quark coupling $1/\lambda_\alpha(t)$ for $\alpha=\pi,\sigma,K$ as a function of the $t$-channel momentum. The data points denote the Euclidean results and the solid lines fit with Pad\'e[4,4] approximants. The pole masses are determined by the $t$-value of $1/\lambda_\alpha(t)=0$, see \labelcref{eq:ZeroPoleMass}.}
	\label{fig:InverseLambdaPoleMasses}
\end{figure}
%
In the present work we concentrate on the pole masses of the pion, the $\sigma$-mode and the kaon. These are the lowest lying resonances in the respective four-quark channels, and the pole mass is the distance of the first pole or singularity from the Euclidean frequency axis. This information can be safely extracted by an analytic continuation with Padé approximants, for more details and the respective stability analysis see \cite{Fu:2022uow, Fu:2024ysj}. Instead of the pole we extract the location of the zero of the inverse four-quark dressings, 
\begin{align}
   \frac{1}{\lambda_\alpha}(t=-m_\alpha^2)=0\,,\qquad \textrm{for}\qquad \alpha=\pi,\sigma,K\,.
\label{eq:ZeroPoleMass}
\end{align}
In the present work we use Padé[4,4] approximants and this Padé approximant is shown in \Cref{fig:InverseLambdaPoleMasses} together with $1/\lambda_\alpha(t)$ for 
$\alpha=\pi,\sigma,K$. The pion and $K$ masses are used for determining the physical point with 
\begin{align} 
\frac{m_\pi}{f_\pi}=\frac{137}{93}\,,\qquad \frac{m_K}{f_\pi} = \frac{494}{93}\,,
\label{eq:PhysPoint}
\end{align}
see \Cref{tab:Parameters}. In contradistinction, the pole mass of the $\sigma$-mode is a prediction and we find 
\begin{align}
    m_\sigma=515.2\,\textrm{MeV}\,.
    \label{eq:mPoleSigma}
\end{align}
The two observables \labelcref{eq:PhysPoint} used to determine the physical point also require the knowledge of the pion decay constant $f_\pi$, see \cite{Maris:1997hd}. It can be computed from the light quark propagator and the Bethe-Salpeter amplitude $h_\pi(p)$ of the pion. Similarly, the kaon decay constant $f_K$ can be computed from the light and strange quark propagators and the BSE amplitude $h_K(p)$ of the kaon. The pion and kaon decay constants are given by
\begin{subequations} 
\label{eq:Pi+K-DecayConstants}
\begin{align}
    \bra{0}J_{5\mu}^{a}(x)\ket{\pi^{b}}=&\mathrm{i} P_{\mu}f_{\pi}\delta^{ab}\,,\nonumber\\[2ex]
    \bra{0}J_{5\mu}^{a}(x)\ket{K^{b}}=&\mathrm{i} P_{\mu}f_{K}\delta^{ab}\,,\label{eq:def-fpifK}
\end{align}
with the pion and kaon matrix elements of the axial currents, 
\begin{align}
  &\bra{0}J_{5\mu}^{i}(x)\ket{\pi^{j}}\nonumber\\[2ex]
  =&\frac{\delta^{ij}}{2}\int \frac{d^{4}q}{(2\pi)^{4}} \mathrm{Tr}\,\Biggl[\gamma_{\mu}\gamma_{5}\, G_{l}(q+P)\, h_{\pi}(q)\,\gamma_{5}\, G_{l}(q)\Biggr]\,, 
  \label{eq:fpi-expr} 
\end{align}
and 
\begin{align}
  &\bra{0}J_{5\mu}^{i}(x)\ket{K^{j}}\nonumber\\[2ex]
  =&\frac{\delta^{ij}}{4}\int \frac{d^{4}q}{(2\pi)^{4}} \mathrm{Tr}\,\Biggl[\gamma_{\mu}\gamma_{5}\, G_{l}(q+P)\, h_{K}(q)\,\gamma_{5}\, G_{s}(q)\nonumber\\[2ex]
  &\quad\quad\quad\quad+\gamma_{\mu}\gamma_{5}\, G_{l}(q)\, h_{K}(q)\,\gamma_{5}\, G_{s}(q+P)
  \Biggr]\,. 
  \label{eq:fK-expr} 
\end{align}
\end{subequations}
In \labelcref{eq:fpi-expr,eq:fK-expr}, the traces are in colour and Dirac spaces. The indices $i,j$ label the adjoint representation of the  flavor group, with $i, j=1,\,2,\,3$ for the pion and $4,\,5,\,6,\, 7$ for the kaon. Here the BS amplitudes are defined at the mass poles of the mesons, which are given by,
\begin{align}
    h_{\pi/K}(q)=h_{\pi/K}(q;P^{2}=-m_{\pi/K}^{2})\,,
\end{align}
where the integrated momentum $q$ includes the dependence on both the momentum magnitude and the angle. The extraction of the BS amplitude will be discussed in detail in \labelcref{eq:BSa}. Note that $f_K$ is a prediction, as the physical point is determined with the ratios \labelcref{eq:PhysPoint}, that only require $f_\pi$, together with $m_\pi$ and $m_K$, see \Cref{tab:Parameters}.  

%
\begin{figure}[t]
\includegraphics[width=0.48\textwidth]{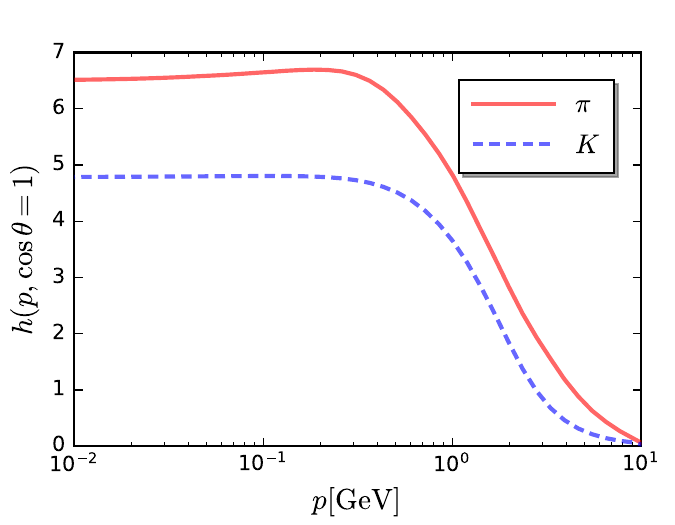}
\caption{Bethe-Salpeter amplitudes of the pion (red solid line) and the kaon (blue dashed line) as a function of the magnitude of the quark momentum at fixed angle $\cos\theta=1$. }
\label{fig:BS-amplitudeptheta1}
\end{figure}
%
The computation of \labelcref{eq:fpi-expr,eq:fK-expr} requires the quark propagators in \Cref{fig:Quark} and the BS amplitudes $h_\pi$ and $h_K$. 
They are extracted from the residue of the relevant four-quark dressing at the pole of bound state. 
For a detailed discussion we refer to \cite{Fu:2022uow, Fu:2024ysj}, and we get  
\begin{align}
   h_{\alpha}(p,\cos \theta)=\lim_{P^2\to -m_\alpha^2} \sqrt{\lambda_{\alpha}(p_1,...,p_4)\,\left(P^2+m_\alpha^2\right)}\,,
   \label{eq:BSa}
\end{align}
where $P,\bar p, \bar p^\prime$ are related to $p_1,...,p_4$ with \labelcref{eq:Pp-p1234}, and we evaluate $\lambda_\alpha(p_1,...,p_4)$ on the momentum configuration 
\begin{align}
    P_\mu=&\sqrt{P^{2}}\,\Big(1,\,0,\,0,\,0 \Big)\,,\nonumber\\[2ex]
    \bar p_\mu=&\sqrt{p^{2}}\, \Big(\cos \theta,\, \sin \theta,\,0,\,0 \Big)\,,\nonumber\\[2ex]
    \bar p^\prime_\mu=&-\sqrt{p^{2}}\, \Big(\cos \theta,\, \sin \theta,\,0,\,0 \Big)\,.
    \label{eq:4quark-mom-sym}
\end{align}
This configuration differs from that used to compute the flows of $\lambda_\alpha(s,t,u)$, see \labelcref{eq:4quark-mom}. Hence we cannot directly use our results for $\lambda_\alpha(s,t,u)$. Instead, we follow the procedure detailed in \cite{Fu:2024ysj}: 
the flow for $\lambda_\alpha(p_1,...,p_4)$ is evaluated for the configuration \labelcref{eq:4quark-mom-sym}, and we use our results for the four-quark dressings $\lambda_\alpha(s,t,u)$ discussed in \Cref{sec:4qCorrelations} on the right hand side of the flow in the diagrams. This offers quantitative results, given the small angular dependence discussed in \Cref{sec:4qCorrelations}. 
%
\begin{figure}[t]
	\includegraphics[width=0.48\textwidth]{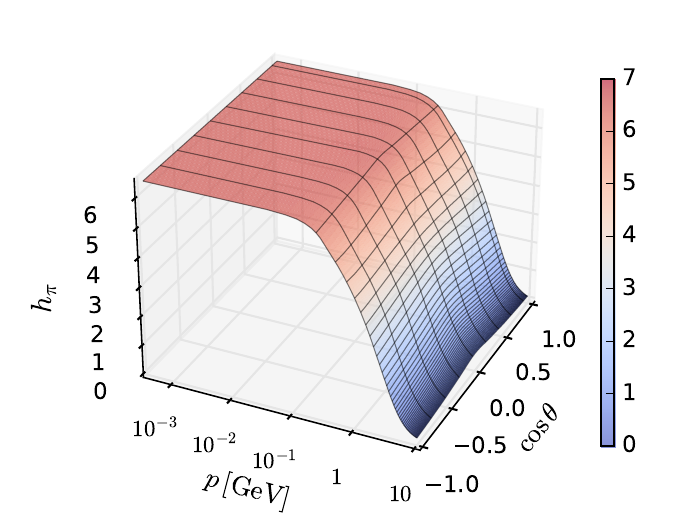}
	\caption{Bethe-Salpeter amplitude of the pion as a function of the magnitude of
		the quark momentum and the angle between the quark and meson momenta. }
	\label{fig:BS-amplitudeptheta}
\end{figure}
%
%
\begin{figure}[b]
	\includegraphics[width=0.48\textwidth]{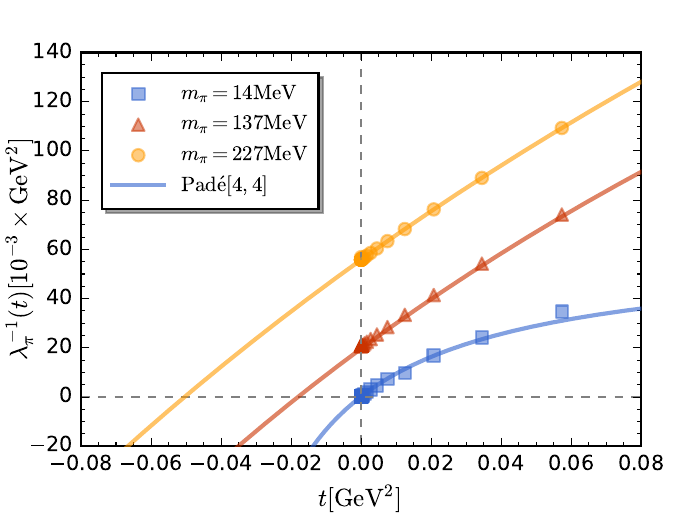}
	\caption{Inverse four-quark coupling in the $\pi$ channel, $1/\lambda_{\pi}(t)$, as a function of the Mandelstam variable $t=P^2$ with $s=0$ and $u=0$. Results for several different values of $m_{\pi}$ are compared. Data points denote the results calculated in the flow equation in the Euclidean $t>0$ region. The solid lines stand for results of analytic continuation from $t>0$ to $t<0$ based on the fit of the Pad\'e[4,4] approximants.}
	\label{fig:mpiPole}
\end{figure}
%
The dependence of the Bethe-Salpeter amplitudes of pion and kaon on the quark momentum is shown in \Cref{fig:BS-amplitudeptheta1}, with the momenta of quark and meson in the same direction $\cos\theta=1$. One can see that with the increase of the momentum the pion BS amplitude increases a bit and then decreases at around $p\sim 300$ MeV, and then decreases significantly at $p\sim 1$ GeV. The kaon BS amplitude does not show this trend of small increase, but begins to decrease later, at about $p\sim 500$ MeV. In \Cref{fig:BS-amplitudeptheta} we show the angular dependence of the pion BS amplitude. We conclude that the angular dependence of the pion BS amplitude in QCD is more prominent, of compared to that obtained in low energy effective theories, see \cite{Fu:2024ysj}. We also find that the momentum dependence of the BS amplitude is softened for $\cos\theta=0$. In turn, for angles $\theta\neq \pi/2$, the momentum dependence becomes steeper. The present results pave the way for calculations of the pion and kaon distribution amplitudes and distribution functions. This is work under completion and we shall report on it in the near future.

\begin{figure}[t]
	\includegraphics[width=0.48\textwidth]{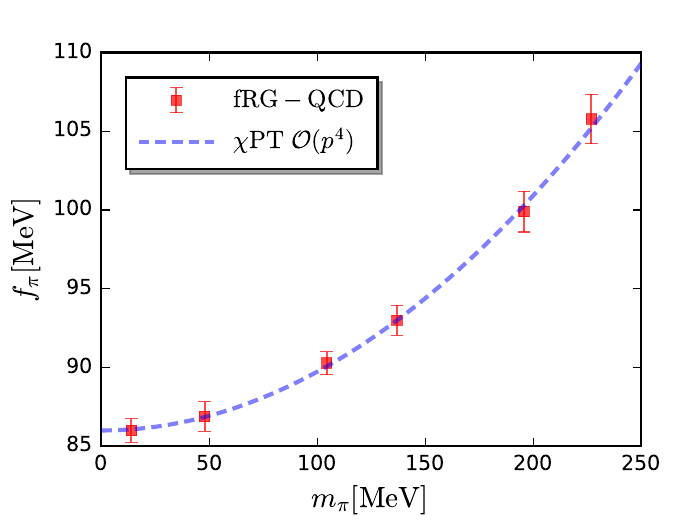}
	\caption{Pion decay constant as a function of pion mass in comparison to the result of chiral perturbation theory at the order of $\mathcal{O}(p^4)$ \cite{Gasser:1983yg}. The errors are estimated from the reconstruction of the Pad\'e approximants.} \label{fig:mpi-fpi}
\end{figure}

We close this Section with an evaluation of the pion mass dependence of the pion decay constant $f_\pi(m_\pi)$ with a fixed physical strange current quark mass $m_s=55.9$\,MeV, see \Cref{tab:Parameters}. This result is used to assess the validity regime of chiral perturbation theory. This task requires the computation of the pion pole mass as a function of the current quark mass via the Padé approximants as used in \Cref{fig:InverseLambdaPoleMasses} for the pion, $\sigma$-mode and kaon pole masses. In \Cref{fig:mpiPole} this is illustrated for three different pion masses, $m_\pi = 14,137,227$\,MeV with $m_l=0.74,\,2.12,\,3.24$\,MeV. These are the smallest and largest pion mass used, as well as the physical one. The pion decay constants are computed readily with \labelcref{eq:fpi-expr} and are shown in dependence on $m_\pi$ in \Cref{fig:mpi-fpi} (red data points). The errors are from the reconstruction of the Pad\'e approximants. With an increasing pion mass, the fit error of the Pad\'e approximants increases as well. Now we use the results to assess the validity range of chiral perturbation theory: In \Cref{fig:mpi-fpi} we compare our results to that in chiral perturbation theory ($\chi$PT) at the order of $\mathcal{O}(p^4)$ in \Cref{fig:mpi-fpi}, see e.g.~\cite{Gasser:1983yg, Guo:2015xva, Gao:2022dln}. We find that the $\chi$PT results agree with ours up to the large pion masses considered here.

%
%

\section{Conclusions and outlook}
\label{sec:conclusion}

In the present work we have studied the full infrared dynamics of 2+1 flavour QCD with the functional renormalisation group approach. The present work completes a series of works initiated in \cite{Fu:2022uow,Fu:2024ysj} on the infrared dynamics of QCD. This series aims at establishing a first-principles fRG approach to QCD for the computation of timelike observables such as parton distribution functions as well as the phase structure of QCD. 

For the first time, we have solved the complete dynamics of the system in a  \textit{self-contained} and \textit{self-consistent} fRG approach to \textit{first principle} 2+1 flavour QCD without using any external input. 
Here, \textit{self-contained} refers to the fact that all correlation functions are computed within the fRG approach themselves. 
While the use of external input such as gluon correlation functions obtained from other functional computations or from lattice data is a major advantage of functional approaches, a self-contained computation constitutes an important benchmark test and step towards the reliable computation of timelike observables as well as the phase structure of QCD within the fRG. 
\textit{Self-consistent} refers to the fact that all correlation functions computed are also fed-back into the diagrams. 
Finally, it is a \textit{first principle} approach as it comprises the full dynamics of QCD with reliable systematic error control, and the only input in the present computation are the fundamental parameters of QCD: the light and strange current quark masses at the initial cutoff scale. 
The procedure is detailed in \Cref{sec:InitialConditions}, while \Cref{sec:QCD-EffAction} discusses the quantitative approximation used. 
In particular, \Cref{sec:Wrapup+SystematicError} comprises a detailed systematic error analysis with a very conservative overall systematic error estimate of 10\%.   

Our results are presented in \Cref{sec:QCDCorrelations} and \Cref{sec:meson-part}, and we refer the reader to the comprehensive discussions there. 
In particular, we have computed ghost, gluon and quark two-point correlation functions in quantitative agreement with lattice data and other functional results, see \Cref{fig:YM-twopoint,fig:Quark}. 
We also show the ghost-gluon, three-gluon and quark-gluon couplings, see \Cref{fig:InterfaceCouplings,fig:alpha}. 
In particular, \Cref{sec:4qCorrelations} contains our results on the dressings $\lambda_\alpha(s,t, u)$ of the $\alpha=\pi,\sigma,K,\kappa$ tensor channels, where $s,t,u$ are the Mandelstam variables.  

The correlation functions discussed in \Cref{sec:QCDCorrelations} have been put to work in \Cref{sec:meson-part} for the computation of the $\pi,\sigma,K$ pole masses, the pion and $K$-meson BS amplitudes and the pion and $K$-meson decay constants. 
Partially these results have been used for the determination of the current quark masses at the physical point, but they also constitute a proof of principle of the potential of the present approach. 
Applications to the phase structure of QCD are under way, and works on parton distribution functions and further timelike observables are under completion. 
We hope to report on the respective results in the near future.

\section{Acknowledgements}
We thank Franz Sattler,  Joannis Papavassiliou and Gernot Eichmann for discussions. This work is done within the fQCD collaboration \cite{fQCD} and we thank its members for discussions and collaborations on related projects. This work is supported by the National Natural Science Foundation of China under Grant Nos.\ 12447102, 12175030, the Deutsche Forschungsgemeinschaft (DFG, German Research Foundation) under Germany’s Excellence Strategy EXC 2181/1 - 390900948 (the Heidelberg STRUCTURES Excellence Cluster) and the Collaborative Research Centre SFB 1225 (ISOQUANT). It is also supported by EMMI. JMP acknowledges support by the Chinese Academy of Sciences President's International Fellowship Initiative Grant No.~2024PG0023.

\appendix

\section{Classical QCD action}
\label{app:SQCD}

In this Appendix we detail the classical gauge-fixed QCD action in a general covariant gauge, as well as introducing some notation. The classical gauge-fixed action reads
\begin{align}
	S_\textrm{QCD}= S_A[A] +S_\textrm{gf}[A] +S_\textrm{gh}[A,c, \bar c] +S_q[A,q,\bar q] \,.
	\label{eq:SQCD}
\end{align}
with 
\begin{align}\nonumber
	S_A[A] =            & \, \frac14 \int_x F_{\mu\nu}^a F_{\mu\nu}^a\,,                           \\[2ex]
	S_q [A,q,\bar q]  = & \, \int_x \,\bar q \left( \gamma_\mu D_\mu+m_q\right)\,q \,. 
	\label{eq:SA+Sq}
\end{align}
The fieldstrength tensor in \labelcref{eq:SA+Sq} is given by 
\begin{align}
	F_{\mu\nu}=F^a_{\mu\nu}t^a \,,\quad 	F_{\mu\nu}^a =\partial_\mu A^a_\nu -\partial_\nu A_\mu^a +g_s f^{abc} A_\mu^b A^c_\nu\,. 
	\label{eq:Fstrength}
\end{align}
with the su($N_c$) Lie algebra
\begin{align}
	[t^a\,,\,t^b] = { i } f^{abc} t^c\,. 
	\label{eq:LieAlg}
\end{align}
Here, $t^a $ are the generators of su($N_c$) in a given representation, and $f^{ abc}$ are its structure constants. Ghosts and gluons carry the adjoint representation of the gauge group. For both, the colour and flavour groups we will use the notation 
\begin{align}
	\left(t_\textrm{ad}^a\right)^{bc} = \left( T_{c}^{a}\right)^{bc} \,,\qquad \left( t^{a}_\textrm{f}\right)^{BC} = \left(T^a\right)^{BC} \,,
\end{align}
where lowercase letters $b,c$ label indices in the adjoint representation and capital letters $B,C$ label indices in the fundamental representation.
Furthermore, the generators in the fundamental and adjoint representations are normalised to 
\begin{align}
	\textrm{tr}_f\,T^a T^b = \frac12  \delta^{ab}\,,\qquad \qquad  {\textrm{tr}}_\textrm{ad} \,T_{c}^{a} T_{c}^{b} =  N_c \delta^{ab}\,,
	\label{eq:LieNorm}
\end{align}
where the traces are taken in the respective representation. In order to distinguish the gauge and flavour group generators we shall also use $t_c$, $t_{c,\textrm{ad}}$ and $T_c$ for that of the colour gauge group. 

The 2+1 flavour quark field in \labelcref{eq:SQCD} is given by $q=(l,s)$ with the light quark field $l=(u,d)$. We use a flavour-diagonal mass matrix in the isospin-symmetric approximation with $m_u=m_d=m_l$, 
\begin{align} 
	m_q=\textrm{diag}(m_l,m_l,m_s)\,.
\label{eq:mqall}
\end{align} 
The covariant derivative is given by
\begin{align}
	D_\mu = \partial_\mu - { i }\,g_s\,A_\mu\,,
	\label{eq:CovDer}
\end{align}
and the quarks live in the fundamental representation of the gauge group. Hence, in component notation, the pure matter sector of the QCD action reads 
\begin{align} 
	S_q [A,q,\bar q]  = & \, \int_x \,\bar q^B  \left( \gamma_\mu D^{BC}_\mu+m_q  \right)\,q^C \,,  
	\label{eq:SqComponents} 
\end{align}
with 
\begin{align}
D_\mu^{BC} = \delta^{BC} \partial_\mu - { i } g_s A^{BC}_\mu \,, 
\end{align}
where we have suppressed the Dirac and flavour indices. 

Finally, the gauge-fixing part of the classical action for a general covariant gauge consists of the gauge-fixing term $S_\textrm{gf}$ and the corresponding ghost action $S_\textrm{gh}$,
\begin{align}
	S_\textrm{gf}= \frac{1}{2\xi} \int_x\, (\partial_\mu A_\mu^a)^2 \,,\qquad  S_\textrm{gh}=	\int_x \, \bar c^{\,a}\left(-\partial_\mu D_\mu^{ab}\right) c^b \,,
	\label{eq:Sgauge}
\end{align}
with the covariant derivative in the adjoint representation 
\begin{align} 
D_{\mu}^{ab}=\partial_{\mu}\delta^{ab}-g_s f^{abc}A_{\mu}^{c}\,. 
\end{align}
Here we have used \labelcref{eq:CovDer} and 
\begin{align} 
(t^a_\textrm{ad})^{bc} = -\imag f^{abc}\,. 
\end{align}
In all explicit computations in the present work we use the Landau gauge, i.e. $\xi\to 0$.

\section{Truncation of the pure glue sector}
\label{app:truncation-gauge}

In this Appendix we discuss the approximation of the pure glue sector of the effective action, extending the analysis in \Cref{sec:GammaGlue}. We do this in a two-step procedure. Firstly, we drop all higher order gauge-field correlation functions: $\Gamma^{(n)}_{A^n}\approx 0$ for $n>4$. 
Moreover we reduce the full tensor structure of the remaining vertices to the classical ones, only taking into account the primitively diverging vertices. Then, the full expression for the pure glue sector of the effective action \labelcref{eq:GammaGlue} is given by 
\begin{widetext} 
\begin{align}\nonumber
	\Gamma_{\textrm{glue},k}[A,c,\bar c]  =&\,
	\frac12 \int\limits_p \, A^a_\mu(p) \, p^2 \left[  Z_A(p)\,\Pi^\perp_{\mu\nu}(p) 
		+\frac{ 1}{\xi}Z^\parallel_A(p)\,\Pi^\parallel_{\mu\nu}(p) \right] \, A^a_\nu(-p) + 	\int\limits_p   Z_c(p)  \bar c^{\,a}(p) p^2 \delta^{ab}c^b(-p)  \\[2ex]\nonumber
	&\hspace{-2.3cm}+ \int\limits_{p_1,p_2} \left[Z^{\frac{1}{2}}_{c}(p_{1})Z^{\frac{1}{2}}_{c}(p_{2})Z^{\frac{1}{2}}_{A}(p_{3})\right]\lambda_{A \bar c c}(p_1,p_2)  \left[ {\cal T}_{A\bar c c}^{(1)}(p_1,p_2)\right]^{a_1 a_2 a_3}_{\mu}
	\bar c^{a_2} (p_2) c^{a_1}(p_1) A^{a_3}_\mu(-(p_1+p_2)) \\[3ex] 
	&\hspace{-2.3cm}  + \frac{1}{3!} \int\limits_{p_1,p_2}  \left[\prod_{i=1}^3 Z^{\frac12}_{A}(p_{i})\right] \lambda_{A^3}(p_1,p_2)  \left[ {\cal T}_{A^3}^{(1)}(p_1,p_2)\right]^{a_1 a_2 a_3}_{\mu_1\mu_2\mu_3}
	\prod_{i=1}^3 A^{a_i}_{\mu_i}(p_i)  \nonumber   \\[3ex]  
	&\hspace{-2.3cm}  
 +  \frac{1}{4!}\int\limits_{p_1,p_2,p_3}  \left[\prod_{i=1}^4 Z^{\frac12}_{A}(p_{i})\right]\lambda_{A^4}(p_1,p_2,p_3)   \left[ {\cal T}_{A^4}^{(1)}(p_1,p_2,p_3)\right]^{a_1 a_2 a_3 a_4}_{\mu_1\mu_2\mu_3\mu_4}
	\prod_{i=1}^4 A^{a_i}_{\mu_i}(p_i) 
	     \,, 
	\label{eq:GammaGlueDetailed}
\end{align}
\end{widetext} 
with the transversal and longitudinal projection operators 
\begin{align}
	\Pi_{\mu\nu}^\perp(p)     = \delta_{\mu\nu} - \frac{p_\mu p_\nu}{p^2}\,,\qquad 
	\Pi_{\mu\nu}^\parallel(p)  = \frac{p_\mu p_\nu}{p^2}\,. 
	\label{eq:ProjectionOps}
\end{align}
The tensors ${\cal T}^{(1)}_{A^3}\,,\,{\cal T}^{(1)}_{A^4}\,,\, {\cal T}^{(1)}_{A \bar c c }$ refer to the classical tensor structures. In general, we define these tensor structures by the respective derivatives of the classical QCD action \labelcref{eq:SA+Sq}, evaluated at vanishing fields and $g_s=1$, 
\begin{align}\nonumber 
   & {\cal T}^{(1)}_{\Phi_{i_1}\cdots \Phi_{i_n}}(p_1,...,p_{n-1})(2\pi)^4 \delta(p_1+\cdots +p_n) &\\[2ex] 
    &\hspace{1cm}
    =\left[\frac{\delta^n S_\textrm{QCD}^{(n)}[\Phi]}{\delta \Phi_{i_1}(p_1) \cdots \delta\Phi_{i_n}(p_n)}\right][\Phi=0;g_s = 1]\,.  
\end{align}
The relative signs in \labelcref{eq:GammaGlueDetailed}, are chosen such that all dressings reduce to their classical counterparts, if the full effective action is reduced to the classical one,  
\begin{align}
    \lambda_{A^3}\,,\, \lambda_{A\bar c c}\to g_s\,,\qquad \lambda_{A^4}\to g_s^2\,.
\end{align}
As discussed in \Cref{sec:GammaGlue}, we compute the full momentum dependence of the two-point functions. The transversal two-point function of the gluon and the ghost two-point function is derived from \labelcref{eq:GammaGlueDetailed} as 
\begin{align}
    \left[\Gamma_{AA}^{(2)}\,\Pi^\perp \right]_{\mu\nu}^{ab}(p)&=Z_{A}(p)\,p^{2}\,\delta^{ab}\, \Pi_{\perp}^{\mu\nu}(p)\,,\nonumber\\[2ex] \left(\Gamma_{c\bar c}\right)^{ab}(p)&=Z_{c}(p)p^{2}\delta^{ab}\,.   \label{eq:YM-twopoint}
\end{align}
The transverse projection operator $\Pi^\perp$ is defined in \labelcref{eq:ProjectionOps}. The dressings $Z_{A}(p)$ and $Z_{c}(p)$ are the wave functions of the gluon and ghost respectively. 

We now proceed to the vertices, where we use a further approximation. This constitutes the second step in our approximation procedure. Instead of the full momentum dependence of the vertices we only consider the symmetry point approximation exemplified in \Cref{sec:GammaGlue} with the ghost gluon vertex. We use 
\begin{align}
\lambda_{\Phi_{i_1}\cdots \Phi_{i_n}}(p_1,...,p_{n-1}) \to \lambda_{\Phi_{i_1}\cdots \Phi_{i_n}}(\bar p)
\label{eq:GenSymPointAverage}
\end{align}
with $\bar p$ and the maximally symmetric simplex configuration defined in \labelcref{eq:SymPoint-npoint,eq:MomentumConservation}. 

With these preparations we get from  \labelcref{eq:GammaGlueDetailed,eq:GenSymPointAverage} the final expression for the gluon three-point function,  
\begin{align}\nonumber 
    \left[\Gamma_{A^{3}}^{(3)}\right]_{\mu\nu\rho}^{abc}(p_{1},p_{2},p_{3})=&\,(2 \pi)^4 \delta\left(p_1+p_2+p_3\right)\\[2ex]
 &\hspace{-1.5cm}\times \left[\prod_{i=1}^3 Z^{\frac12}_{A}(p_{i})\right]\lambda_{A^3} (\bar{p})\,{\cal T}^{(1)}_{A^3}(p_1,p_2)\,,
\label{eq:GammaA3}
\end{align}
with 
\begin{align}\nonumber 
\left[{\cal T}^{(1)}_{A^3}\right]^{abc}_{\mu\nu\rho}(p_1,p_2) =&\,-\mathrm{i}f^{abc}\,\Biggl[\delta_{\mu\nu}(p_{1}-p_{2})_{\rho}\\[2ex] 
&\hspace{-1cm}+\delta_{\nu\rho}(p_{2}-p_{3})_{\mu}+\delta_{\mu\rho}(p_{3}-p_{1})_{\nu}\Biggr]\,, 
\end{align}
with $p_3=-p_1-p_2$. We use the same approximation for the four-gluon vertex, and with  \labelcref{eq:GammaGlueDetailed,eq:GenSymPointAverage} we arrive at  
\begin{align}\nonumber
\Gamma_{A^4}^{(4)}(p_1,p_2,p_3,p_4)=&\,(2 \pi)^4 \delta\left(\sum_{i=1}^4 p_i\right)\,\prod_{i=1}^4 Z^{\frac{1}{2}}_{A}(p_{i}) \\[2ex]
& \hspace{.5cm}\times \lambda_{A^{4}}(\bar{p})\,{\cal T}^{(1)}_{A^4}(p_1,p_2,p_3)\,,
\end{align}
with the classical tensor structure ${\cal T}^{(1)}_{A^4}$,
\begin{align} \nonumber 
\left[{\cal T}_{A^4}^{(1)}\right]_{\mu\nu\rho\sigma}^{abcd}(p_1,p_2,p_3)
=&\, \Biggl[f^{eab}f^{ecd}(\delta_{\mu\rho}\delta_{\nu\sigma}-\delta_{\mu\sigma}\delta_{\nu\rho})
\\[2ex]\nonumber
&\hspace{-1cm}+f^{eac}f^{ebd}(\delta_{\mu\nu}\delta_{\rho\sigma}-\delta_{\mu\sigma}\delta_{\nu\rho})\\[2ex]
&\hspace{-1cm}+f^{ead}f^{ebc}(\delta_{\mu\nu}\delta_{\rho\sigma}-\delta_{\mu\rho}\delta_{\nu\sigma})\Biggr]\,.
\end{align}
It follows from the STIs, that all the avatars of the strong couplings agree for perturbative and semi-perturbative momenta, 
\begin{align}
    \alpha_i(p) \approx  \alpha_j(p) \,,\quad \textrm{for} \quad p\gtrsim 5\,\textrm{GeV}\,,
 \label{eq:STIalphas}
 \end{align}
for $i,j=A^3,A^4, A \bar c c, A\bar q q$ and 
\begin{align}
 \alpha_{i}(p)= \frac{\lambda_{i}^2(p)}{4 \pi}\,, \qquad \alpha_{A^4}(p)= \frac{\lambda_{A^4}(p)}{4 \pi} \,,
\end{align}
with $i=A^3, A \bar c c, A\bar q q$. \labelcref{eq:STIalphas} holds true for $p\gtrsim 5$\,GeV, as for these momenta 
the scattering kernels, that separate these couplings, approach unity. This has been confirmed numerically in two-flavour QCD in \cite{Mitter:2014wpa,Cyrol:2017ewj}. Indeed, for the purely gluonic vertices it even holds true for $p\gtrsim 1.5$\,GeV, below which the gluonic dynamics decouples. Therefore we shall use 
\begin{align}
    \lambda_{A^4}(\bar p) = \left[\lambda_{A^3}(\bar p)\right]^2\,, 
\label{eq:lambdaA34}
\end{align}
for all momenta. 

The ghost-gluon vertex and the respective symmetric point approximation have been discussed in \Cref{sec:GammaGlue}, see in particular \labelcref{eq:SymPoint3Average,eq:lambdaAbarccbarp,eq:SymPoint-npoint,eq:MomentumConservation}. We derive from  \labelcref{eq:GammaGlueDetailed,eq:GenSymPointAverage}
\begin{align}\nonumber 
    \left[\Gamma_{A\bar{c}c }^{(3)}\right]_{\mu}^{abc}(p_{1},p_{2},p_{3})&=(2 \pi)^4 \delta\left(p_1+p_2+p_3\right)\\[2ex]
    &\hspace{-3cm} \times \left[Z^{\frac{1}{2}}_{c}(p_{1})Z^{\frac{1}{2}}_{c}(p_{2})Z^{\frac{1}{2}}_{A}(p_{3})\right]\, \lambda_{A\bar{c}c}(\bar{p})\,{\cal T}^{(1)}_{A\bar c c}(p_1,p_2)\,,\label{eq:YM-ccA}
\end{align}
with 
\begin{align}
\left[{\cal T}^{(1)}_{A\bar c c}\right]_{\mu}^{abc}(p_1,p_2) =\mathrm{i}f^{abc} (p_2)_{\mu} \,.
\label{eq:TAbarcc}
\end{align}
This concludes the discussion of the pure glue sector of the effective action.

\section{Truncation of the glue-matter interface}
\label{app:truncation-qqA}

The glue-matter interface of the QCD effective action includes the two-point functions of light quarks and strange quarks, quark-gluon vertices and higher quark-gluon interactions. Together with the pure matter sector discussed in \Cref{app:truncation-matter} they are closely related to the spontaneous chiral symmetry breaking and the dynamical emergence of the mass of visible matter. In \labelcref{eq:GammaInter} in \Cref{sec:GammaInterface} we have only written down the quark-gluon interaction with the classical tensor structure. Here we provide the full approximation including all quark-gluon terms as well as the higher order scatterings we take into account. 
\begin{widetext}
\begin{align}
    \Gamma_{d,k}=& \int_{p}\bar{q}(-p)\, Z_{q}(p)\,\Big[\mathrm{i}\slashed{p}+ M_{q}(p)\Big]\,q(p)+\int_{p,q}\bar q(q)\,\left[ Z_{A}^{\frac{1}{2}}(p) Z_{q}^{\frac{1}{2}}(q)Z_{q}^{\frac{1}{2}}(r)\sum_{i=1}^{8}\lambda_{A\bar{q}{q}}^{(i)}(p,q)\left[\mathcal{T}^{(i)}_{A\bar{q}q}(p,q)\right]^a_{\mu}\right]\, A^a_\mu(p) q(r)\nonumber\\[2ex]
    &+ \int_{p,q,r}\bar q(r)\,\Bigl[ Z_{A}^{\frac{1}{2}}(p) Z_{A}^{\frac{1}{2}}(q) Z_{q}^{\frac{1}{2}}(r)Z_{q}^{\frac{1}{2}}(s) \lambda_{AA\bar{q}{q}}(p,q,r)\left[\mathcal{T}_{AA\bar{q}q}(p,q,r)\right]^{ab}_{\mu\nu}\Bigr]\, A^a_\mu(p) A^b_\nu(q)q(s)+\cdots \,,
    \label{eq:GammaInterApp}
\end{align}
\end{widetext}
As discussed in \Cref{sec:GammaInterface}, we only take into account the tensor structures ${\cal T}^{(1,4,7)}(p,q)$ in the numerics as the other tensor structures give negligible contributions. This has been checked explicitly, see also e.g.~\cite{Mitter:2014wpa,Cyrol:2017ewj}. This, and the higher order terms are discussed below \labelcref{eq:Sigmamunun}. 

In the current work we use the isospin symmetric approximation with $Z_u=Z_d=Z_l$ and $M_u=M_d=M_l$. 
Then, the two-point correlation function of the quarks in \labelcref{eq:GammaInterApp} can be represented concisely in a diagonal matrix form, 
\begin{align}\nonumber 
    \Gamma_{\bar{q}q}^{(2)}(p)=&\,\textrm{diag}\bigl(Z_l(p)\,,\,Z_l(p)\,,\,Z_s(p)\bigr)\, \mathrm{i}\slashed{p} \\[2ex]
    &+ 
    \textrm{diag}\bigl(M_l(p)\,,\,M_l(p)\,,\,M_s(p)\bigr)\,,\label{eq:quark-twopoint}
\end{align}
The quark-gluon vertex plays the crucial rôle in the dynamics of strong chiral symmetry breaking. In the Landau gauge, the transverse part of the quark-gluon vertex is given by the flavour-diagonal correlation function 
\begin{align}\nonumber 
&   \Pi_{\mu\nu}^\bot(p) \left(\Gamma_{A \bar{q}q }^{(3)}\right)_{\nu}^{a}(p,q,r)\\[2ex]
    &\hspace{1cm}=(2 \pi)^4\delta(p+q+r)\,
    Z_{A}^{\frac{1}{2}}(p) Z_{q}^{\frac{1}{2}}(q)Z_{q}^{\frac{1}{2}}(r)
    \nonumber\\[2ex]
    &\hspace{1.4cm}\times \Pi_{\mu\nu}^\bot(p)\sum_{i=1}^{8}\lambda_{A\bar{q}{q}}^{(i)}(\bar{p})\left[\mathcal{T}^{(i)}_{A\bar{q}q}(p,q)\right]^a_{\nu}\,,
    \label{eq:Gamma-qqA}
\end{align}
with the transverse projection operator $ \Pi_{\mu\nu}^{\perp}(p)$ in \labelcref{eq:ProjectionOps}. In \labelcref{eq:Gamma-qqA} we have suppressed the color, flavour and Dirac indices of the quarks. The average momentum $\bar p$ is defined in \labelcref{eq:SymPoint-npoint} and the tensor structure ${\cal T}_{A\bar q q}^{(i)}$ with $i=1,...,8$ factorise, 
\begin{align}
\left[\mathcal{T}^{(i)}_{A\bar{q}q}(p,q)\right]^a_{\mu} = T_c^a \left[\mathcal{T}^{(i)}_{A\bar{q}q}(p,q)\right]_{\mu}\,, 
\end{align}
with the generators $T_c^a$ of the gauge group in the fundamental representation. The tensor structures ${\cal T}^{(i)}_{A\bar q q}$ with $i=1,...,8$ are given by 
\begin{align}\nonumber 
    \left[\mathcal{T}^{(1)}_{A\bar{q}q}(p,q)\right]_{\mu}
    =&\,\mathrm{i}\gamma_{\mu}\,,\\[2ex]\nonumber
    \left[\mathcal{T}^{(2)}_{A\bar{q}q}(p,q)\right]_{\mu}
    =&\,(q-r)_{\mu},\\[2ex]\nonumber
    \left[\mathcal{T}^{(3)}_{A\bar{q}q}(p,q)\right]_{\mu}
    =&\,\mathrm{i}\sigma_{\mu\alpha}\,(q-r)_{\alpha}\,,\\[2ex]\nonumber
  \left[\mathcal{T}^{(4)}_{A\bar{q}q}(p,q)\right]_{\mu}
  =&\,\mathrm{i}\sigma_{\mu\alpha} p_{\alpha}\,,\\[2ex]\nonumber
   \left[\mathcal{T}^{(5)}_{A\bar{q}q}(p,q)\right]_{\mu}
   =&\,\mathrm{i}(q-r)_{\mu}\,\slashed{p}\,,\\[2ex]\nonumber
    \left[\mathcal{T}^{(6)}_{A\bar{q}q}(p,q)\right]_{\mu}
    =&\,\left[ \Pi_{\mu\alpha} ^{\perp}(p,q-r)+\delta_{\mu\alpha}\right]\,\mathrm{i}\gamma_{\alpha}\,p_\beta (q-r)_\beta\,,\\[2ex]\nonumber
   \left[\mathcal{T}^{(7)}_{A\bar{q}q}(p,q)\right]_{\mu}
   =&\,\frac{{\textrm{1}}}{3}\, \Big\{ \sigma_{\alpha\beta}\gamma_\mu + \sigma_{\beta\mu}\gamma_\alpha + \sigma_{\mu\alpha}\gamma_\beta \Big\}\\[2ex] \nonumber 
   &\,\times \,\left({p+q}\right)_\alpha (p-q)_\beta
	\,,\\[2ex]      
  \left[\mathcal{T}^{(8)}_{A\bar{q}q}(p,q)\right]_{\mu}=&\,\Pi_{\mu\alpha}^{\perp}(p,q-r)\,\mathrm{i}\sigma_{\alpha\beta}p_{\beta}\,p_\sigma(q-r)_\sigma 
\,,
\label{eq:qqA-tensor}
\end{align}
with 
\begin{align}
\sigma_{\mu\nu}=\frac{\mathrm{i}}{2}[\gamma_{\mu},\gamma_{\nu}]\,, \qquad \Pi_{\mu\nu}^{\perp}(p,q)=\delta_{\mu\nu}-\frac{p_{\mu}q_{\nu}}{p_\rho q_\rho}\,, 
\label{eq:Sigmamunun}
\end{align}
see also \cite{Ihssen:2024miv}. For the tensor basis in \labelcref{eq:qqA-tensor}, the first one $\mathcal{T}^{(1)}_{A\bar{q}q}$ is the classical channel in QCD action, which dominates the chiral symmetry breaking. The remainings are non-classical, which vanish on the UV scale and contribute to the fluctuations in RG running. Previous functional QCD studies have revealed that only $\mathcal{T}^{(1,4,7)}_{A\bar{q}q}$ have a sizable impact on the results, and the other non-classical channels $\mathcal{T}^{(2,3,5,6,8)}_{A\bar{q}q}$ are negligible. See \cite{Williams:2014iea,Mitter:2014wpa, Williams:2015cvx,Cyrol:2017ewj,Gao:2020fbl}.  Based on the Lorentz symmetry analysis of Dirac structure, channel $\mathcal{T}^{(4)}_{A\bar{q}q}$ is chiral symmetry breaking, while channel $\mathcal{T}^{(7)}_{A\bar{q}q}$ respects the chiral symmetry. In this work, for the light-quark-gluon vertex, we consider the classical channel $\mathcal{T}^{(1)}_{A\bar{q}q}$ and non-classical channels $\mathcal{T}^{(4)}_{A\bar{q}q}$ and $\mathcal{T}^{(7)}_{A\bar{q}q}$. For the strange-quark-gluon vertex, the classical channel $\mathcal{T}^{(1)}_{A\bar{q}q}$ is included in the computation.

We close this Appendix with a discussion of the higher order scatterings of gluons with a quark--anti-quark pair. This has been studied in great detail in \cite{Mitter:2014wpa} (quenched) and \cite{Cyrol:2017ewj} (unquenched) for two-flavour QCD. 
In \cite{Mitter:2014wpa} relations between the dressings of the different quark-gluon tensor structure have been found numerically, which suggests a good expansion of the quark-gluon terms in local gauge-invariant operators of the type 
\begin{align}
\bar q \slashed{D} q\,,\quad \bar q\, T_{\mu\nu} D_\mu D_\nu q\,,\quad \bar q\, T_{\mu\nu\rho } D_\mu D_\nu D_\rho q\,,\  \cdots \,. 
\label{eq:QuarGluonInvariants}
\end{align}
\labelcref{eq:QuarGluonInvariants} includes $\bar q\, \slashed{D}^n\, q$, $\bar q \,\sigma_{\mu\nu} F_{\mu\nu} q$ and further local invariants. This allows us to infer the dressings of higher order scattering terms with $n$ gluons from lower order ones. The reliability of this gauge-consistent expansion has been corroborated in \cite{Cyrol:2017ewj}, where 18 tensor structures for the two-gluon--quark--anti-quark vertex (three derived from $\bar q\, \slashed{D}^2\, q$ and 15 derived from $ \bar q\, \slashed{D}^3\, q$), have been considered explicitly. For the three-gluon--quark--anti-quark vertex five tensor structures have been considered, derived from $\int_x \bar q\, \slashed{D}^3\, q$. The latter explicit analysis confirmed the gauge-consistent approximation used in \cite{Mitter:2014wpa}, including efficient importance ordering. 

Accordingly, in the present work we use the approach developed in \cite{Mitter:2014wpa} and only consider the most relevant higher order term, a chiral symmetry breaking term in \labelcref{eq:QuarGluonInvariants}, related to the ${\cal T}^{(2)}_{A\bar q q}$ and ${\cal T}^{(4)}_{A\bar q q}$ tensor structures: it has been found in \cite{Mitter:2014wpa}, that the dressings $\lambda_{A\bar q q}^{(2)}\,,\,\lambda_{A\bar q q}^{(4)}$ satisfy the relation 
\begin{align} 
\lambda_{A\bar q q}^{(2)}(p)\approx  \frac12 \lambda_{A\bar q q}^{(4)}(p)\,. 
\label{eq:lambda2=12lambda4}
\end{align}
In view of the invariants \labelcref{eq:QuarGluonInvariants} this suggests the existence of a local operator $ \bar q \,T_{\mu\nu}\, D_\mu D_\nu\,q$ with the tensor structure 
\begin{align}
T_{\mu\nu} = \delta_{\mu\nu}-2 \,\imag\, \sigma_{\mu\nu} \,,
\label{eq:Tmunu}
\end{align}
with \labelcref{eq:Sigmamunun} and the dressing $\lambda_4$. From this term we derive a two-gluon--quark--anti-quark vertex with the tensor structure 
\begin{align}
   \left[{\cal T}_{AA\bar{q}q}\right]^{ab}_{\mu\nu} =\delta_{\mu\nu}\left\{ T^a_c \,,\, T^b_c\right\}+2\,\sigma_{\mu\nu}\, f^{abc} T_c^c \,.
\end{align}
%
%
%
and a symmetric point dressing 
\begin{align}
\lambda_{AA\bar{q}q}(p)=\frac{1}{\sqrt{p^2}}\,\lambda_{A\bar{q}q}^{(4)}(p)\,.
\end{align}
The impact of the other symmetry-breaking channels is subleading, see also \cite{Mitter:2014wpa, Cyrol:2017ewj}. In summary this leads us to the approximation of the two-quark-two-gluon vertex used in the present work, 
\begin{align}\nonumber 
&   \Pi_{\mu\rho}^\bot(p)\Pi_{\nu\sigma}^\bot(p) \left(\Gamma_{AA \bar{q}q }^{(4)}\right)^{ab}_{\rho\sigma}(p,q,r,s)\\[2ex]
    &\hspace{0.5cm}=(2 \pi)^4\delta(p+q+r+s)\,
    Z_{A}^{\frac{1}{2}}(p) Z_{A}^{\frac{1}{2}}(q) Z_{q}^{\frac{1}{2}}(r)Z_{q}^{\frac{1}{2}}(s)
    \nonumber\\[2ex]
    &\hspace{0.9cm}\times \Pi_{\mu\rho}^\bot(p)\Pi_{\nu\sigma}^\bot(p)\lambda_{AA\bar{q}{q}}(\bar{p})\left[\mathcal{T}_{AA\bar{q}q}(p,q)\right]^{ab}_{\rho\sigma}\,.
    \label{eq:Gamma-qqAA}
\end{align}
This concludes our analysis of the glue-matter interface of the effective action.  

\section{Truncation of the pure matter sector}
\label{app:truncation-matter}

Here we discuss the tensor structure and the symmetry relations of $2+1$ flavour Fierz-complete basis for the four-quark vertex, see also the review \cite{Klevansky:1992qe}. The two-flavour basis has been summarised and used in our previous works \cite{Fu:2022uow, Fu:2024ysj} and related fRG works \cite{Mitter:2014wpa, Cyrol:2017ewj, Braun:2020mhk}.

We have already discussed in \Cref{app:truncation-qqA}, that in the present isospin symmetric approximation, the mass matrix is diagonal, see \labelcref{eq:quark-twopoint}. 
For the $2+1$ flavour QCD, the quark mass can be represented by a diagonal mass matrix 
\begin{align}
    M_q=\left(\begin{array}{ccc}
    M_l & 0 & 0 \\
    0 & M_l & 0 \\
    0 & 0 & M_s
    \end{array}\right)\,,\label{eq:Nf2+1-mq}
\end{align}
here we set the masses of the up and down quark equal to the light quark mass $m_{l}$. This sector is invariant under the transformation of the $\mathrm{SU}_{\mathrm{V}}(N_{f}=2)$ flavour symmetry. However, the strange quark mass $M_{s}$ is significantly bigger than the $M_{l}$ and this leads to breaking the $\mathrm{SU}_{\mathrm{V}}(N_{f}=3)$ flavour group. If we only maintain the smallest symmetry $\mathrm{U}_\mathrm{V}(1) \otimes \mathrm{SU}_\mathrm{V}(N_{f}=2) \otimes \mathrm{SU}(N_{c})$, a full basis in $2+1$ flavour QCD is provided by
\begin{align}
    \left\{\begin{array}{c}
    \mathbb{1}_D \otimes \mathbb{1}_D \\[2ex]
    \gamma_5 \otimes \gamma_5 \\[2ex]
    \gamma_\mu \otimes \gamma_\mu \\[2ex]
    \gamma_\mu \gamma_5 \otimes \gamma_\mu \gamma_5 \\[2ex]
    \sigma_{\mu \nu} \otimes \sigma_{\mu \nu}
    \end{array}\right\}\left\{\begin{array}{c}
    \mathbb{1}_{f} \otimes \mathbb{1}_{f} \\[2ex]
    \sum_{i=1}^3 T^{i} \otimes T^{i} \\[2ex]
    \sum_{i=4}^7 T^{i} \otimes T^{i} \\[2ex]
    T^{8} \otimes T^{8} \\[2ex]
    \mathbb{1}_{f} \otimes T^{8}
    \end{array}\right\}\left\{\begin{array}{c}
    \mathbb{1}_c \otimes \mathbb{1}_c \\[2ex]
    T_c^a \otimes T_c^a
    \end{array}\right\}\,,
    \label{eq:25Channels}
\end{align}
with $\sigma_{\mu\nu}$ in \labelcref{eq:Sigmamunun}. Here, $T^{i}$ and $T_c^{a}$ denote the generators in the fundamental representation of the flavour group and the colour group respectively. Because of the Fierz transformation, there are 26 linear independent four-quark interaction channels. Here we clarify them according to the group structure in the flavour space. The tensor structures in the basis with the flavor group being the identity element $T^{0}$ include,
\begin{align}
    {\cal T}^{(S-P)_{\text{SIN}}}_{ijkm}=&\,(T^0)_{ij}(T^0)_{km} -(\gamma_{5}T^0)_{ij}(\gamma_{5}T^0)_{km}\,,\nonumber\\[1ex] 
    {\cal T}^{(S+P)_{\text{SIN}}}_{ijkm}=&\,(T^0)_{ij}(T^0)_{km} +(\gamma_{5}T^0)_{ij}(\gamma_{5}T^0)_{km}\,,\nonumber\\[1ex]
    {\cal T}^{(S+P)^{\text{adj}}_{\text{SIN}}}_{ijkm}=&\,(T^0 T_{c}^{a})_{ij}(T^0 T_{c}^{a})_{km} \nonumber\\[1ex]&\,+(\gamma_{5}T^0 T_{c}^{a})_{ij}(\gamma_{5}T^0 T_{c}^{a})_{km}\,,\nonumber\\[1ex]
    {\cal T}^{(V+A)_{\text{SIN}}}_{ijkm}=&\,(\gamma_{\mu}T^0)_{ij}(\gamma_{\mu}T^0)_{km} \nonumber\\[1ex]&\,-(\gamma_{\mu}\gamma_{5}T^0)_{ij}(\gamma_{\mu}\gamma_{5}T^0)_{km}\,,\nonumber\\[1ex]
    {\cal T}^{(V-A)^{\text{adj}}_{\text{SIN}}}_{ijkm}=&\,(\gamma_{\mu}T^0 T_{c}^{a})_{ij}(\gamma_{\mu}T^0 T_{c}^{a})_{km} \nonumber\\[1ex]&\,+(\gamma_{\mu}\gamma_{5}T^0 T_{c}^{a})_{ij}(\gamma_{\mu}\gamma_{5}T^0 T_{c}^{a})_{km}\,.
\label{eq:4qBasisI}
\end{align}
The basis that includes $T^{(1-3)}$, which describes the interaction between light quarks, is given by
\begin{align}
    {\cal T}^{(S-P)_{\text{ISO}}}_{ijkm}=&\,(T^{(1-3)})_{ij}(T^{(1-3)})_{km} \nonumber\\[1ex]&\,-(\gamma_{5}T^{(1-3)})_{ij}(\gamma_{5}T^{(1-3)})_{km}\,,\nonumber\\[1ex] 
    {\cal T}^{(S+P)_{\text{ISO}}}_{ijkm}=&\,(T^{(1-3)})_{ij}(T^{(1-3)})_{km} \nonumber\\[1ex]&\,+(\gamma_{5}T^{(1-3)})_{ij}(\gamma_{5}T^{(1-3)})_{km}\,,\nonumber\\[1ex]
    {\cal T}^{(S+P)^{\text{adj}}_{\text{ISO}}}_{ijkm}=&\,(T^{(1-3)} T_{c}^{a})_{ij}(T^{(1-3)} T_{c}^{a})_{km} \nonumber\\[1ex]&\,+(\gamma_{5}T^{(1-3)} T_{c}^{a})_{ij}(\gamma_{5}T^{(1-3)} T_{c}^{a})_{km}\,,\nonumber\\[1ex]
    {\cal T}^{(V+A)_{\text{ISO}}}_{ijkm}=&\,(\gamma_{\mu}T^{(1-3)})_{ij}(\gamma_{\mu}T^{(1-3)})_{km} \nonumber\\[1ex]&\,-(\gamma_{\mu}\gamma_{5}T^{(1-3)})_{ij}(\gamma_{\mu}\gamma_{5}T^{(1-3)})_{km}\,,\nonumber\\[1ex]
    {\cal T}^{(V-A)^{\text{adj}}_{\text{ISO}}}_{ijkm}=&\,(\gamma_{\mu}T^{(1-3)} T_{c}^{a})_{ij}(\gamma_{\mu}T^{(1-3)} T_{c}^{a})_{km} \nonumber\\[1ex]&\,+(\gamma_{\mu}\gamma_{5}T^{(1-3)} T_{c}^{a})_{ij}(\gamma_{\mu}\gamma_{5}T^{(1-3)} T_{c}^{a})_{km}\,.
\label{eq:4qBasisII}
\end{align}
The basis, that includes $T^{(4-7)}$ and describes the interaction between light and strange quarks, is given by
\begin{align}
    {\cal T}^{(S-P)_{\text{CRO}}}_{ijkm}=&\,(T^{(4-7)})_{ij}(T^{(4-7)})_{km} \nonumber\\[1ex]&\,-(\gamma_{5}T^{(4-7)})_{ij}(\gamma_{5}T^{(4-7)})_{km}\,,\nonumber\\[1ex] 
    {\cal T}^{(S+P)_{\text{CRO}}}_{ijkm}=&\,(T^{(4-7)})_{ij}(T^{(4-7)})_{km} \nonumber\\[1ex]&\,+(\gamma_{5}T^{(4-7)})_{ij}(\gamma_{5}T^{(4-7)})_{km}\,,\nonumber\\[1ex]
    {\cal T}^{(S+P)^{\text{adj}}_{\text{CRO}}}_{ijkm}=&\,(T^{(4-7)} T_{c}^{a})_{ij}(T^{(4-7)} T_{c}^{a})_{km} \nonumber\\[1ex]&\,+(\gamma_{5}T^{(4-7)} T_{c}^{a})_{ij}(\gamma_{5}T^{(4-7)} T_{c}^{a})_{km}\,,\nonumber\\[1ex]
    {\cal T}^{(V+A)_{\text{CRO}}}_{ijkm}=&\,(\gamma_{\mu}T^{(4-7)})_{ij}(\gamma_{\mu}T^{(4-7)})_{km} \nonumber\\[1ex]&\,-(\gamma_{\mu}\gamma_{5}T^{(4-7)})_{ij}(\gamma_{\mu}\gamma_{5}T^{(4-7)})_{km}\,,\nonumber\\[1ex]
    {\cal T}^{(V-A)^{\text{adj}}_{\text{CRO}}}_{ijkm}=&\,(\gamma_{\mu}T^{(4-7)} T_{c}^{a})_{ij}(\gamma_{\mu}T^{(4-7)} T_{c}^{a})_{km} \nonumber\\[1ex]&\,+(\gamma_{\mu}\gamma_{5}T^{(4-7)} T_{c}^{a})_{ij}(\gamma_{\mu}\gamma_{5}T^{(4-7)} T_{c}^{a})_{km}\,.
\label{eq:4qBasisIII}
\end{align}
The basis that includes $T^{8}$ is given by
\begin{align}
    {\cal T}^{(S-P)_{\text{HYP}}}_{ijkm}=&\,(T^{8})_{ij}(T^{8})_{km} -(\gamma_{5}T^{8})_{ij}(\gamma_{5}T^{8})_{km}\,,\nonumber\\[1ex] 
    {\cal T}^{(S+P)_{\text{HYP}}}_{ijkm}=&\,(T^{8})_{ij}(T^{8})_{km} +(\gamma_{5}T^{8})_{ij}(\gamma_{5}T^{8})_{km}\,,\nonumber\\[1ex]
    {\cal T}^{(S+P)^{\text{adj}}_{\text{HYP}}}_{ijkm}=&\,(T^{8} T_{c}^{a})_{ij}(T^{8} T_{c}^{a})_{km} \nonumber\\[1ex]&\,+(\gamma_{5}T^{8} T_{c}^{a})_{ij}(\gamma_{5}T^{8} T_{c}^{a})_{km}\,,\nonumber\\[1ex]
    {\cal T}^{(V+A)_{\text{HYP}}}_{ijkm}=&\,(\gamma_{\mu}T^{8})_{ij}(\gamma_{\mu}T^{8})_{km} \nonumber\\[1ex]&\,-(\gamma_{\mu}\gamma_{5}T^{8})_{ij}(\gamma_{\mu}\gamma_{5}T^{8})_{km}\,,\nonumber\\[1ex]
    {\cal T}^{(V-A)^{\text{adj}}_{\text{HYP}}}_{ijkm}=&\,(\gamma_{\mu}T^{8} T_{c}^{a})_{ij}(\gamma_{\mu}T^{8} T_{c}^{a})_{km} \nonumber\\[1ex]&\,+(\gamma_{\mu}\gamma_{5}T^{8} T_{c}^{a})_{ij}(\gamma_{\mu}\gamma_{5}T^{8} T_{c}^{a})_{km}\,,
\label{eq:4qBasisIV}
\end{align}
and the basis which involves the mixing of $T^{0}$ and $T^{8}$ is given by
\begin{align}
    {\cal T}^{(S-P)_{\text{MIX}}}_{ijkm}=&\,(T^{0})_{ij}(T^{8})_{km} + (T^{8})_{ij}(T^{0})_{km} \nonumber\\[1ex]&\,-(\gamma_{5}T^{0})_{ij}(\gamma_{5}T^{8})_{km}-(\gamma_{5}T^{8})_{ij}(\gamma_{5}T^{0})_{km}\,,\nonumber\\[1ex] 
    {\cal T}^{(S+P)_{\text{MIX}}}_{ijkm}=&\,(T^{0})_{ij}(T^{8})_{km} + (T^{8})_{ij}(T^{0})_{km} \nonumber\\[1ex]&\,+(\gamma_{5}T^{0})_{ij}(\gamma_{5}T^{8})_{km}+(\gamma_{5}T^{8})_{ij}(\gamma_{5}T^{0})_{km}\,,\nonumber\\[1ex] 
    {\cal T}^{(S+P)^{\text{adj}}_{\text{MIX}}}_{ijkm}=&\,(T^{0} T_{c}^{a})_{ij}(T^{8} T_{c}^{a})_{km} + (T^{8} T_{c}^{a})_{ij}(T^{0} T_{c}^{a})_{km} \nonumber\\[1ex]&\,+(\gamma_{5}T^{0} T_{c}^{a})_{ij}(\gamma_{5}T^{8} T_{c}^{a})_{km}\nonumber\\[1ex]&\,+(\gamma_{5}T^{8} T_{c}^{a})_{ij}(\gamma_{5}T^{0} T_{c}^{a})_{km}\,,\nonumber\\[1ex]
    {\cal T}^{(V+A)_{\text{MIX}}}_{ijkm}=&\,(\gamma_{\mu}T^{0})_{ij}(\gamma_{\mu}T^{8})_{km}+(\gamma_{\mu}T^{8})_{ij}(\gamma_{\mu}T^{0})_{km} \nonumber\\[1ex]&-(\gamma_{\mu}\gamma_{5}T^{0})_{ij}(\gamma_{\mu}\gamma_{5}T^{8})_{km}\nonumber\\[1ex]&\,-(\gamma_{\mu}\gamma_{5}T^{8})_{ij}(\gamma_{\mu}\gamma_{5}T^{0})_{km}\,,\nonumber\\[1ex]
    {\cal T}^{(V-A)^{\text{adj}}_{\text{MIX}}}_{ijkm}=&\,(\gamma_{\mu}\gamma_{5}T^{0} T_{c}^{a})_{ij}(\gamma_{\mu}\gamma_{5}T^{8} T_{c}^{a})_{km}\nonumber\\[1ex]&\,+(\gamma_{\mu}\gamma_{5}T^{8} T_{c}^{a})_{ij}(\gamma_{\mu}\gamma_{5}T^{0} T_{c}^{a})_{km}\nonumber\\[1ex]&\,+(\gamma_{\mu}T^{0} T_{c}^{a})_{ij}(\gamma_{\mu}T^{8} T_{c}^{a})_{km} \nonumber\\[1ex]&\,+ (\gamma_{\mu}T^{8} T_{c}^{a})_{ij}(\gamma_{\mu}T^{0} T_{c}^{a})_{km}\,.
\label{eq:4qBasisV}
\end{align}
At last, in addition to the channels listed above, we also need to take into account the channel related to the $U_{A}(1)$ symmetry breaking ’t Hooft determinant, which reads
\begin{align}
     {\cal T}^{A}_{ijkm}=&\,(T^0)_{ij}(T^0)_{km} +(\gamma_{5}T^0)_{ij}(\gamma_{5}T^0)_{km}\nonumber\\[1ex]
     &-\,(T^{(4-7)})_{ij}(T^{(4-7)})_{km} \nonumber\\[1ex]&\,-(\gamma_{5}T^{(4-7)})_{ij}(\gamma_{5}T^{(4-7)})_{km}\,.
\end{align}
In a final step we change the basis \labelcref{eq:4Tensors2+1} guided by a quest for optimising the approximation: we want to only keep the four 
tensors with the lowest lying resonances, related to the $\sigma,\pi,\kappa,K$.mesons, see the discussion in \Cref{sec:Gamma4q}. However, the basis \labelcref{eq:4Tensors2+1} is not diagonal and the flows of the respective dressings mix different channels even if the basis is diagonalised. The respective optimisation task will be discussed elsewhere, \cite{TensorBases2025}, and goes beyond the scope of the present work. Hence we simply 
define combinations of tensors in the complete basis \labelcref{eq:4Tensors2+1} that increase significantly the overlap with the resonant channels and in turn decrease the overlap of the other tensors with the low lying resonances. This leads us to $\sigma,\pi,\kappa,K$-channels defined by 
\begin{align}
    {\cal T}^{\sigma}_{ijkm}&=\left({\cal T}^{(S+P)_{\mathrm{SIN}}}_{ijkm}+{\cal T}^{(S-P)_{\mathrm{SIN}}}_{ijkm}\right)/2\,,\nonumber\\[1ex]
    {\cal T}^{\pi}_{ijkm}&=\left({\cal T}^{(S+P)_{\mathrm{ISO}}}_{ijkm}-{\cal T}^{(S-P)_{\mathrm{ISO}}}_{ijkm}\right)/2\,,\nonumber\\[1ex]
    {\cal T}^{\kappa}_{ijkm}&=\left({\cal T}^{(S+P)_{\mathrm{CRO}}}_{ijkm}+{\cal T}^{(S-P)_{\mathrm{CRO}}}_{ijkm}\right)/2\,,\nonumber\\[1ex]
    {\cal T}^{K}_{ijkm}&=\left({\cal T}^{(S+P)_{\mathrm{CRO}}}_{ijkm}-{\cal T}^{(S-P)_{\mathrm{CRO}}}_{ijkm}\right)/2\,, 
\label{eq:4qBasis-sigmaPiKkappa}
\end{align}
with the explicit form provided in \labelcref{eq:4Tensors2+1} in \Cref{sec:Gamma4q}. Note that such a choice is not unique and the optimisation task requires an analysis of the complete basis. 

Finally, we discuss the separation in symmetric and anti-symmetric parts of given tensors. While this extends the given momentum-independent basis in a natural way, it also lacks the full systematics of taking into account all $p^2$-tensor elements, for further discussions see \cite{Fu:2024ysj}. From the definition of the four-quark scattering vertex, 
\begin{align}
 \Gamma^{(4)}_{\bar q_i q_j \bar q_k q_m}(\boldsymbol{p})=\frac{\delta^4 \Gamma_k[q,\bar q] }{ \delta \bar q_i(p_1)\delta q_j(p_2) \delta \bar q_k(p_3)\delta q_m(p_4)} \,, 
 \label{eq:DefofGam4}
\end{align}
we arrive at 
\begin{align}\nonumber 
 \Gamma_{4q,k}^{(4)}[q,\bar q] =&-4\, (2 \pi)^4\prod_{i=1}^4 Z_{q}^{\frac{1}{2}}(p_{i}) \,\delta\left(\sum_{i=1}^4 p_i\right)\\[2ex]
 & \times\sum_{\alpha}\left[\lambda_{\alpha}(\boldsymbol{p}){\cal T}^{(\alpha)}_{ijkm}-\lambda_{\alpha}(\boldsymbol{p}'){\cal T}^{(\alpha)}_{kjim}\right]\,, 
 \label{eq:Gamma4q-1}
\end{align}
with $\boldsymbol{p}=(p_{1},p_{2},p_{3},p_{4})$ defined in \labelcref{eq:boldp} and $\boldsymbol{p}'=(p_{3},p_{2},p_{1},p_{4})$. \Cref{eq:Gamma4q-1} comprises the crossing symmetry of the vertex that arises from the anti-commutation of anti-quark derivatives.  
Accordingly, we can split a vertex contribution from a given tensor structure $ {\cal T}^{(\alpha-)}$ into two parts, proportional to  
the symmetric and anti-symmetric components of the tensor, 
\begin{align}\nonumber 
    {\cal T}^{(\alpha-)}_{ijkm}=&\,\frac{1}{2}\left({\cal T}^{(\alpha)}_{ijkm}-{\cal T}^{(\alpha)}_{kjim}\right)\,,\\[2ex]
     {\cal T}^{(\alpha+)}_{ijkm}=&\,\frac{1}{2}\left({\cal T}^{(\alpha)}_{ijkm}+{\cal T}^{(\alpha)}_{kjim}\right)\,. 
    \label{eq:Tensor-SymAsym}
\end{align}
The respective symmetric and anti-symmetric dressings are given by 
\begin{align}\nonumber 
 \lambda_{\alpha}^{+}(p_1,&p_2,p_3,p_4)\\[2ex]\nonumber 
 \equiv&\frac{1}{2}\Big[\lambda_{\alpha}(p_1,p_2,p_3,p_4)+\lambda_{\alpha}(p_3,p_2,p_1,p_4)\Big]\,,\\[2ex]\nonumber 
 \lambda_{\alpha}^{-}(p_1,&p_2,p_3,p_4)\\[2ex]
 \equiv&\frac{1}{2}\Big[\lambda_{\alpha}(p_1,p_2,p_3,p_4)-\lambda_{\alpha}(p_3,p_2,p_1,p_4)\Big]\,.
 \label{eq:lambdaM}
\end{align}
Note that the anti-symmetric dressings $\lambda_{\alpha}^{-}$ are proportional to momentum-squared times regular terms and hence the respective tensor structure if a momentum-dependent one. As indicated above, we have simply taken into account the small part of all momentum-dependent tensors of the order momentum squared, suggested by crossing symmetry. 

In summary, \labelcref{eq:Tensor-SymAsym} and \labelcref{eq:lambdaM} allow us to rewrite \labelcref{eq:Gamma4q-1} as  
\begin{align}\nonumber 
 \Gamma_{4q,k}^{(4)}[q,\bar q] =&-4\, (2 \pi)^4\prod_{i=1}^4 Z_{q}^{\frac{1}{2}}(p_{i}) \,\delta\left(\sum_{i=1}^4 p_i\right)\\[2ex]
 & \hspace{-0.2cm}\times\sum_{\alpha}\left[\lambda_{\alpha}^{+}(\boldsymbol{p})\,{\cal T}^{(\alpha-)}_{ijkm}-\lambda_{\alpha}^{-}(\boldsymbol{p})\,{\cal T}^{(\alpha+)}_{kjim}\right]\,. 
 \label{eq:Gamma4q-final}
\end{align}
In this work, we reduce the tensor structures, considered in the computation to $\alpha=\alpha\in\{\sigma,\pi,\kappa, K$.  Within the crossing-symmetric procedure introduced above these are eight tensors with the dressings $\lambda_{\alpha}^{+}(\boldsymbol{p})$ and $\lambda_{\alpha}^{-}(\boldsymbol{p})$.

\section{Independence on the initial scale}
\label{app:UVLimit}

In the present computation we start the flow at the initial cutoff scale $\Lambda= 35.7$\,GeV with the initial effective action 
\begin{align} 
\Gamma_\Lambda[\Phi]=S_\textrm{QCD}[\Phi] + \int_x m^2_{\textrm{scal},\Lambda} \,(A^a_\mu)^2 \,. 
\label{eq:GammaInitial}
\end{align}
Here, the initial mass parameter of the gluon leads to the scaling solution at $k=0$. We emphasise that it is not a free mass parameter and the mass parameter only reflects the breaking of gauge symmetry due to the regulators, encoded in the mSTIs. It is a function of the three relevant parameters of 2+1 flavour QCD,
\begin{align}
m^2_{\textrm{scal},\Lambda}=m^2_{\textrm{scal},\Lambda}(\alpha_{s;\Lambda},m_l,m_s)\,. 
\label{eq:InitialScalingMass}
\end{align}
This has been discussed in detail in \cite{Fischer:2008uz, Cyrol:2016tym, Cyrol:2017ewj}, see also the reviews \cite{Dupuis:2020fhh, Fu:2022gou}. The strong coupling in the classical action and the light and strange current quark masses are presented in \Cref{tab:Parameters}. 

This initial condition is only approximate as $\Gamma_\Lambda[\Phi]$ already comprises the quantum fluctuations with $k>\Lambda$, leading to non-trivial momentum dependences in $\Gamma_\Lambda[\Phi]$. The latter is needed for guaranteeing the full $\Lambda$-independence of the full effective action, called \textrm{RG-consistency}, see \cite{Pawlowski:2005xe,Braun:2018svj}: The cutoff dependence of the initial effective action satisfies the flow equation itself. The flow of the irrelevant operators can be expanded in terms of inverse powers of the initial scale $\Lambda^2$ and this expansion shows a rapid convergence for sufficiently large cutoffs. 
However, since all momentum dependences in the initial effective action are suppressed with powers of $p^2/\Lambda^2$ for the momenta under investigation here $p\ll \Lambda$, and the quantitative nature of the approximation \labelcref{eq:GammaInitial} is achieved rapidly by increasing the initial cutoff scale, or rather decreasing the initial strong coupling. This has been tested thoroughly in particular in \cite{Cyrol:2017ewj}, where an initial cutoff $\Lambda\approx 10^2 $\,GeV has been used, and even higher ones have been tested. Moreover, a respective analysis in pure Yang-Mills theory has been performed in \cite{Cyrol:2016tym}.

\begin{table}[t]
  \begin{center}
  \begin{tabular}{|c|c|c|c|c}
    \hline & & & \\[-2ex]
    $\Lambda$\,[GeV]  & $\alpha_{s}$ & \,$(m_{l},m_{s})$\,[MeV]\,  & \,$({m_\pi}/{f_\pi},{m_K}/{f_\pi})\, $ \\[1ex]
    \hline & & &   \\[-2ex]
    333.3 & \, 0.1146  \, & (2,\,40) & \, (${149}/{93},\,{481}/{93})$ \, \\[1ex]
    \hline & & &   \\[-2ex]
    90 & \, 0.1268  \, & (2.19,\,43.70) & \, (${149}/{93},\,{481}/{93})$ \, \\[1ex]
    \hline & & &   \\[-2ex]
    30 & \, 0.1438  \, & (2.47,\,48.79) & \, (${149}/{93},\,{481}/{93})$ \, \\[1ex]
    
    \hline
  \end{tabular}
  \caption{Initial conditions and parameters at different initial scales. Specifically, the initial conditions for $\Lambda=90\,$GeV and $\Lambda=30\,$GeV are obtained from the results of $\Lambda=333.3\,$GeV.}
  \label{tab:UV-Parameters}
  \end{center}\vspace{-0.5cm}
\end{table}
%

\begin{figure}[t]
	\includegraphics[width=0.48\textwidth]{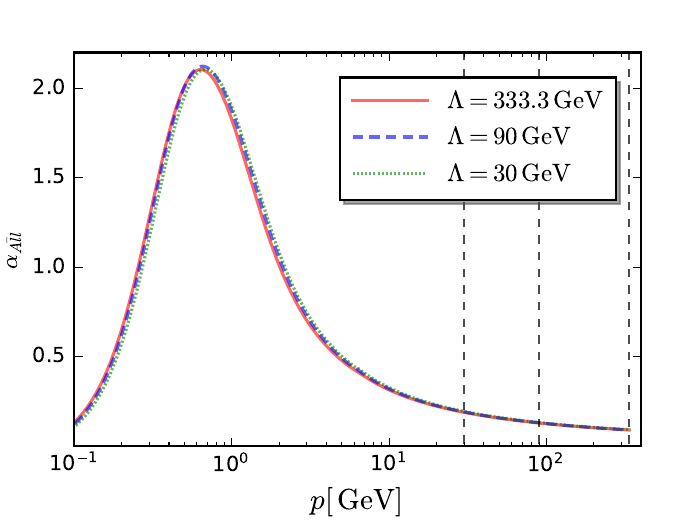}
	\caption{Quark-gluon avatars of $\alpha_{A\bar l l}(p)$ at $k=0$ as functions of the momentum obtained from flows from different initial cutoff scales $\Lambda=30\,,90$\,GeV and $\Lambda=333.3$\,GeV. The latter setup is described in \labelcref{eq:InitialCondsLargeLambda}. The initial conditions for the smaller $\Lambda$ is described around \labelcref{eq:DerivedInitialConds}.}
	\label{fig:alpha-UV}
\end{figure}

\begin{figure}[b]
	\includegraphics[width=0.48\textwidth]{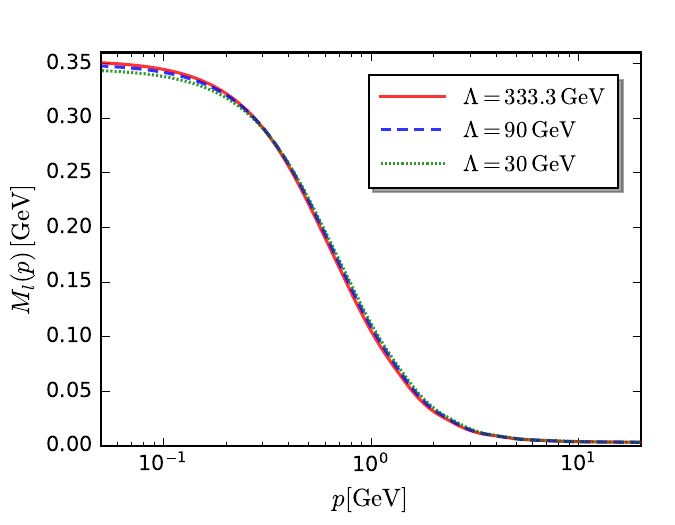}
	\caption{Light quark mass function $M_l(p)$ at $k=0$ as functions of the momentum obtained from flows from different initial cutoff scales $\Lambda=30\,,90$\,GeV and $\Lambda=333.3$\,GeV. The latter setup is described in \labelcref{eq:InitialCondsLargeLambda}. The initial conditions for the smaller $\Lambda$ is described around \labelcref{eq:DerivedInitialConds}.}
	\label{fig:Mq-UV}
\end{figure}

In short, these analyses entail that the current initial cutoff scale $\Lambda=35.7$\,GeV is sufficiently large to lead to converged results for the infrared momenta under investigation within the current error estimate of 10\%. Still, for the sake of completeness, we augment the current computations with an explicit analysis of the $\Lambda$-independence of the results. We shall concentrate on the pure matter sector with its novel ingredient in comparison to \cite{Cyrol:2017ewj}, the $s$ quark and the three-momentum channel approximation for the four-quark vertex. Given the well-tested convergence of the pure glue sector, we do not assess this part again here. 
\begin{subequations}   \label{eq:InitialCondsLargeLambda}
For this test of the pure matter sector we initiate the flow at an asymptotically large initial scale or rather a small coupling,  
\begin{align}
\alpha_{s,\Lambda_\textrm{UV}} = \frac{1.44}{4 \pi} \approx 0.1146\,,\qquad \Lambda_\textrm{UV}= 333.3\,\mathrm{GeV}\,, 
\label{eq:Largeinitial}
\end{align}
with the current quark masses 
\begin{align}
    m_{l,\Lambda_\textrm{UV}}= 2\,\textrm{MeV}\,,\qquad m_{s,\Lambda_\textrm{UV}} = 40\,\textrm{MeV}\,.
\end{align}
\end{subequations}
These units are measured in the pion decay constant, whose value we simply define with $f_\pi=93$\,GeV. The respective strong coupling and light quark mass function is shown in  \Cref{fig:alpha-UV,fig:Mq-UV} together with results obtained with initial cutoffs $\Lambda=30,90$\,GeV. The latter cases are discussed below in more detail. 

Evidently, the initial conditions for \labelcref{eq:Largeinitial} are not subject to a full fine-tuning of the physical units. In \Cref{tab:UV-Parameters} we have listed the respective ratios of pion and kaon masses with the pion decay constant. They are close but not identical to the physical ones. While full fine-tuning is possible and straightforward, it is tedious. Indeed, this is the, purely numerical, reason for not re-assessing the glue sector again: the quadratic fine-tuning of the scaling mass parameter \labelcref{eq:InitialScalingMass} is the most costly part of the setup. However, in this Appendix we are only interested in the cutoff independence of our approach. This can be fully tested by an analysis in the vicinity of the physical point, and hence we refrain from performing the full fine-tuning. We also would like to add, that the required numerical accuracy of the solver is increasing successively with $\Lambda$. While this can be readily done, it slows down the solver considerably, while adding nothing to the infrared accuracy of our results, for further discussions see \cite{Cyrol:2017ewj}. 

%
%

%
%
Now we initiate further flows at the smaller cutoff scales $\Lambda=90,30$\,GeV by using the results of the initial conditions \labelcref{eq:InitialCondsLargeLambda} in the following way: We read out an initial condition for the classical initial action \labelcref{eq:GammaInitial} of further initial cutoff scales $\Lambda_i< \Lambda$ from 
\begin{subequations}   \label{eq:InitialConditions-Lambdai}
\begin{align}\nonumber 
\alpha_{A\bar q q,\Lambda_i}= &\,\alpha_{A\bar q q,\Lambda_i}(p=0)\,,\\[2ex]
m_{q,\Lambda_i}=&\,M_{q,\Lambda_i}(p=0)\,,
\label{eq:DerivedInitialConds}
\end{align}    
with the respective scaling mass parameter \labelcref{eq:InitialScalingMass} 
\begin{align}
m^2_{\textrm{scal},\Lambda_i}=m^2_{\textrm{scal},\Lambda_i}(\alpha_{s;\Lambda_i},m_{l,\Lambda_i},m_{s,\Lambda_i})\,. 
\label{eq:InitialScalingMass-i}    
\end{align}
\end{subequations}
The right hand sides of \labelcref{eq:InitialConditions-Lambdai} are the vanishing momentum results of the full momentum-dependent coupling and constituent quark masses obtained from the UV setup 
\labelcref{eq:Largeinitial}. In short, we define a classical initial condition in the matter sector by reducing the full momentum-dependent quark-gluon vertex and mass function to classical ones, dropping the entire four-quark sector and re-initiating the flow. We keep the full momentum dependences of the glue sector as this facilitates the quadratic fine-tuning problem \labelcref{eq:InitialScalingMass-i}. If this reduction leads to agreeing results with the UV setup \labelcref{eq:InitialCondsLargeLambda}, this proves two properties: (1) the independence of our results of the initial cutoff and (ii) the quantitative accuracy of the approximation \labelcref{eq:GammaInitial}. Note that (i) entails that we can safely take the limit $\Lambda\to \infty$, the current setup is UV safe and only has the three relevant parameters of 2+1 flavour QCD, and only two of them, the masses, determine the physical point. 

%
\begin{figure}[t]
	\includegraphics[width=0.48\textwidth]{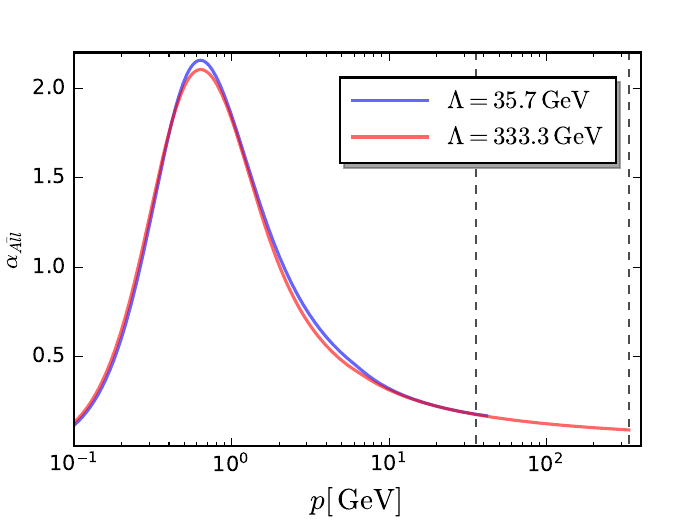}
	\caption{Quark-gluon avatars of $\alpha_{A\bar l l}(p)$ at $k=0$ as functions of the momentum obtained from flows from different initial cutoff scales $\Lambda=30\,,90$\,GeV and $\Lambda=333.3$\,GeV. The setups are described in \labelcref{eq:InitialConditions,eq:InitialCondsLargeLambda}.}
	\label{fig:alpha-UV-v1}
\end{figure}
%
The results obtained within this procedure show a quantitative agreement for all the tested initial cutoff scales, and we illustrate this with the two relevant correlation functions, the light quark-gluon coupling $\alpha_{A,\bar l l}(p)$ and the light constituent quark mass at $k=0$, see \Cref{fig:alpha-UV,fig:Mq-UV}. Moreover, in \Cref{tab:UV-Parameters} we also list the ratios $m_\pi/f_\pi$ and $m_K/f_\pi$. These observables and all correlation functions obtained from the initial confitions at $\Lambda=30,90$\,GeV agree quantitatively with that obtained from the UV initial condition $\Lambda=333.3$\,GeV. Note also that the small deviations also reflect the fact that the current simple read-out procedure overestimates the error and the results converge even more if the initial conditions are fully tuned to the same physics observables $m_\pi/f_\pi$ and $m_K/f_\pi$ at $k=0$. 

Finally, we compare the light quark-gluon avatar of the strong coupling from the computation with $\Lambda=333.3$\,GeV with that at the physical point with $\Lambda=35.7$\,GeV: In \Cref{fig:alpha-UV-v1} we show $\alpha_{A\bar l l}(p)$ obtained from both initial conditions. Note that the results from the initial cutoff scale $\Lambda_\textrm{UV}=333.3$\,GeV can be obtained with a flow from $\Lambda=35.7$\,GeV by using the full effective action $\Gamma_{k=35.7\,\textrm{GeV}}$, obtained from the $\Lambda_\textrm{UV}$ flow. Since the light quark  masses and the respective pion mass differ by less than 10\%, the difference in the strong coupling is dominated by the lack of momentum dependence in the initial condition \labelcref{eq:GammaInitial} for $\Lambda=35.7$\,GeV as well from dropping the initial four-quark couplings. The differences are negligible. 

We conclude that the current analysis proves impressively the first principle nature of the approach. We rush to add that respective analyses, both numerically and conceptually, have been presented in the literature before for QCD in the past three decades. In the context of the specific approach here see in particular \cite{Braun:2014ata, Mitter:2014wpa, Cyrol:2016tym, Cyrol:2017ewj, Cyrol:2017qkl, Fu:2019hdw, Ihssen:2024miv}, and for a complete survey we refer to the reviews \cite{Dupuis:2020fhh, Fu:2022gou}.

\section{Flow Equations}
\label{app:Flow}

By now there exists a plethora of fRG works in QCD that include technical details on the derivation and the explicit form of the flows for the correlation functions. The present work draws in particular from \cite{Braun:2014ata,Mitter:2014wpa, Rennecke:2015eba, Cyrol:2017ewj, Corell:2018yil, Fu:2019hdw, Braun:2023qak, Ihssen:2024miv} and we refer to the discussions there. Moreover, we use smooth regulators, see \labelcref{eq:Regs} which are more adapted to the advanced momentum-dependent approximation used in the present work. For respective discussions on the optimisation of these approximations see \cite{Pawlowski:2005xe, Pawlowski:2015mlf, Zorbach:2024zjx, Ihssen:2024miv}. With these regulators and momentum-dependent correlation functions, all flows can only be represented in terms of momentum integrals, and are represented by quite lengthy expressions. The derivations can be done with \cite{Huber:2011qr, Huber:2019dkb}, a general code framework for solving fRGs can be found in \cite{Sattler:2024ozv}. Below we sketch the derivation of these flows and depict them in a diagrammatic form. 

The flow equations for all correlation functions are derived from that for the effective action, the Wetterich equation \cite{Wetterich:1992yh}. For QCD it reads 
\begin{align}
    \partial_t \Gamma_k[\Phi] = \frac12 \Tr \, G_k[\Phi]\, \partial_t R_k\,,
    \label{eq:FunFlow}
\end{align}
with the propagator matrix in field space 
\begin{align} 
G_{k,\Phi_{i_1}\Phi_{i_2}}[\Phi] = \left[\frac{1}{\Gamma_k^{(2}[\Phi]+ R_k}\right]_{\Phi_{i_1}\Phi_{i_2}}\,, 
\label{eq:Propagators}
\end{align}
with is derived from the path integral, where the regulator term 
\begin{align} 
\Delta S_k[\Phi] = \frac12 \int_p \Phi_{i}(-p)  R_{k,\Phi_i \Phi_j}(p)   \Phi_j(p)\,,
\end{align} 
with $\Phi=(A_\mu ,c ,\bar c, q, \bar q)$, \labelcref{eq:Phi} is added to the classical QCD action \labelcref{eq:SQCD}. The regulator matrix $R_k$ in field space is given by, 
\begin{subequations}
\label{eq:Regs}    
\begin{align}
R_k =\left( \begin{array}{ccccc} 
R_A & 0 & 0    & 0 & 0\\[1ex] 
 0  & 0 & -R_c & 0 & 0 \\[1ex]
 0  & R_c & 0 & 0 & 0 \\[1ex]
 0 & 0 & 0 & 0 &  -R_q\\[1ex] 
  0 & 0 & 0 & R_q &  0
 \end{array} \right) 
\label{eq:RegMatrix} 
\end{align}
with the entries 
\begin{align} \nonumber 
R_A(p) = &\, \Pi^\perp(p) \hat Z_A(p)\,p^2\, r_A(x)\,,\\[1ex]\nonumber 
R_c(p)= &\,Z_c(p)\,p^2\,r_c(x) \,, \\[1ex]
R_c(p)= &\,Z_q(p)\,p^2\,r_q(x) \,.  
\label{eq:RegsAll}
\end{align}
and the uniform shape function 
\begin{align}
r_i(x) = r(x)\,,\quad \textrm{with}\quad  r(x) = k^2 e^{-x}\,,\quad x=\frac{p^2}{k^2}\,,  
\end{align}
for all regulators $i= A,c,q$. 
\end{subequations} 
Here $\hat Z_{A}$ is from the reparametrisation of the gluon two-point function, which is given by
\begin{align}
\left(\Gamma_{AA}^{(2)}\,\Pi^\perp \right)_{\mu\nu}^{ab}(p)&=Z_{A}(p)\,p^{2}\,\delta^{ab}\, \Pi_{\perp}^{\mu\nu}(p)\nonumber\\[2ex]&=\hat Z_{A}(p)(p^2+m_{A}^2)\,\delta^{ab}\, \Pi_{\perp}^{\mu\nu}(p)\,.
\label{eq:GamAAreparameter}
\end{align}
At $p=0$, $\hat Z_{A}(0)\equiv 1$ and the flow of the gluon two-point function is entirely contained in the flow of the gluon mass for any $k$. Note that the second line in \labelcref{eq:GamAAreparameter} only 
comprises a convenient reparametrisation of the gluon two-point function. 

With these preparations the flow equations for all correlation functions are simply obtained by taking the respective field derivatives of the flow \labelcref{eq:FunFlow}. 
The diagrammatic representations of all flow equations are provided in \Cref{app:2point,app:3point,app:4point} in  \Cref{fig:twopoint-feyn,fig:threepoint-feyn,fig:fourpoint-feyn} for the two-, three-, and four-point functions respectively. 

\vfill

\newpage 

\begin{widetext} 

\newpage 

\subsection{Flow of the two-point functions}
\label{app:2point}

We consider the flow of all 2+1 flavour QCD two-point functions, see  
\Cref{sec:GammaGlue,sec:GammaInterface}
 and \Cref{app:truncation-gauge,app:truncation-qqA}. The diagrammatic approximation is depicted in \Cref{fig:twopoint-feyn}:  All propagators and vertices on the right hand side are computed from their flows and fed back self-consistently. This also entails that we have dropped diagrams with higher order couplings which are set to zero: 
\begin{itemize} 
\item[(1)] tadpoles in the flows of all two-point functions containing the following four-point functions: two-gluon--ghost--anti-ghost coupling, the quark--anti-quark--ghost--anti-ghost, and the four-strange quark vertices.  
\end{itemize}
Here $\tilde{\partial}_{t}$ is defined as the $t$-derivative which only acts on the regulator in the flow equation. It's the same definition in \Cref{fig:threepoint-feyn} and \Cref{fig:fourpoint-feyn}.
%
\begin{figure*}[h]
	\centering
	\includegraphics[width=0.7\textwidth]{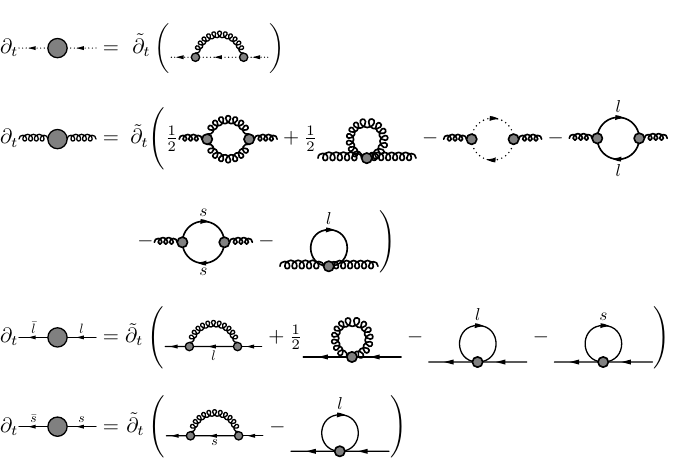}
	\caption{Diagrammatic representation of the flow equations of two-point functions. For the sake of brevity we only depict one diagram of whole classes of diagrams with the same vertices. The operator $\tilde \partial_t$ only hits the $k$-dependence in the explicit regulators terms in the propagators. }
	\label{fig:twopoint-feyn}
\end{figure*}
\vfill 
\ 
\eject 
\ 
\subsection{Flow of the three-point functions}
\label{app:3point}

We consider the flow of the ghost-gluon, three-gluon and quark-gluon three-point functions in 2+1 flavour QCD within the approximation details in \Cref{sec:GammaGlue,sec:GammaInterface}
 and \Cref{app:truncation-gauge,app:truncation-qqA}. The diagrammatic approximation is depicted in \Cref{fig:threepoint-feyn}:  All propagators and vertices on the right hand side are computed from their flows and fed back self-consistently. This also entails that we have dropped diagrams with higher order couplings which are set to zero: 
\begin{itemize} 
\item[(1)] diagrams in the flows of all three-point functions considered here containing the following four-point functions: two-gluon--ghost--anti-ghost coupling, the quark--anti-quark--ghost--anti-ghost, and the four-strange quark vertices.  
\item[(2)] All tadpoles: all five-point vertices are set to zero in the present approximation.  
\end{itemize}
%
%
\begin{figure*}[h]
	\centering
	\includegraphics[width=0.7\textwidth]{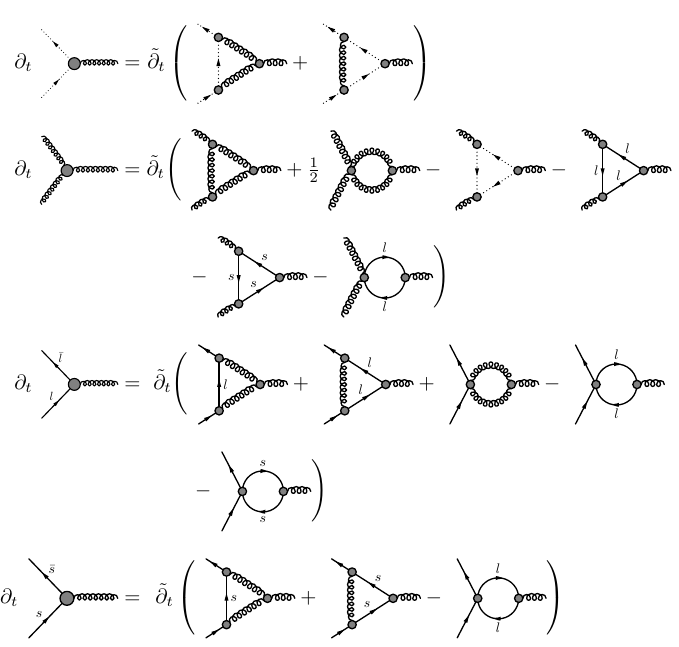}
	\caption{Diagrammatic representation of the flow equations of three-point functions. For the sake of brevity we only depict one diagram of whole classes of diagrams with the same vertices. The operator $\tilde \partial_t$ only hits the $k$-dependence in the explicit regulators terms in the propagators. }
	\label{fig:threepoint-feyn}
\end{figure*}
%
\vfill 

\eject 

\subsection{Flow of the four-point functions}
\label{app:4point}

We consider the flow of the four-gluon and the four- quark four-point functions in 2+1 flavour QCD within the approximation detailed in \Cref{sec:Gamma4q}
 and \Cref{app:truncation-matter}. The diagrammatic approximation is depicted in \Cref{fig:fourpoint-feyn}:  All propagators and vertices on the right hand side are computed from their flows and fed back self-consistently. This also entails that we have dropped diagrams with higher order couplings which are set to zero: 
\begin{itemize} 
\item[(1)] diagrams in the flows of all four-point functions considered containing the following four-point functions: two-gluon--ghost--anti-ghost coupling, the quark--anti-quark--ghost--anti-ghost, and the four-strange quark vertices.   
\item[(2)] All diagrams with a three-point and a five-point function: all five-point vertices are set to zero in the present approximation.  
\item[(3)] All tadpoles: all six-point vertices are set to zero in the present approximation.  
\end{itemize}
%
%
\begin{figure*}[h]
	\centering
	\includegraphics[width=0.7\textwidth]{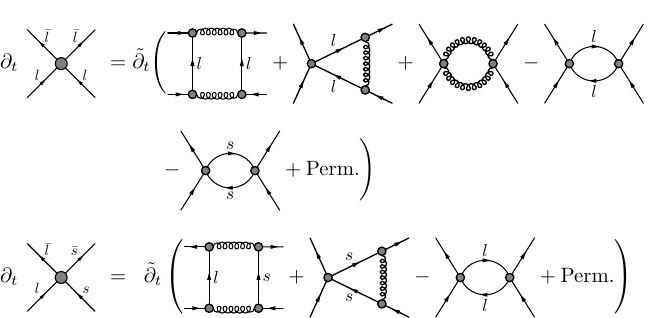}
	\caption{Diagrammatic representation of the flow equations of four-point functions. For the sake of brevity we only depict one diagram of whole classes of diagrams with the same vertices, see e.g.~\Cref{fig:stu-equ} for all different fish diagrams. The operator $\tilde \partial_t$ only hits the $k$-dependence in the explicit regulators terms in the propagators. }
	\label{fig:fourpoint-feyn}
\end{figure*}
%

\vfill 


\end{widetext} 

\ 
\newpage  
\ 
\newpage 

\bibliography{ref-lib}

\begin{thebibliography}{81}%
\makeatletter
\providecommand \@ifxundefined [1]{%
 \@ifx{#1\undefined}
}%
\providecommand \@ifnum [1]{%
 \ifnum #1\expandafter \@firstoftwo
 \else \expandafter \@secondoftwo
 \fi
}%
\providecommand \@ifx [1]{%
 \ifx #1\expandafter \@firstoftwo
 \else \expandafter \@secondoftwo
 \fi
}%
\providecommand \natexlab [1]{#1}%
\providecommand \enquote  [1]{``#1''}%
\providecommand \bibnamefont  [1]{#1}%
\providecommand \bibfnamefont [1]{#1}%
\providecommand \citenamefont [1]{#1}%
\providecommand \href@noop [0]{\@secondoftwo}%
\providecommand \href [0]{\begingroup \@sanitize@url \@href}%
\providecommand \@href[1]{\@@startlink{#1}\@@href}%
\providecommand \@@href[1]{\endgroup#1\@@endlink}%
\providecommand \@sanitize@url [0]{\catcode `\\12\catcode `\$12\catcode
  `\&12\catcode `\#12\catcode `\^12\catcode `\_12\catcode `\%12\relax}%
\providecommand \@@startlink[1]{}%
\providecommand \@@endlink[0]{}%
\providecommand \url  [0]{\begingroup\@sanitize@url \@url }%
\providecommand \@url [1]{\endgroup\@href {#1}{\urlprefix }}%
\providecommand \urlprefix  [0]{URL }%
\providecommand \Eprint [0]{\href }%
\providecommand \doibase [0]{https://doi.org/}%
\providecommand \selectlanguage [0]{\@gobble}%
\providecommand \bibinfo  [0]{\@secondoftwo}%
\providecommand \bibfield  [0]{\@secondoftwo}%
\providecommand \translation [1]{[#1]}%
\providecommand \BibitemOpen [0]{}%
\providecommand \bibitemStop [0]{}%
\providecommand \bibitemNoStop [0]{.\EOS\space}%
\providecommand \EOS [0]{\spacefactor3000\relax}%
\providecommand \BibitemShut  [1]{\csname bibitem#1\endcsname}%
\let\auto@bib@innerbib\@empty
\bibitem [{\citenamefont {Fu}\ \emph {et~al.}(2023)\citenamefont {Fu},
  \citenamefont {Huang}, \citenamefont {Pawlowski},\ and\ \citenamefont
  {Tan}}]{Fu:2022uow}%
  \BibitemOpen
  \bibfield  {author} {\bibinfo {author} {\bibfnamefont {W.-j.}\ \bibnamefont
  {Fu}}, \bibinfo {author} {\bibfnamefont {C.}~\bibnamefont {Huang}}, \bibinfo
  {author} {\bibfnamefont {J.~M.}\ \bibnamefont {Pawlowski}},\ and\ \bibinfo
  {author} {\bibfnamefont {Y.-y.}\ \bibnamefont {Tan}},\ }\bibfield  {title}
  {\bibinfo {title} {{Four-quark scatterings in QCD I}},\ }\href
  {https://doi.org/10.21468/SciPostPhys.14.4.069} {\bibfield  {journal}
  {\bibinfo  {journal} {SciPost Phys.}\ }\textbf {\bibinfo {volume} {14}},\
  \bibinfo {pages} {069} (\bibinfo {year} {2023})},\ \Eprint
  {https://arxiv.org/abs/2209.13120} {arXiv:2209.13120 [hep-ph]} \BibitemShut
  {NoStop}%
\bibitem [{\citenamefont {Fu}\ \emph {et~al.}(2024)\citenamefont {Fu},
  \citenamefont {Huang}, \citenamefont {Pawlowski},\ and\ \citenamefont
  {Tan}}]{Fu:2024ysj}%
  \BibitemOpen
  \bibfield  {author} {\bibinfo {author} {\bibfnamefont {W.-j.}\ \bibnamefont
  {Fu}}, \bibinfo {author} {\bibfnamefont {C.}~\bibnamefont {Huang}}, \bibinfo
  {author} {\bibfnamefont {J.~M.}\ \bibnamefont {Pawlowski}},\ and\ \bibinfo
  {author} {\bibfnamefont {Y.-y.}\ \bibnamefont {Tan}},\ }\bibfield  {title}
  {\bibinfo {title} {{Four-quark scatterings in QCD II}},\ }\href
  {https://doi.org/10.21468/SciPostPhys.17.5.148} {\bibfield  {journal}
  {\bibinfo  {journal} {SciPost Phys.}\ }\textbf {\bibinfo {volume} {17}},\
  \bibinfo {pages} {148} (\bibinfo {year} {2024})},\ \Eprint
  {https://arxiv.org/abs/2401.07638} {arXiv:2401.07638 [hep-ph]} \BibitemShut
  {NoStop}%
\bibitem [{\citenamefont {Eichmann}\ \emph {et~al.}(2016)\citenamefont
  {Eichmann}, \citenamefont {Sanchis-Alepuz}, \citenamefont {Williams},
  \citenamefont {Alkofer},\ and\ \citenamefont {Fischer}}]{Eichmann:2016yit}%
  \BibitemOpen
  \bibfield  {author} {\bibinfo {author} {\bibfnamefont {G.}~\bibnamefont
  {Eichmann}}, \bibinfo {author} {\bibfnamefont {H.}~\bibnamefont
  {Sanchis-Alepuz}}, \bibinfo {author} {\bibfnamefont {R.}~\bibnamefont
  {Williams}}, \bibinfo {author} {\bibfnamefont {R.}~\bibnamefont {Alkofer}},\
  and\ \bibinfo {author} {\bibfnamefont {C.~S.}\ \bibnamefont {Fischer}},\
  }\bibfield  {title} {\bibinfo {title} {{Baryons as relativistic three-quark
  bound states}},\ }\href {https://doi.org/10.1016/j.ppnp.2016.07.001}
  {\bibfield  {journal} {\bibinfo  {journal} {Prog. Part. Nucl. Phys.}\
  }\textbf {\bibinfo {volume} {91}},\ \bibinfo {pages} {1} (\bibinfo {year}
  {2016})},\ \Eprint {https://arxiv.org/abs/1606.09602} {arXiv:1606.09602
  [hep-ph]} \BibitemShut {NoStop}%
\bibitem [{\citenamefont {Fischer}(2019)}]{Fischer:2018sdj}%
  \BibitemOpen
  \bibfield  {author} {\bibinfo {author} {\bibfnamefont {C.~S.}\ \bibnamefont
  {Fischer}},\ }\bibfield  {title} {\bibinfo {title} {{QCD at finite
  temperature and chemical potential from Dyson-Schwinger equations}},\ }\href
  {https://doi.org/10.1016/j.ppnp.2019.01.002} {\bibfield  {journal} {\bibinfo
  {journal} {Prog. Part. Nucl. Phys.}\ }\textbf {\bibinfo {volume} {105}},\
  \bibinfo {pages} {1} (\bibinfo {year} {2019})},\ \Eprint
  {https://arxiv.org/abs/1810.12938} {arXiv:1810.12938 [hep-ph]} \BibitemShut
  {NoStop}%
\bibitem [{\citenamefont {Dupuis}\ \emph {et~al.}(2021)\citenamefont {Dupuis},
  \citenamefont {Canet}, \citenamefont {Eichhorn}, \citenamefont {Metzner},
  \citenamefont {Pawlowski}, \citenamefont {Tissier},\ and\ \citenamefont
  {Wschebor}}]{Dupuis:2020fhh}%
  \BibitemOpen
  \bibfield  {author} {\bibinfo {author} {\bibfnamefont {N.}~\bibnamefont
  {Dupuis}}, \bibinfo {author} {\bibfnamefont {L.}~\bibnamefont {Canet}},
  \bibinfo {author} {\bibfnamefont {A.}~\bibnamefont {Eichhorn}}, \bibinfo
  {author} {\bibfnamefont {W.}~\bibnamefont {Metzner}}, \bibinfo {author}
  {\bibfnamefont {J.~M.}\ \bibnamefont {Pawlowski}}, \bibinfo {author}
  {\bibfnamefont {M.}~\bibnamefont {Tissier}},\ and\ \bibinfo {author}
  {\bibfnamefont {N.}~\bibnamefont {Wschebor}},\ }\bibfield  {title} {\bibinfo
  {title} {{The nonperturbative functional renormalization group and its
  applications}},\ }\href {https://doi.org/10.1016/j.physrep.2021.01.001}
  {\bibfield  {journal} {\bibinfo  {journal} {Phys. Rept.}\ }\textbf {\bibinfo
  {volume} {910}},\ \bibinfo {pages} {1} (\bibinfo {year} {2021})},\ \Eprint
  {https://arxiv.org/abs/2006.04853} {arXiv:2006.04853 [cond-mat.stat-mech]}
  \BibitemShut {NoStop}%
\bibitem [{\citenamefont {Fu}(2022)}]{Fu:2022gou}%
  \BibitemOpen
  \bibfield  {author} {\bibinfo {author} {\bibfnamefont {W.-j.}\ \bibnamefont
  {Fu}},\ }\bibfield  {title} {\bibinfo {title} {{QCD at finite temperature and
  density within the fRG approach: an overview}},\ }\href
  {https://doi.org/10.1088/1572-9494/ac86be} {\bibfield  {journal} {\bibinfo
  {journal} {Commun. Theor. Phys.}\ }\textbf {\bibinfo {volume} {74}},\
  \bibinfo {pages} {097304} (\bibinfo {year} {2022})},\ \Eprint
  {https://arxiv.org/abs/2205.00468} {arXiv:2205.00468 [hep-ph]} \BibitemShut
  {NoStop}%
\bibitem [{\citenamefont {Mitter}\ \emph {et~al.}(2015)\citenamefont {Mitter},
  \citenamefont {Pawlowski},\ and\ \citenamefont
  {Strodthoff}}]{Mitter:2014wpa}%
  \BibitemOpen
  \bibfield  {author} {\bibinfo {author} {\bibfnamefont {M.}~\bibnamefont
  {Mitter}}, \bibinfo {author} {\bibfnamefont {J.~M.}\ \bibnamefont
  {Pawlowski}},\ and\ \bibinfo {author} {\bibfnamefont {N.}~\bibnamefont
  {Strodthoff}},\ }\bibfield  {title} {\bibinfo {title} {{Chiral symmetry
  breaking in continuum QCD}},\ }\href
  {https://doi.org/10.1103/PhysRevD.91.054035} {\bibfield  {journal} {\bibinfo
  {journal} {Phys. Rev.}\ }\textbf {\bibinfo {volume} {D91}},\ \bibinfo {pages}
  {054035} (\bibinfo {year} {2015})},\ \Eprint
  {https://arxiv.org/abs/1411.7978} {arXiv:1411.7978 [hep-ph]} \BibitemShut
  {NoStop}%
\bibitem [{\citenamefont {Williams}(2015)}]{Williams:2014iea}%
  \BibitemOpen
  \bibfield  {author} {\bibinfo {author} {\bibfnamefont {R.}~\bibnamefont
  {Williams}},\ }\bibfield  {title} {\bibinfo {title} {{The quark-gluon vertex
  in Landau gauge bound-state studies}},\ }\href
  {https://doi.org/10.1140/epja/i2015-15057-4} {\bibfield  {journal} {\bibinfo
  {journal} {Eur. Phys. J. A}\ }\textbf {\bibinfo {volume} {51}},\ \bibinfo
  {pages} {57} (\bibinfo {year} {2015})},\ \Eprint
  {https://arxiv.org/abs/1404.2545} {arXiv:1404.2545 [hep-ph]} \BibitemShut
  {NoStop}%
\bibitem [{\citenamefont {Williams}\ \emph {et~al.}(2016)\citenamefont
  {Williams}, \citenamefont {Fischer},\ and\ \citenamefont
  {Heupel}}]{Williams:2015cvx}%
  \BibitemOpen
  \bibfield  {author} {\bibinfo {author} {\bibfnamefont {R.}~\bibnamefont
  {Williams}}, \bibinfo {author} {\bibfnamefont {C.~S.}\ \bibnamefont
  {Fischer}},\ and\ \bibinfo {author} {\bibfnamefont {W.}~\bibnamefont
  {Heupel}},\ }\bibfield  {title} {\bibinfo {title} {{Light mesons in QCD and
  unquenching effects from the 3PI effective action}},\ }\href
  {https://doi.org/10.1103/PhysRevD.93.034026} {\bibfield  {journal} {\bibinfo
  {journal} {Phys. Rev.}\ }\textbf {\bibinfo {volume} {D93}},\ \bibinfo {pages}
  {034026} (\bibinfo {year} {2016})},\ \Eprint
  {https://arxiv.org/abs/1512.00455} {arXiv:1512.00455 [hep-ph]} \BibitemShut
  {NoStop}%
\bibitem [{\citenamefont {Cyrol}\ \emph
  {et~al.}(2018{\natexlab{a}})\citenamefont {Cyrol}, \citenamefont {Mitter},
  \citenamefont {Pawlowski},\ and\ \citenamefont {Strodthoff}}]{Cyrol:2017ewj}%
  \BibitemOpen
  \bibfield  {author} {\bibinfo {author} {\bibfnamefont {A.~K.}\ \bibnamefont
  {Cyrol}}, \bibinfo {author} {\bibfnamefont {M.}~\bibnamefont {Mitter}},
  \bibinfo {author} {\bibfnamefont {J.~M.}\ \bibnamefont {Pawlowski}},\ and\
  \bibinfo {author} {\bibfnamefont {N.}~\bibnamefont {Strodthoff}},\ }\bibfield
   {title} {\bibinfo {title} {{Nonperturbative quark, gluon, and meson
  correlators of unquenched QCD}},\ }\href
  {https://doi.org/10.1103/PhysRevD.97.054006} {\bibfield  {journal} {\bibinfo
  {journal} {Phys. Rev.}\ }\textbf {\bibinfo {volume} {D97}},\ \bibinfo {pages}
  {054006} (\bibinfo {year} {2018}{\natexlab{a}})},\ \Eprint
  {https://arxiv.org/abs/1706.06326} {arXiv:1706.06326 [hep-ph]} \BibitemShut
  {NoStop}%
\bibitem [{\citenamefont {Gao}\ \emph {et~al.}(2021)\citenamefont {Gao},
  \citenamefont {Papavassiliou},\ and\ \citenamefont
  {Pawlowski}}]{Gao:2021wun}%
  \BibitemOpen
  \bibfield  {author} {\bibinfo {author} {\bibfnamefont {F.}~\bibnamefont
  {Gao}}, \bibinfo {author} {\bibfnamefont {J.}~\bibnamefont {Papavassiliou}},\
  and\ \bibinfo {author} {\bibfnamefont {J.~M.}\ \bibnamefont {Pawlowski}},\
  }\bibfield  {title} {\bibinfo {title} {{Fully coupled functional equations
  for the quark sector of QCD}},\ }\href
  {https://doi.org/10.1103/PhysRevD.103.094013} {\bibfield  {journal} {\bibinfo
   {journal} {Phys. Rev. D}\ }\textbf {\bibinfo {volume} {103}},\ \bibinfo
  {pages} {094013} (\bibinfo {year} {2021})},\ \Eprint
  {https://arxiv.org/abs/2102.13053} {arXiv:2102.13053 [hep-ph]} \BibitemShut
  {NoStop}%
\bibitem [{\citenamefont {Aguilar}\ \emph {et~al.}(2024)\citenamefont
  {Aguilar}, \citenamefont {Ferreira}, \citenamefont {Oliveira}, \citenamefont
  {Papavassiliou},\ and\ \citenamefont {Linhares}}]{Aguilar:2024ciu}%
  \BibitemOpen
  \bibfield  {author} {\bibinfo {author} {\bibfnamefont {A.~C.}\ \bibnamefont
  {Aguilar}}, \bibinfo {author} {\bibfnamefont {M.~N.}\ \bibnamefont
  {Ferreira}}, \bibinfo {author} {\bibfnamefont {B.~M.}\ \bibnamefont
  {Oliveira}}, \bibinfo {author} {\bibfnamefont {J.}~\bibnamefont
  {Papavassiliou}},\ and\ \bibinfo {author} {\bibfnamefont {G.~T.}\
  \bibnamefont {Linhares}},\ }\bibfield  {title} {\bibinfo {title} {{Infrared
  properties of the quark-gluon vertex in general kinematics}},\ }\href
  {https://doi.org/10.1140/epjc/s10052-024-13605-9} {\bibfield  {journal}
  {\bibinfo  {journal} {Eur. Phys. J. C}\ }\textbf {\bibinfo {volume} {84}},\
  \bibinfo {pages} {1231} (\bibinfo {year} {2024})},\ \Eprint
  {https://arxiv.org/abs/2408.15370} {arXiv:2408.15370 [hep-ph]} \BibitemShut
  {NoStop}%
\bibitem [{\citenamefont {Wetterich}(1993)}]{Wetterich:1992yh}%
  \BibitemOpen
  \bibfield  {author} {\bibinfo {author} {\bibfnamefont {C.}~\bibnamefont
  {Wetterich}},\ }\bibfield  {title} {\bibinfo {title} {{Exact evolution
  equation for the effective potential}},\ }\href
  {https://doi.org/10.1016/0370-2693(93)90726-X} {\bibfield  {journal}
  {\bibinfo  {journal} {Phys. Lett.}\ }\textbf {\bibinfo {volume} {B301}},\
  \bibinfo {pages} {90} (\bibinfo {year} {1993})}\BibitemShut {NoStop}%
\bibitem [{\citenamefont {Braun}\ \emph {et~al.}(2016)\citenamefont {Braun},
  \citenamefont {Fister}, \citenamefont {Pawlowski},\ and\ \citenamefont
  {Rennecke}}]{Braun:2014ata}%
  \BibitemOpen
  \bibfield  {author} {\bibinfo {author} {\bibfnamefont {J.}~\bibnamefont
  {Braun}}, \bibinfo {author} {\bibfnamefont {L.}~\bibnamefont {Fister}},
  \bibinfo {author} {\bibfnamefont {J.~M.}\ \bibnamefont {Pawlowski}},\ and\
  \bibinfo {author} {\bibfnamefont {F.}~\bibnamefont {Rennecke}},\ }\bibfield
  {title} {\bibinfo {title} {{From Quarks and Gluons to Hadrons: Chiral
  Symmetry Breaking in Dynamical QCD}},\ }\href
  {https://doi.org/10.1103/PhysRevD.94.034016} {\bibfield  {journal} {\bibinfo
  {journal} {Phys. Rev.}\ }\textbf {\bibinfo {volume} {D94}},\ \bibinfo {pages}
  {034016} (\bibinfo {year} {2016})},\ \Eprint
  {https://arxiv.org/abs/1412.1045} {arXiv:1412.1045 [hep-ph]} \BibitemShut
  {NoStop}%
\bibitem [{\citenamefont {Rennecke}(2015)}]{Rennecke:2015eba}%
  \BibitemOpen
  \bibfield  {author} {\bibinfo {author} {\bibfnamefont {F.}~\bibnamefont
  {Rennecke}},\ }\bibfield  {title} {\bibinfo {title} {{Vacuum structure of
  vector mesons in QCD}},\ }\href {https://doi.org/10.1103/PhysRevD.92.076012}
  {\bibfield  {journal} {\bibinfo  {journal} {Phys. Rev.}\ }\textbf {\bibinfo
  {volume} {D92}},\ \bibinfo {pages} {076012} (\bibinfo {year} {2015})},\
  \Eprint {https://arxiv.org/abs/1504.03585} {arXiv:1504.03585 [hep-ph]}
  \BibitemShut {NoStop}%
\bibitem [{\citenamefont {Cyrol}\ \emph {et~al.}(2016)\citenamefont {Cyrol},
  \citenamefont {Fister}, \citenamefont {Mitter}, \citenamefont {Pawlowski},\
  and\ \citenamefont {Strodthoff}}]{Cyrol:2016tym}%
  \BibitemOpen
  \bibfield  {author} {\bibinfo {author} {\bibfnamefont {A.~K.}\ \bibnamefont
  {Cyrol}}, \bibinfo {author} {\bibfnamefont {L.}~\bibnamefont {Fister}},
  \bibinfo {author} {\bibfnamefont {M.}~\bibnamefont {Mitter}}, \bibinfo
  {author} {\bibfnamefont {J.~M.}\ \bibnamefont {Pawlowski}},\ and\ \bibinfo
  {author} {\bibfnamefont {N.}~\bibnamefont {Strodthoff}},\ }\bibfield  {title}
  {\bibinfo {title} {{Landau gauge Yang-Mills correlation functions}},\ }\href
  {https://doi.org/10.1103/PhysRevD.94.054005} {\bibfield  {journal} {\bibinfo
  {journal} {Phys. Rev.}\ }\textbf {\bibinfo {volume} {D94}},\ \bibinfo {pages}
  {054005} (\bibinfo {year} {2016})},\ \Eprint
  {https://arxiv.org/abs/1605.01856} {arXiv:1605.01856 [hep-ph]} \BibitemShut
  {NoStop}%
\bibitem [{\citenamefont {Corell}\ \emph {et~al.}(2018)\citenamefont {Corell},
  \citenamefont {Cyrol}, \citenamefont {Mitter}, \citenamefont {Pawlowski},\
  and\ \citenamefont {Strodthoff}}]{Corell:2018yil}%
  \BibitemOpen
  \bibfield  {author} {\bibinfo {author} {\bibfnamefont {L.}~\bibnamefont
  {Corell}}, \bibinfo {author} {\bibfnamefont {A.~K.}\ \bibnamefont {Cyrol}},
  \bibinfo {author} {\bibfnamefont {M.}~\bibnamefont {Mitter}}, \bibinfo
  {author} {\bibfnamefont {J.~M.}\ \bibnamefont {Pawlowski}},\ and\ \bibinfo
  {author} {\bibfnamefont {N.}~\bibnamefont {Strodthoff}},\ }\bibfield  {title}
  {\bibinfo {title} {{Correlation functions of three-dimensional Yang-Mills
  theory from the FRG}},\ }\href {https://doi.org/10.21468/SciPostPhys.5.6.066}
  {\bibfield  {journal} {\bibinfo  {journal} {SciPost Phys.}\ }\textbf
  {\bibinfo {volume} {5}},\ \bibinfo {pages} {066} (\bibinfo {year} {2018})},\
  \Eprint {https://arxiv.org/abs/1803.10092} {arXiv:1803.10092 [hep-ph]}
  \BibitemShut {NoStop}%
\bibitem [{\citenamefont {Fu}\ \emph {et~al.}(2020)\citenamefont {Fu},
  \citenamefont {Pawlowski},\ and\ \citenamefont {Rennecke}}]{Fu:2019hdw}%
  \BibitemOpen
  \bibfield  {author} {\bibinfo {author} {\bibfnamefont {W.-j.}\ \bibnamefont
  {Fu}}, \bibinfo {author} {\bibfnamefont {J.~M.}\ \bibnamefont {Pawlowski}},\
  and\ \bibinfo {author} {\bibfnamefont {F.}~\bibnamefont {Rennecke}},\
  }\bibfield  {title} {\bibinfo {title} {{QCD phase structure at finite
  temperature and density}},\ }\href
  {https://doi.org/10.1103/PhysRevD.101.054032} {\bibfield  {journal} {\bibinfo
   {journal} {Phys. Rev. D}\ }\textbf {\bibinfo {volume} {101}},\ \bibinfo
  {pages} {054032} (\bibinfo {year} {2020})},\ \Eprint
  {https://arxiv.org/abs/1909.02991} {arXiv:1909.02991 [hep-ph]} \BibitemShut
  {NoStop}%
\bibitem [{\citenamefont {Gao}\ and\ \citenamefont
  {Pawlowski}(2021)}]{Gao:2020fbl}%
  \BibitemOpen
  \bibfield  {author} {\bibinfo {author} {\bibfnamefont {F.}~\bibnamefont
  {Gao}}\ and\ \bibinfo {author} {\bibfnamefont {J.~M.}\ \bibnamefont
  {Pawlowski}},\ }\bibfield  {title} {\bibinfo {title} {{Chiral phase structure
  and critical end point in QCD}},\ }\href
  {https://doi.org/10.1016/j.physletb.2021.136584} {\bibfield  {journal}
  {\bibinfo  {journal} {Phys. Lett. B}\ }\textbf {\bibinfo {volume} {820}},\
  \bibinfo {pages} {136584} (\bibinfo {year} {2021})},\ \Eprint
  {https://arxiv.org/abs/2010.13705} {arXiv:2010.13705 [hep-ph]} \BibitemShut
  {NoStop}%
\bibitem [{\citenamefont {Gao}\ and\ \citenamefont
  {Pawlowski}(2020)}]{Gao:2020qsj}%
  \BibitemOpen
  \bibfield  {author} {\bibinfo {author} {\bibfnamefont {F.}~\bibnamefont
  {Gao}}\ and\ \bibinfo {author} {\bibfnamefont {J.~M.}\ \bibnamefont
  {Pawlowski}},\ }\bibfield  {title} {\bibinfo {title} {{QCD phase structure
  from functional methods}},\ }\href
  {https://doi.org/10.1103/PhysRevD.102.034027} {\bibfield  {journal} {\bibinfo
   {journal} {Phys. Rev. D}\ }\textbf {\bibinfo {volume} {102}},\ \bibinfo
  {pages} {034027} (\bibinfo {year} {2020})},\ \Eprint
  {https://arxiv.org/abs/2002.07500} {arXiv:2002.07500 [hep-ph]} \BibitemShut
  {NoStop}%
\bibitem [{\citenamefont {Lu}\ \emph {et~al.}(2024)\citenamefont {Lu},
  \citenamefont {Gao}, \citenamefont {Liu},\ and\ \citenamefont
  {Pawlowski}}]{Lu:2023mkn}%
  \BibitemOpen
  \bibfield  {author} {\bibinfo {author} {\bibfnamefont {Y.}~\bibnamefont
  {Lu}}, \bibinfo {author} {\bibfnamefont {F.}~\bibnamefont {Gao}}, \bibinfo
  {author} {\bibfnamefont {Y.-X.}\ \bibnamefont {Liu}},\ and\ \bibinfo {author}
  {\bibfnamefont {J.~M.}\ \bibnamefont {Pawlowski}},\ }\bibfield  {title}
  {\bibinfo {title} {{QCD equation of state and thermodynamic observables from
  computationally minimal Dyson-Schwinger equations}},\ }\href
  {https://doi.org/10.1103/PhysRevD.110.014036} {\bibfield  {journal} {\bibinfo
   {journal} {Phys. Rev. D}\ }\textbf {\bibinfo {volume} {110}},\ \bibinfo
  {pages} {014036} (\bibinfo {year} {2024})},\ \Eprint
  {https://arxiv.org/abs/2310.18383} {arXiv:2310.18383 [hep-ph]} \BibitemShut
  {NoStop}%
\bibitem [{\citenamefont {Ihssen}\ \emph {et~al.}(2024)\citenamefont {Ihssen},
  \citenamefont {Pawlowski}, \citenamefont {Sattler},\ and\ \citenamefont
  {Wink}}]{Ihssen:2024miv}%
  \BibitemOpen
  \bibfield  {author} {\bibinfo {author} {\bibfnamefont {F.}~\bibnamefont
  {Ihssen}}, \bibinfo {author} {\bibfnamefont {J.~M.}\ \bibnamefont
  {Pawlowski}}, \bibinfo {author} {\bibfnamefont {F.~R.}\ \bibnamefont
  {Sattler}},\ and\ \bibinfo {author} {\bibfnamefont {N.}~\bibnamefont
  {Wink}},\ }\bibfield  {title} {\bibinfo {title} {{Towards quantitative
  precision in functional QCD I}},\ }\href@noop {} {\  (\bibinfo {year}
  {2024})},\ \Eprint {https://arxiv.org/abs/2408.08413} {arXiv:2408.08413
  [hep-ph]} \BibitemShut {NoStop}%
\bibitem [{\citenamefont {Huber}(2020{\natexlab{a}})}]{Huber:2018ned}%
  \BibitemOpen
  \bibfield  {author} {\bibinfo {author} {\bibfnamefont {M.~Q.}\ \bibnamefont
  {Huber}},\ }\bibfield  {title} {\bibinfo {title} {{Nonperturbative properties
  of Yang\textendash{}Mills theories}},\ }\href
  {https://doi.org/10.1016/j.physrep.2020.04.004} {\bibfield  {journal}
  {\bibinfo  {journal} {Phys. Rept.}\ }\textbf {\bibinfo {volume} {879}},\
  \bibinfo {pages} {1} (\bibinfo {year} {2020}{\natexlab{a}})},\ \Eprint
  {https://arxiv.org/abs/1808.05227} {arXiv:1808.05227 [hep-ph]} \BibitemShut
  {NoStop}%
\bibitem [{\citenamefont {Huber}(2020{\natexlab{b}})}]{Huber:2020keu}%
  \BibitemOpen
  \bibfield  {author} {\bibinfo {author} {\bibfnamefont {M.~Q.}\ \bibnamefont
  {Huber}},\ }\bibfield  {title} {\bibinfo {title} {{Correlation functions of
  Landau gauge Yang-Mills theory}},\ }\href
  {https://doi.org/10.1103/PhysRevD.101.114009} {\bibfield  {journal} {\bibinfo
   {journal} {Phys. Rev. D}\ }\textbf {\bibinfo {volume} {101}},\ \bibinfo
  {pages} {114009} (\bibinfo {year} {2020}{\natexlab{b}})},\ \Eprint
  {https://arxiv.org/abs/2003.13703} {arXiv:2003.13703 [hep-ph]} \BibitemShut
  {NoStop}%
\bibitem [{\citenamefont {Eichmann}\ \emph {et~al.}(2021)\citenamefont
  {Eichmann}, \citenamefont {Pawlowski},\ and\ \citenamefont
  {Silva}}]{Eichmann:2021zuv}%
  \BibitemOpen
  \bibfield  {author} {\bibinfo {author} {\bibfnamefont {G.}~\bibnamefont
  {Eichmann}}, \bibinfo {author} {\bibfnamefont {J.~M.}\ \bibnamefont
  {Pawlowski}},\ and\ \bibinfo {author} {\bibfnamefont {J.~a.~M.}\ \bibnamefont
  {Silva}},\ }\bibfield  {title} {\bibinfo {title} {{Mass generation in
  Landau-gauge Yang-Mills theory}},\ }\href
  {https://doi.org/10.1103/PhysRevD.104.114016} {\bibfield  {journal} {\bibinfo
   {journal} {Phys. Rev. D}\ }\textbf {\bibinfo {volume} {104}},\ \bibinfo
  {pages} {114016} (\bibinfo {year} {2021})},\ \Eprint
  {https://arxiv.org/abs/2107.05352} {arXiv:2107.05352 [hep-ph]} \BibitemShut
  {NoStop}%
\bibitem [{\citenamefont {Fischer}\ \emph {et~al.}(2009)\citenamefont
  {Fischer}, \citenamefont {Maas},\ and\ \citenamefont
  {Pawlowski}}]{Fischer:2008uz}%
  \BibitemOpen
  \bibfield  {author} {\bibinfo {author} {\bibfnamefont {C.~S.}\ \bibnamefont
  {Fischer}}, \bibinfo {author} {\bibfnamefont {A.}~\bibnamefont {Maas}},\ and\
  \bibinfo {author} {\bibfnamefont {J.~M.}\ \bibnamefont {Pawlowski}},\
  }\bibfield  {title} {\bibinfo {title} {{On the infrared behavior of Landau
  gauge Yang-Mills theory}},\ }\href
  {https://doi.org/10.1016/j.aop.2009.07.009} {\bibfield  {journal} {\bibinfo
  {journal} {Annals Phys.}\ }\textbf {\bibinfo {volume} {324}},\ \bibinfo
  {pages} {2408} (\bibinfo {year} {2009})},\ \Eprint
  {https://arxiv.org/abs/0810.1987} {arXiv:0810.1987 [hep-ph]} \BibitemShut
  {NoStop}%
\bibitem [{\citenamefont {Denz}\ \emph {et~al.}(2018)\citenamefont {Denz},
  \citenamefont {Pawlowski},\ and\ \citenamefont {Reichert}}]{Denz:2016qks}%
  \BibitemOpen
  \bibfield  {author} {\bibinfo {author} {\bibfnamefont {T.}~\bibnamefont
  {Denz}}, \bibinfo {author} {\bibfnamefont {J.~M.}\ \bibnamefont
  {Pawlowski}},\ and\ \bibinfo {author} {\bibfnamefont {M.}~\bibnamefont
  {Reichert}},\ }\bibfield  {title} {\bibinfo {title} {{Towards apparent
  convergence in asymptotically safe quantum gravity}},\ }\href
  {https://doi.org/10.1140/epjc/s10052-018-5806-0} {\bibfield  {journal}
  {\bibinfo  {journal} {Eur. Phys. J. C}\ }\textbf {\bibinfo {volume} {78}},\
  \bibinfo {pages} {336} (\bibinfo {year} {2018})},\ \Eprint
  {https://arxiv.org/abs/1612.07315} {arXiv:1612.07315 [hep-th]} \BibitemShut
  {NoStop}%
\bibitem [{\citenamefont {von Smekal}\ \emph {et~al.}(1997)\citenamefont {von
  Smekal}, \citenamefont {Alkofer},\ and\ \citenamefont
  {Hauck}}]{vonSmekal:1997ohs}%
  \BibitemOpen
  \bibfield  {author} {\bibinfo {author} {\bibfnamefont {L.}~\bibnamefont {von
  Smekal}}, \bibinfo {author} {\bibfnamefont {R.}~\bibnamefont {Alkofer}},\
  and\ \bibinfo {author} {\bibfnamefont {A.}~\bibnamefont {Hauck}},\ }\bibfield
   {title} {\bibinfo {title} {{The Infrared behavior of gluon and ghost
  propagators in Landau gauge QCD}},\ }\href
  {https://doi.org/10.1103/PhysRevLett.79.3591} {\bibfield  {journal} {\bibinfo
   {journal} {Phys. Rev. Lett.}\ }\textbf {\bibinfo {volume} {79}},\ \bibinfo
  {pages} {3591} (\bibinfo {year} {1997})},\ \Eprint
  {https://arxiv.org/abs/hep-ph/9705242} {arXiv:hep-ph/9705242} \BibitemShut
  {NoStop}%
\bibitem [{\citenamefont {Boucaud}\ \emph {et~al.}(2009)\citenamefont
  {Boucaud}, \citenamefont {De~Soto}, \citenamefont {Leroy}, \citenamefont
  {Le~Yaouanc}, \citenamefont {Micheli}, \citenamefont {Pene},\ and\
  \citenamefont {Rodriguez-Quintero}}]{Boucaud:2008gn}%
  \BibitemOpen
  \bibfield  {author} {\bibinfo {author} {\bibfnamefont {P.}~\bibnamefont
  {Boucaud}}, \bibinfo {author} {\bibfnamefont {F.}~\bibnamefont {De~Soto}},
  \bibinfo {author} {\bibfnamefont {J.~P.}\ \bibnamefont {Leroy}}, \bibinfo
  {author} {\bibfnamefont {A.}~\bibnamefont {Le~Yaouanc}}, \bibinfo {author}
  {\bibfnamefont {J.}~\bibnamefont {Micheli}}, \bibinfo {author} {\bibfnamefont
  {O.}~\bibnamefont {Pene}},\ and\ \bibinfo {author} {\bibfnamefont
  {J.}~\bibnamefont {Rodriguez-Quintero}},\ }\bibfield  {title} {\bibinfo
  {title} {{Ghost-gluon running coupling, power corrections and the
  determination of Lambda(MS-bar)}},\ }\href
  {https://doi.org/10.1103/PhysRevD.79.014508} {\bibfield  {journal} {\bibinfo
  {journal} {Phys. Rev. D}\ }\textbf {\bibinfo {volume} {79}},\ \bibinfo
  {pages} {014508} (\bibinfo {year} {2009})},\ \Eprint
  {https://arxiv.org/abs/0811.2059} {arXiv:0811.2059 [hep-ph]} \BibitemShut
  {NoStop}%
\bibitem [{\citenamefont {von Smekal}\ \emph {et~al.}(2009)\citenamefont {von
  Smekal}, \citenamefont {Maltman},\ and\ \citenamefont
  {Sternbeck}}]{vonSmekal:2009ae}%
  \BibitemOpen
  \bibfield  {author} {\bibinfo {author} {\bibfnamefont {L.}~\bibnamefont {von
  Smekal}}, \bibinfo {author} {\bibfnamefont {K.}~\bibnamefont {Maltman}},\
  and\ \bibinfo {author} {\bibfnamefont {A.}~\bibnamefont {Sternbeck}},\
  }\bibfield  {title} {\bibinfo {title} {{The Strong coupling and its running
  to four loops in a minimal MOM scheme}},\ }\href
  {https://doi.org/10.1016/j.physletb.2009.10.030} {\bibfield  {journal}
  {\bibinfo  {journal} {Phys. Lett. B}\ }\textbf {\bibinfo {volume} {681}},\
  \bibinfo {pages} {336} (\bibinfo {year} {2009})},\ \Eprint
  {https://arxiv.org/abs/0903.1696} {arXiv:0903.1696 [hep-ph]} \BibitemShut
  {NoStop}%
\bibitem [{\citenamefont {Blossier}\ \emph {et~al.}(2012)\citenamefont
  {Blossier}, \citenamefont {Boucaud}, \citenamefont {Brinet}, \citenamefont
  {De~Soto}, \citenamefont {Du}, \citenamefont {Morenas}, \citenamefont {Pene},
  \citenamefont {Petrov},\ and\ \citenamefont
  {Rodriguez-Quintero}}]{Blossier:2012ef}%
  \BibitemOpen
  \bibfield  {author} {\bibinfo {author} {\bibfnamefont {B.}~\bibnamefont
  {Blossier}}, \bibinfo {author} {\bibfnamefont {P.}~\bibnamefont {Boucaud}},
  \bibinfo {author} {\bibfnamefont {M.}~\bibnamefont {Brinet}}, \bibinfo
  {author} {\bibfnamefont {F.}~\bibnamefont {De~Soto}}, \bibinfo {author}
  {\bibfnamefont {X.}~\bibnamefont {Du}}, \bibinfo {author} {\bibfnamefont
  {V.}~\bibnamefont {Morenas}}, \bibinfo {author} {\bibfnamefont
  {O.}~\bibnamefont {Pene}}, \bibinfo {author} {\bibfnamefont {K.}~\bibnamefont
  {Petrov}},\ and\ \bibinfo {author} {\bibfnamefont {J.}~\bibnamefont
  {Rodriguez-Quintero}},\ }\bibfield  {title} {\bibinfo {title} {{The Strong
  running coupling at $\tau$ and $Z_0$ mass scales from lattice QCD}},\ }\href
  {https://doi.org/10.1103/PhysRevLett.108.262002} {\bibfield  {journal}
  {\bibinfo  {journal} {Phys. Rev. Lett.}\ }\textbf {\bibinfo {volume} {108}},\
  \bibinfo {pages} {262002} (\bibinfo {year} {2012})},\ \Eprint
  {https://arxiv.org/abs/1201.5770} {arXiv:1201.5770 [hep-ph]} \BibitemShut
  {NoStop}%
\bibitem [{\citenamefont {Zafeiropoulos}\ \emph {et~al.}(2019)\citenamefont
  {Zafeiropoulos}, \citenamefont {Boucaud}, \citenamefont {De~Soto},
  \citenamefont {Rodr\'\i{}guez-Quintero},\ and\ \citenamefont
  {Segovia}}]{Zafeiropoulos:2019flq}%
  \BibitemOpen
  \bibfield  {author} {\bibinfo {author} {\bibfnamefont {S.}~\bibnamefont
  {Zafeiropoulos}}, \bibinfo {author} {\bibfnamefont {P.}~\bibnamefont
  {Boucaud}}, \bibinfo {author} {\bibfnamefont {F.}~\bibnamefont {De~Soto}},
  \bibinfo {author} {\bibfnamefont {J.}~\bibnamefont
  {Rodr\'\i{}guez-Quintero}},\ and\ \bibinfo {author} {\bibfnamefont
  {J.}~\bibnamefont {Segovia}},\ }\bibfield  {title} {\bibinfo {title} {{Strong
  Running Coupling from the Gauge Sector of Domain Wall Lattice QCD with
  Physical Quark Masses}},\ }\href
  {https://doi.org/10.1103/PhysRevLett.122.162002} {\bibfield  {journal}
  {\bibinfo  {journal} {Phys. Rev. Lett.}\ }\textbf {\bibinfo {volume} {122}},\
  \bibinfo {pages} {162002} (\bibinfo {year} {2019})},\ \Eprint
  {https://arxiv.org/abs/1902.08148} {arXiv:1902.08148 [hep-ph]} \BibitemShut
  {NoStop}%
\bibitem [{\citenamefont {Gao}\ \emph {et~al.}(2024)\citenamefont {Gao},
  \citenamefont {Miramontes}, \citenamefont {Papavassiliou},\ and\
  \citenamefont {Pawlowski}}]{Gao:2024gdj}%
  \BibitemOpen
  \bibfield  {author} {\bibinfo {author} {\bibfnamefont {F.}~\bibnamefont
  {Gao}}, \bibinfo {author} {\bibfnamefont {A.~S.}\ \bibnamefont {Miramontes}},
  \bibinfo {author} {\bibfnamefont {J.}~\bibnamefont {Papavassiliou}},\ and\
  \bibinfo {author} {\bibfnamefont {J.~M.}\ \bibnamefont {Pawlowski}},\
  }\bibfield  {title} {\bibinfo {title} {{Heavy-light mesons from a
  flavour-dependent interaction}},\ }\href@noop {} {\  (\bibinfo {year}
  {2024})},\ \Eprint {https://arxiv.org/abs/2411.19680} {arXiv:2411.19680
  [hep-ph]} \BibitemShut {NoStop}%
\bibitem [{\citenamefont {Eichmann}(2011)}]{Eichmann:2011vu}%
  \BibitemOpen
  \bibfield  {author} {\bibinfo {author} {\bibfnamefont {G.}~\bibnamefont
  {Eichmann}},\ }\bibfield  {title} {\bibinfo {title} {{Nucleon electromagnetic
  form factors from the covariant Faddeev equation}},\ }\href
  {https://doi.org/10.1103/PhysRevD.84.014014} {\bibfield  {journal} {\bibinfo
  {journal} {Phys. Rev. D}\ }\textbf {\bibinfo {volume} {84}},\ \bibinfo
  {pages} {014014} (\bibinfo {year} {2011})},\ \Eprint
  {https://arxiv.org/abs/1104.4505} {arXiv:1104.4505 [hep-ph]} \BibitemShut
  {NoStop}%
\bibitem [{\citenamefont {Braun}\ \emph {et~al.}(2023)\citenamefont {Braun}
  \emph {et~al.}}]{Braun:2023qak}%
  \BibitemOpen
  \bibfield  {author} {\bibinfo {author} {\bibfnamefont {J.}~\bibnamefont
  {Braun}} \emph {et~al.},\ }\bibfield  {title} {\bibinfo {title} {{Soft modes
  in hot QCD matter}},\ }\href@noop {} {\  (\bibinfo {year} {2023})},\ \Eprint
  {https://arxiv.org/abs/2310.19853} {arXiv:2310.19853 [hep-ph]} \BibitemShut
  {NoStop}%
\bibitem [{\citenamefont {Pawlowski}(2007)}]{Pawlowski:2005xe}%
  \BibitemOpen
  \bibfield  {author} {\bibinfo {author} {\bibfnamefont {J.~M.}\ \bibnamefont
  {Pawlowski}},\ }\bibfield  {title} {\bibinfo {title} {{Aspects of the
  functional renormalisation group}},\ }\href
  {https://doi.org/10.1016/j.aop.2007.01.007} {\bibfield  {journal} {\bibinfo
  {journal} {Annals Phys.}\ }\textbf {\bibinfo {volume} {322}},\ \bibinfo
  {pages} {2831} (\bibinfo {year} {2007})},\ \Eprint
  {https://arxiv.org/abs/hep-th/0512261} {arXiv:hep-th/0512261 [hep-th]}
  \BibitemShut {NoStop}%
\bibitem [{\citenamefont {Blaizot}\ \emph {et~al.}(2011)\citenamefont
  {Blaizot}, \citenamefont {Pawlowski},\ and\ \citenamefont
  {Reinosa}}]{Blaizot:2010zx}%
  \BibitemOpen
  \bibfield  {author} {\bibinfo {author} {\bibfnamefont {J.-P.}\ \bibnamefont
  {Blaizot}}, \bibinfo {author} {\bibfnamefont {J.~M.}\ \bibnamefont
  {Pawlowski}},\ and\ \bibinfo {author} {\bibfnamefont {U.}~\bibnamefont
  {Reinosa}},\ }\bibfield  {title} {\bibinfo {title} {{Exact renormalization
  group and $\Phi$-derivable approximations}},\ }\href
  {https://doi.org/10.1016/j.physletb.2010.12.058} {\bibfield  {journal}
  {\bibinfo  {journal} {Phys. Lett. B}\ }\textbf {\bibinfo {volume} {696}},\
  \bibinfo {pages} {523} (\bibinfo {year} {2011})},\ \Eprint
  {https://arxiv.org/abs/1009.6048} {arXiv:1009.6048 [hep-ph]} \BibitemShut
  {NoStop}%
\bibitem [{\citenamefont {Fu}(2013)}]{Fu:2013sku}%
  \BibitemOpen
  \bibfield  {author} {\bibinfo {author} {\bibfnamefont {W.-J.}\ \bibnamefont
  {Fu}},\ }\bibfield  {title} {\bibinfo {title} {{A new resummation scheme in
  scalar field theories}},\ }\href
  {https://doi.org/10.1140/epjc/s10052-013-2411-0} {\bibfield  {journal}
  {\bibinfo  {journal} {Eur. Phys. J. C}\ }\textbf {\bibinfo {volume} {73}},\
  \bibinfo {pages} {2411} (\bibinfo {year} {2013})},\ \Eprint
  {https://arxiv.org/abs/1211.1110} {arXiv:1211.1110 [hep-ph]} \BibitemShut
  {NoStop}%
\bibitem [{\citenamefont {Carrington}(2013)}]{Carrington:2012ea}%
  \BibitemOpen
  \bibfield  {author} {\bibinfo {author} {\bibfnamefont {M.~E.}\ \bibnamefont
  {Carrington}},\ }\bibfield  {title} {\bibinfo {title} {{Renormalization group
  flow equations connected to the $n$-particle-irreducible effective action}},\
  }\href {https://doi.org/10.1103/PhysRevD.87.045011} {\bibfield  {journal}
  {\bibinfo  {journal} {Phys. Rev. D}\ }\textbf {\bibinfo {volume} {87}},\
  \bibinfo {pages} {045011} (\bibinfo {year} {2013})},\ \Eprint
  {https://arxiv.org/abs/1211.4127} {arXiv:1211.4127 [hep-ph]} \BibitemShut
  {NoStop}%
\bibitem [{\citenamefont {Carrington}\ \emph {et~al.}(2015)\citenamefont
  {Carrington}, \citenamefont {Fu}, \citenamefont {Pickering},\ and\
  \citenamefont {Pulver}}]{Carrington:2014lba}%
  \BibitemOpen
  \bibfield  {author} {\bibinfo {author} {\bibfnamefont {M.~E.}\ \bibnamefont
  {Carrington}}, \bibinfo {author} {\bibfnamefont {W.-J.}\ \bibnamefont {Fu}},
  \bibinfo {author} {\bibfnamefont {D.}~\bibnamefont {Pickering}},\ and\
  \bibinfo {author} {\bibfnamefont {J.~W.}\ \bibnamefont {Pulver}},\ }\bibfield
   {title} {\bibinfo {title} {{Renormalization group methods and the 2PI
  effective action}},\ }\href {https://doi.org/10.1103/PhysRevD.91.025003}
  {\bibfield  {journal} {\bibinfo  {journal} {Phys. Rev. D}\ }\textbf {\bibinfo
  {volume} {91}},\ \bibinfo {pages} {025003} (\bibinfo {year} {2015})},\
  \Eprint {https://arxiv.org/abs/1404.0710} {arXiv:1404.0710 [hep-ph]}
  \BibitemShut {NoStop}%
\bibitem [{\citenamefont {Carrington}\ \emph {et~al.}(2018)\citenamefont
  {Carrington}, \citenamefont {Friesen}, \citenamefont {Meggison},
  \citenamefont {Phillips}, \citenamefont {Pickering},\ and\ \citenamefont
  {Sohrabi}}]{Carrington:2017lry}%
  \BibitemOpen
  \bibfield  {author} {\bibinfo {author} {\bibfnamefont {M.~E.}\ \bibnamefont
  {Carrington}}, \bibinfo {author} {\bibfnamefont {S.~A.}\ \bibnamefont
  {Friesen}}, \bibinfo {author} {\bibfnamefont {B.~A.}\ \bibnamefont
  {Meggison}}, \bibinfo {author} {\bibfnamefont {C.~D.}\ \bibnamefont
  {Phillips}}, \bibinfo {author} {\bibfnamefont {D.}~\bibnamefont
  {Pickering}},\ and\ \bibinfo {author} {\bibfnamefont {K.}~\bibnamefont
  {Sohrabi}},\ }\bibfield  {title} {\bibinfo {title} {{2PI effective theory at
  next-to-leading order using the functional renormalization group}},\ }\href
  {https://doi.org/10.1103/PhysRevD.97.036005} {\bibfield  {journal} {\bibinfo
  {journal} {Phys. Rev. D}\ }\textbf {\bibinfo {volume} {97}},\ \bibinfo
  {pages} {036005} (\bibinfo {year} {2018})},\ \Eprint
  {https://arxiv.org/abs/1711.09135} {arXiv:1711.09135 [hep-th]} \BibitemShut
  {NoStop}%
\bibitem [{\citenamefont {Carrington}\ \emph {et~al.}(2019)\citenamefont
  {Carrington}, \citenamefont {Friesen}, \citenamefont {Phillips},\ and\
  \citenamefont {Pickering}}]{Carrington:2019fwp}%
  \BibitemOpen
  \bibfield  {author} {\bibinfo {author} {\bibfnamefont {M.~E.}\ \bibnamefont
  {Carrington}}, \bibinfo {author} {\bibfnamefont {S.~A.}\ \bibnamefont
  {Friesen}}, \bibinfo {author} {\bibfnamefont {C.~D.}\ \bibnamefont
  {Phillips}},\ and\ \bibinfo {author} {\bibfnamefont {D.}~\bibnamefont
  {Pickering}},\ }\bibfield  {title} {\bibinfo {title} {{Renormalization of the
  4PI effective action using the functional renormalization group}},\ }\href
  {https://doi.org/10.1103/PhysRevD.99.074002} {\bibfield  {journal} {\bibinfo
  {journal} {Phys. Rev. D}\ }\textbf {\bibinfo {volume} {99}},\ \bibinfo
  {pages} {074002} (\bibinfo {year} {2019})},\ \Eprint
  {https://arxiv.org/abs/1901.00840} {arXiv:1901.00840 [hep-th]} \BibitemShut
  {NoStop}%
\bibitem [{\citenamefont {Blaizot}\ \emph {et~al.}(2021)\citenamefont
  {Blaizot}, \citenamefont {Pawlowski},\ and\ \citenamefont
  {Reinosa}}]{Blaizot:2021ikl}%
  \BibitemOpen
  \bibfield  {author} {\bibinfo {author} {\bibfnamefont {J.-P.}\ \bibnamefont
  {Blaizot}}, \bibinfo {author} {\bibfnamefont {J.~M.}\ \bibnamefont
  {Pawlowski}},\ and\ \bibinfo {author} {\bibfnamefont {U.}~\bibnamefont
  {Reinosa}},\ }\bibfield  {title} {\bibinfo {title} {{Functional
  renormalization group and 2PI effective action formalism}},\ }\href
  {https://doi.org/10.1016/j.aop.2021.168549} {\bibfield  {journal} {\bibinfo
  {journal} {Annals Phys.}\ }\textbf {\bibinfo {volume} {431}},\ \bibinfo
  {pages} {168549} (\bibinfo {year} {2021})},\ \Eprint
  {https://arxiv.org/abs/2102.13628} {arXiv:2102.13628 [hep-th]} \BibitemShut
  {NoStop}%
\bibitem [{\citenamefont {Boucaud}\ \emph {et~al.}(2018)\citenamefont
  {Boucaud}, \citenamefont {De~Soto}, \citenamefont {Raya}, \citenamefont
  {Rodr{\'\i}guez-Quintero},\ and\ \citenamefont
  {Zafeiropoulos}}]{Boucaud:2018xup}%
  \BibitemOpen
  \bibfield  {author} {\bibinfo {author} {\bibfnamefont {P.}~\bibnamefont
  {Boucaud}}, \bibinfo {author} {\bibfnamefont {F.}~\bibnamefont {De~Soto}},
  \bibinfo {author} {\bibfnamefont {K.}~\bibnamefont {Raya}}, \bibinfo {author}
  {\bibfnamefont {J.}~\bibnamefont {Rodr{\'\i}guez-Quintero}},\ and\ \bibinfo
  {author} {\bibfnamefont {S.}~\bibnamefont {Zafeiropoulos}},\ }\bibfield
  {title} {\bibinfo {title} {{Discretization effects on renormalized
  gauge-field Green's functions, scale setting, and the gluon mass}},\ }\href
  {https://doi.org/10.1103/PhysRevD.98.114515} {\bibfield  {journal} {\bibinfo
  {journal} {Phys. Rev.}\ }\textbf {\bibinfo {volume} {D98}},\ \bibinfo {pages}
  {114515} (\bibinfo {year} {2018})},\ \Eprint
  {https://arxiv.org/abs/1809.05776} {arXiv:1809.05776 [hep-ph]} \BibitemShut
  {NoStop}%
\bibitem [{\citenamefont {Braun}\ \emph {et~al.}(2010)\citenamefont {Braun},
  \citenamefont {Gies},\ and\ \citenamefont {Pawlowski}}]{Braun:2007bx}%
  \BibitemOpen
  \bibfield  {author} {\bibinfo {author} {\bibfnamefont {J.}~\bibnamefont
  {Braun}}, \bibinfo {author} {\bibfnamefont {H.}~\bibnamefont {Gies}},\ and\
  \bibinfo {author} {\bibfnamefont {J.~M.}\ \bibnamefont {Pawlowski}},\
  }\bibfield  {title} {\bibinfo {title} {{Quark Confinement from Color
  Confinement}},\ }\href {https://doi.org/10.1016/j.physletb.2010.01.009}
  {\bibfield  {journal} {\bibinfo  {journal} {Phys.Lett.}\ }\textbf {\bibinfo
  {volume} {B684}},\ \bibinfo {pages} {262} (\bibinfo {year} {2010})},\ \Eprint
  {https://arxiv.org/abs/0708.2413} {arXiv:0708.2413 [hep-th]} \BibitemShut
  {NoStop}%
\bibitem [{\citenamefont {Fister}\ and\ \citenamefont
  {Pawlowski}(2013)}]{Fister:2013bh}%
  \BibitemOpen
  \bibfield  {author} {\bibinfo {author} {\bibfnamefont {L.}~\bibnamefont
  {Fister}}\ and\ \bibinfo {author} {\bibfnamefont {J.~M.}\ \bibnamefont
  {Pawlowski}},\ }\bibfield  {title} {\bibinfo {title} {{Confinement from
  Correlation Functions}},\ }\href {https://doi.org/10.1103/PhysRevD.88.045010}
  {\bibfield  {journal} {\bibinfo  {journal} {Phys.Rev.}\ }\textbf {\bibinfo
  {volume} {D88}},\ \bibinfo {pages} {045010} (\bibinfo {year} {2013})},\
  \Eprint {https://arxiv.org/abs/1301.4163} {arXiv:1301.4163 [hep-ph]}
  \BibitemShut {NoStop}%
\bibitem [{\citenamefont {Chang}\ \emph {et~al.}(2021)\citenamefont {Chang},
  \citenamefont {Liu}, \citenamefont {Raya}, \citenamefont
  {Rodr\'\i{}guez-Quintero},\ and\ \citenamefont {Yang}}]{Chang:2021vvx}%
  \BibitemOpen
  \bibfield  {author} {\bibinfo {author} {\bibfnamefont {L.}~\bibnamefont
  {Chang}}, \bibinfo {author} {\bibfnamefont {Y.-B.}\ \bibnamefont {Liu}},
  \bibinfo {author} {\bibfnamefont {K.}~\bibnamefont {Raya}}, \bibinfo {author}
  {\bibfnamefont {J.}~\bibnamefont {Rodr\'\i{}guez-Quintero}},\ and\ \bibinfo
  {author} {\bibfnamefont {Y.-B.}\ \bibnamefont {Yang}},\ }\bibfield  {title}
  {\bibinfo {title} {{Linking continuum and lattice quark mass functions via an
  effective charge}},\ }\href {https://doi.org/10.1103/PhysRevD.104.094509}
  {\bibfield  {journal} {\bibinfo  {journal} {Phys. Rev. D}\ }\textbf {\bibinfo
  {volume} {104}},\ \bibinfo {pages} {094509} (\bibinfo {year} {2021})},\
  \Eprint {https://arxiv.org/abs/2105.06596} {arXiv:2105.06596 [hep-lat]}
  \BibitemShut {NoStop}%
\bibitem [{\citenamefont {Bowman}\ \emph {et~al.}(2005)\citenamefont {Bowman},
  \citenamefont {Heller}, \citenamefont {Leinweber}, \citenamefont
  {Parappilly}, \citenamefont {Williams},\ and\ \citenamefont
  {Zhang}}]{Bowman:2005vx}%
  \BibitemOpen
  \bibfield  {author} {\bibinfo {author} {\bibfnamefont {P.~O.}\ \bibnamefont
  {Bowman}}, \bibinfo {author} {\bibfnamefont {U.~M.}\ \bibnamefont {Heller}},
  \bibinfo {author} {\bibfnamefont {D.~B.}\ \bibnamefont {Leinweber}}, \bibinfo
  {author} {\bibfnamefont {M.~B.}\ \bibnamefont {Parappilly}}, \bibinfo
  {author} {\bibfnamefont {A.~G.}\ \bibnamefont {Williams}},\ and\ \bibinfo
  {author} {\bibfnamefont {J.-b.}\ \bibnamefont {Zhang}},\ }\bibfield  {title}
  {\bibinfo {title} {{Unquenched quark propagator in Landau gauge}},\ }\href
  {https://doi.org/10.1103/PhysRevD.71.054507} {\bibfield  {journal} {\bibinfo
  {journal} {Phys. Rev. D}\ }\textbf {\bibinfo {volume} {71}},\ \bibinfo
  {pages} {054507} (\bibinfo {year} {2005})},\ \Eprint
  {https://arxiv.org/abs/hep-lat/0501019} {arXiv:hep-lat/0501019} \BibitemShut
  {NoStop}%
\bibitem [{\citenamefont {Aguilar}\ \emph {et~al.}(2017)\citenamefont
  {Aguilar}, \citenamefont {Cardona}, \citenamefont {Ferreira},\ and\
  \citenamefont {Papavassiliou}}]{Aguilar:2016lbe}%
  \BibitemOpen
  \bibfield  {author} {\bibinfo {author} {\bibfnamefont {A.~C.}\ \bibnamefont
  {Aguilar}}, \bibinfo {author} {\bibfnamefont {J.~C.}\ \bibnamefont
  {Cardona}}, \bibinfo {author} {\bibfnamefont {M.~N.}\ \bibnamefont
  {Ferreira}},\ and\ \bibinfo {author} {\bibfnamefont {J.}~\bibnamefont
  {Papavassiliou}},\ }\bibfield  {title} {\bibinfo {title} {{Non-Abelian
  Ball-Chiu vertex for arbitrary Euclidean momenta}},\ }\href
  {https://doi.org/10.1103/PhysRevD.96.014029} {\bibfield  {journal} {\bibinfo
  {journal} {Phys. Rev. D}\ }\textbf {\bibinfo {volume} {96}},\ \bibinfo
  {pages} {014029} (\bibinfo {year} {2017})},\ \Eprint
  {https://arxiv.org/abs/1610.06158} {arXiv:1610.06158 [hep-ph]} \BibitemShut
  {NoStop}%
\bibitem [{\citenamefont {Aguilar}\ \emph {et~al.}(2012)\citenamefont
  {Aguilar}, \citenamefont {Ibanez}, \citenamefont {Mathieu},\ and\
  \citenamefont {Papavassiliou}}]{Aguilar:2011xe}%
  \BibitemOpen
  \bibfield  {author} {\bibinfo {author} {\bibfnamefont {A.~C.}\ \bibnamefont
  {Aguilar}}, \bibinfo {author} {\bibfnamefont {D.}~\bibnamefont {Ibanez}},
  \bibinfo {author} {\bibfnamefont {V.}~\bibnamefont {Mathieu}},\ and\ \bibinfo
  {author} {\bibfnamefont {J.}~\bibnamefont {Papavassiliou}},\ }\bibfield
  {title} {\bibinfo {title} {{Massless bound-state excitations and the
  Schwinger mechanism in QCD}},\ }\href
  {https://doi.org/10.1103/PhysRevD.85.014018} {\bibfield  {journal} {\bibinfo
  {journal} {Phys. Rev. D}\ }\textbf {\bibinfo {volume} {85}},\ \bibinfo
  {pages} {014018} (\bibinfo {year} {2012})},\ \Eprint
  {https://arxiv.org/abs/1110.2633} {arXiv:1110.2633 [hep-ph]} \BibitemShut
  {NoStop}%
\bibitem [{\citenamefont {Ferreira}\ and\ \citenamefont
  {Papavassiliou}(2025)}]{Ferreira:2025anh}%
  \BibitemOpen
  \bibfield  {author} {\bibinfo {author} {\bibfnamefont {M.~N.}\ \bibnamefont
  {Ferreira}}\ and\ \bibinfo {author} {\bibfnamefont {J.}~\bibnamefont
  {Papavassiliou}},\ }\bibfield  {title} {\bibinfo {title} {{Gluon mass scale
  through the Schwinger mechanism}},\ }\href@noop {} {\  (\bibinfo {year}
  {2025})},\ \Eprint {https://arxiv.org/abs/2501.01080} {arXiv:2501.01080
  [hep-ph]} \BibitemShut {NoStop}%
\bibitem [{\citenamefont {Pelaez}\ \emph {et~al.}(2013)\citenamefont {Pelaez},
  \citenamefont {Tissier},\ and\ \citenamefont {Wschebor}}]{Pelaez:2013cpa}%
  \BibitemOpen
  \bibfield  {author} {\bibinfo {author} {\bibfnamefont {M.}~\bibnamefont
  {Pelaez}}, \bibinfo {author} {\bibfnamefont {M.}~\bibnamefont {Tissier}},\
  and\ \bibinfo {author} {\bibfnamefont {N.}~\bibnamefont {Wschebor}},\
  }\bibfield  {title} {\bibinfo {title} {{Three-point correlation functions in
  Yang-Mills theory}},\ }\href {https://doi.org/10.1103/PhysRevD.88.125003}
  {\bibfield  {journal} {\bibinfo  {journal} {Phys. Rev. D}\ }\textbf {\bibinfo
  {volume} {88}},\ \bibinfo {pages} {125003} (\bibinfo {year} {2013})},\
  \Eprint {https://arxiv.org/abs/1310.2594} {arXiv:1310.2594 [hep-th]}
  \BibitemShut {NoStop}%
\bibitem [{\citenamefont {Aguilar}\ \emph {et~al.}(2014)\citenamefont
  {Aguilar}, \citenamefont {Binosi}, \citenamefont {Iba\~nez},\ and\
  \citenamefont {Papavassiliou}}]{Aguilar:2013vaa}%
  \BibitemOpen
  \bibfield  {author} {\bibinfo {author} {\bibfnamefont {A.~C.}\ \bibnamefont
  {Aguilar}}, \bibinfo {author} {\bibfnamefont {D.}~\bibnamefont {Binosi}},
  \bibinfo {author} {\bibfnamefont {D.}~\bibnamefont {Iba\~nez}},\ and\
  \bibinfo {author} {\bibfnamefont {J.}~\bibnamefont {Papavassiliou}},\
  }\bibfield  {title} {\bibinfo {title} {{Effects of divergent ghost loops on
  the Green\textquoteright{}s functions of QCD}},\ }\href
  {https://doi.org/10.1103/PhysRevD.89.085008} {\bibfield  {journal} {\bibinfo
  {journal} {Phys. Rev. D}\ }\textbf {\bibinfo {volume} {89}},\ \bibinfo
  {pages} {085008} (\bibinfo {year} {2014})},\ \Eprint
  {https://arxiv.org/abs/1312.1212} {arXiv:1312.1212 [hep-ph]} \BibitemShut
  {NoStop}%
\bibitem [{\citenamefont {Eichmann}\ \emph {et~al.}(2014)\citenamefont
  {Eichmann}, \citenamefont {Williams}, \citenamefont {Alkofer},\ and\
  \citenamefont {Vujinovic}}]{Eichmann:2014xya}%
  \BibitemOpen
  \bibfield  {author} {\bibinfo {author} {\bibfnamefont {G.}~\bibnamefont
  {Eichmann}}, \bibinfo {author} {\bibfnamefont {R.}~\bibnamefont {Williams}},
  \bibinfo {author} {\bibfnamefont {R.}~\bibnamefont {Alkofer}},\ and\ \bibinfo
  {author} {\bibfnamefont {M.}~\bibnamefont {Vujinovic}},\ }\bibfield  {title}
  {\bibinfo {title} {{Three-gluon vertex in Landau gauge}},\ }\href
  {https://doi.org/10.1103/PhysRevD.89.105014} {\bibfield  {journal} {\bibinfo
  {journal} {Phys. Rev. D}\ }\textbf {\bibinfo {volume} {89}},\ \bibinfo
  {pages} {105014} (\bibinfo {year} {2014})},\ \Eprint
  {https://arxiv.org/abs/1402.1365} {arXiv:1402.1365 [hep-ph]} \BibitemShut
  {NoStop}%
\bibitem [{\citenamefont {Blum}\ \emph {et~al.}(2014)\citenamefont {Blum},
  \citenamefont {Huber}, \citenamefont {Mitter},\ and\ \citenamefont {von
  Smekal}}]{Blum:2014gna}%
  \BibitemOpen
  \bibfield  {author} {\bibinfo {author} {\bibfnamefont {A.}~\bibnamefont
  {Blum}}, \bibinfo {author} {\bibfnamefont {M.~Q.}\ \bibnamefont {Huber}},
  \bibinfo {author} {\bibfnamefont {M.}~\bibnamefont {Mitter}},\ and\ \bibinfo
  {author} {\bibfnamefont {L.}~\bibnamefont {von Smekal}},\ }\bibfield  {title}
  {\bibinfo {title} {{Gluonic three-point correlations in pure Landau gauge
  QCD}},\ }\href {https://doi.org/10.1103/PhysRevD.89.061703} {\bibfield
  {journal} {\bibinfo  {journal} {Phys. Rev. D}\ }\textbf {\bibinfo {volume}
  {89}},\ \bibinfo {pages} {061703} (\bibinfo {year} {2014})},\ \Eprint
  {https://arxiv.org/abs/1401.0713} {arXiv:1401.0713 [hep-ph]} \BibitemShut
  {NoStop}%
\bibitem [{\citenamefont {Athenodorou}\ \emph {et~al.}(2016)\citenamefont
  {Athenodorou}, \citenamefont {Binosi}, \citenamefont {Boucaud}, \citenamefont
  {De~Soto}, \citenamefont {Papavassiliou}, \citenamefont
  {Rodriguez-Quintero},\ and\ \citenamefont
  {Zafeiropoulos}}]{Athenodorou:2016oyh}%
  \BibitemOpen
  \bibfield  {author} {\bibinfo {author} {\bibfnamefont {A.}~\bibnamefont
  {Athenodorou}}, \bibinfo {author} {\bibfnamefont {D.}~\bibnamefont {Binosi}},
  \bibinfo {author} {\bibfnamefont {P.}~\bibnamefont {Boucaud}}, \bibinfo
  {author} {\bibfnamefont {F.}~\bibnamefont {De~Soto}}, \bibinfo {author}
  {\bibfnamefont {J.}~\bibnamefont {Papavassiliou}}, \bibinfo {author}
  {\bibfnamefont {J.}~\bibnamefont {Rodriguez-Quintero}},\ and\ \bibinfo
  {author} {\bibfnamefont {S.}~\bibnamefont {Zafeiropoulos}},\ }\bibfield
  {title} {\bibinfo {title} {{On the zero crossing of the three-gluon
  vertex}},\ }\href {https://doi.org/10.1016/j.physletb.2016.08.065} {\bibfield
   {journal} {\bibinfo  {journal} {Phys. Lett. B}\ }\textbf {\bibinfo {volume}
  {761}},\ \bibinfo {pages} {444} (\bibinfo {year} {2016})},\ \Eprint
  {https://arxiv.org/abs/1607.01278} {arXiv:1607.01278 [hep-ph]} \BibitemShut
  {NoStop}%
\bibitem [{\citenamefont {Barrios}\ \emph {et~al.}(2022)\citenamefont
  {Barrios}, \citenamefont {Pel\'aez},\ and\ \citenamefont
  {Reinosa}}]{Barrios:2022hzr}%
  \BibitemOpen
  \bibfield  {author} {\bibinfo {author} {\bibfnamefont {N.}~\bibnamefont
  {Barrios}}, \bibinfo {author} {\bibfnamefont {M.}~\bibnamefont {Pel\'aez}},\
  and\ \bibinfo {author} {\bibfnamefont {U.}~\bibnamefont {Reinosa}},\
  }\bibfield  {title} {\bibinfo {title} {{Two-loop three-gluon vertex from the
  Curci-Ferrari model and its leading infrared behavior to all loop orders}},\
  }\href {https://doi.org/10.1103/PhysRevD.106.114039} {\bibfield  {journal}
  {\bibinfo  {journal} {Phys. Rev. D}\ }\textbf {\bibinfo {volume} {106}},\
  \bibinfo {pages} {114039} (\bibinfo {year} {2022})},\ \Eprint
  {https://arxiv.org/abs/2207.10704} {arXiv:2207.10704 [hep-ph]} \BibitemShut
  {NoStop}%
\bibitem [{\citenamefont {Ferreira}\ and\ \citenamefont
  {Papavassiliou}(2023)}]{Ferreira:2023fva}%
  \BibitemOpen
  \bibfield  {author} {\bibinfo {author} {\bibfnamefont {M.~N.}\ \bibnamefont
  {Ferreira}}\ and\ \bibinfo {author} {\bibfnamefont {J.}~\bibnamefont
  {Papavassiliou}},\ }\bibfield  {title} {\bibinfo {title} {{Gauge Sector
  Dynamics in QCD}},\ }\href {https://doi.org/10.3390/particles6010017}
  {\bibfield  {journal} {\bibinfo  {journal} {Particles}\ }\textbf {\bibinfo
  {volume} {6}},\ \bibinfo {pages} {312} (\bibinfo {year} {2023})},\ \Eprint
  {https://arxiv.org/abs/2301.02314} {arXiv:2301.02314 [hep-ph]} \BibitemShut
  {NoStop}%
\bibitem [{\citenamefont {Horak}\ \emph {et~al.}(2020)\citenamefont {Horak},
  \citenamefont {Pawlowski},\ and\ \citenamefont {Wink}}]{Horak:2020eng}%
  \BibitemOpen
  \bibfield  {author} {\bibinfo {author} {\bibfnamefont {J.}~\bibnamefont
  {Horak}}, \bibinfo {author} {\bibfnamefont {J.~M.}\ \bibnamefont
  {Pawlowski}},\ and\ \bibinfo {author} {\bibfnamefont {N.}~\bibnamefont
  {Wink}},\ }\bibfield  {title} {\bibinfo {title} {{Spectral functions in the
  $\phi^4$-theory from the spectral DSE}},\ }\href
  {https://doi.org/10.1103/PhysRevD.102.125016} {\bibfield  {journal} {\bibinfo
   {journal} {Phys. Rev. D}\ }\textbf {\bibinfo {volume} {102}},\ \bibinfo
  {pages} {125016} (\bibinfo {year} {2020})},\ \Eprint
  {https://arxiv.org/abs/2006.09778} {arXiv:2006.09778 [hep-th]} \BibitemShut
  {NoStop}%
\bibitem [{\citenamefont {Horak}\ \emph {et~al.}(2021)\citenamefont {Horak},
  \citenamefont {Papavassiliou}, \citenamefont {Pawlowski},\ and\ \citenamefont
  {Wink}}]{Horak:2021pfr}%
  \BibitemOpen
  \bibfield  {author} {\bibinfo {author} {\bibfnamefont {J.}~\bibnamefont
  {Horak}}, \bibinfo {author} {\bibfnamefont {J.}~\bibnamefont
  {Papavassiliou}}, \bibinfo {author} {\bibfnamefont {J.~M.}\ \bibnamefont
  {Pawlowski}},\ and\ \bibinfo {author} {\bibfnamefont {N.}~\bibnamefont
  {Wink}},\ }\bibfield  {title} {\bibinfo {title} {{Ghost spectral function
  from the spectral Dyson-Schwinger equation}},\ }\href
  {https://doi.org/10.1103/PhysRevD.104.074017} {\bibfield  {journal} {\bibinfo
   {journal} {Phys. Rev. D}\ }\textbf {\bibinfo {volume} {104}},\ \bibinfo
  {pages} {074107} (\bibinfo {year} {2021})},\ \Eprint
  {https://arxiv.org/abs/2103.16175} {arXiv:2103.16175 [hep-th]} \BibitemShut
  {NoStop}%
\bibitem [{\citenamefont {Horak}\ \emph {et~al.}(2022)\citenamefont {Horak},
  \citenamefont {Pawlowski},\ and\ \citenamefont {Wink}}]{Horak:2022myj}%
  \BibitemOpen
  \bibfield  {author} {\bibinfo {author} {\bibfnamefont {J.}~\bibnamefont
  {Horak}}, \bibinfo {author} {\bibfnamefont {J.~M.}\ \bibnamefont
  {Pawlowski}},\ and\ \bibinfo {author} {\bibfnamefont {N.}~\bibnamefont
  {Wink}},\ }\bibfield  {title} {\bibinfo {title} {{On the complex structure of
  Yang-Mills theory}},\ }\href@noop {} {\  (\bibinfo {year} {2022})},\ \Eprint
  {https://arxiv.org/abs/2202.09333} {arXiv:2202.09333 [hep-th]} \BibitemShut
  {NoStop}%
\bibitem [{\citenamefont {Horak}\ \emph {et~al.}(2023)\citenamefont {Horak},
  \citenamefont {Pawlowski},\ and\ \citenamefont {Wink}}]{Horak:2022aza}%
  \BibitemOpen
  \bibfield  {author} {\bibinfo {author} {\bibfnamefont {J.}~\bibnamefont
  {Horak}}, \bibinfo {author} {\bibfnamefont {J.~M.}\ \bibnamefont
  {Pawlowski}},\ and\ \bibinfo {author} {\bibfnamefont {N.}~\bibnamefont
  {Wink}},\ }\bibfield  {title} {\bibinfo {title} {{On the quark spectral
  function in QCD}},\ }\href {https://doi.org/10.21468/SciPostPhys.15.4.149}
  {\bibfield  {journal} {\bibinfo  {journal} {SciPost Phys.}\ }\textbf
  {\bibinfo {volume} {15}},\ \bibinfo {pages} {149} (\bibinfo {year} {2023})},\
  \Eprint {https://arxiv.org/abs/2210.07597} {arXiv:2210.07597 [hep-ph]}
  \BibitemShut {NoStop}%
\bibitem [{\citenamefont {Horak}\ \emph {et~al.}(2024)\citenamefont {Horak},
  \citenamefont {Ihssen}, \citenamefont {Pawlowski}, \citenamefont {Wessely},\
  and\ \citenamefont {Wink}}]{Horak:2023hkp}%
  \BibitemOpen
  \bibfield  {author} {\bibinfo {author} {\bibfnamefont {J.}~\bibnamefont
  {Horak}}, \bibinfo {author} {\bibfnamefont {F.}~\bibnamefont {Ihssen}},
  \bibinfo {author} {\bibfnamefont {J.~M.}\ \bibnamefont {Pawlowski}}, \bibinfo
  {author} {\bibfnamefont {J.}~\bibnamefont {Wessely}},\ and\ \bibinfo {author}
  {\bibfnamefont {N.}~\bibnamefont {Wink}},\ }\bibfield  {title} {\bibinfo
  {title} {{Scalar spectral functions from the spectral functional
  renormalization group}},\ }\href
  {https://doi.org/10.1103/PhysRevD.110.056009} {\bibfield  {journal} {\bibinfo
   {journal} {Phys. Rev. D}\ }\textbf {\bibinfo {volume} {110}},\ \bibinfo
  {pages} {056009} (\bibinfo {year} {2024})},\ \Eprint
  {https://arxiv.org/abs/2303.16719} {arXiv:2303.16719 [hep-th]} \BibitemShut
  {NoStop}%
\bibitem [{\citenamefont {Fukushima}\ \emph {et~al.}(2024)\citenamefont
  {Fukushima}, \citenamefont {Horak}, \citenamefont {Pawlowski}, \citenamefont
  {Wink},\ and\ \citenamefont {Zelle}}]{Fukushima:2023wnl}%
  \BibitemOpen
  \bibfield  {author} {\bibinfo {author} {\bibfnamefont {K.}~\bibnamefont
  {Fukushima}}, \bibinfo {author} {\bibfnamefont {J.}~\bibnamefont {Horak}},
  \bibinfo {author} {\bibfnamefont {J.~M.}\ \bibnamefont {Pawlowski}}, \bibinfo
  {author} {\bibfnamefont {N.}~\bibnamefont {Wink}},\ and\ \bibinfo {author}
  {\bibfnamefont {C.~P.}\ \bibnamefont {Zelle}},\ }\bibfield  {title} {\bibinfo
  {title} {{Nuclear liquid-gas transition in QCD}},\ }\href
  {https://doi.org/10.1103/PhysRevD.110.076022} {\bibfield  {journal} {\bibinfo
   {journal} {Phys. Rev. D}\ }\textbf {\bibinfo {volume} {110}},\ \bibinfo
  {pages} {076022} (\bibinfo {year} {2024})},\ \Eprint
  {https://arxiv.org/abs/2308.16594} {arXiv:2308.16594 [nucl-th]} \BibitemShut
  {NoStop}%
\bibitem [{\citenamefont {Pawlowski}\ and\ \citenamefont
  {Strodthoff}(2015)}]{Pawlowski:2015mia}%
  \BibitemOpen
  \bibfield  {author} {\bibinfo {author} {\bibfnamefont {J.~M.}\ \bibnamefont
  {Pawlowski}}\ and\ \bibinfo {author} {\bibfnamefont {N.}~\bibnamefont
  {Strodthoff}},\ }\bibfield  {title} {\bibinfo {title} {{Real time correlation
  functions and the functional renormalization group}},\ }\href
  {https://doi.org/10.1103/PhysRevD.92.094009} {\bibfield  {journal} {\bibinfo
  {journal} {Phys. Rev. D}\ }\textbf {\bibinfo {volume} {92}},\ \bibinfo
  {pages} {094009} (\bibinfo {year} {2015})},\ \Eprint
  {https://arxiv.org/abs/1508.01160} {arXiv:1508.01160 [hep-ph]} \BibitemShut
  {NoStop}%
\bibitem [{\citenamefont {Tan}\ \emph {et~al.}(2022)\citenamefont {Tan},
  \citenamefont {Chen},\ and\ \citenamefont {Fu}}]{Tan:2021zid}%
  \BibitemOpen
  \bibfield  {author} {\bibinfo {author} {\bibfnamefont {Y.-y.}\ \bibnamefont
  {Tan}}, \bibinfo {author} {\bibfnamefont {Y.-r.}\ \bibnamefont {Chen}},\ and\
  \bibinfo {author} {\bibfnamefont {W.-j.}\ \bibnamefont {Fu}},\ }\bibfield
  {title} {\bibinfo {title} {{Real-time dynamics of the $O(4)$ scalar theory
  within the fRG approach}},\ }\href
  {https://doi.org/10.21468/SciPostPhys.12.1.026} {\bibfield  {journal}
  {\bibinfo  {journal} {SciPost Phys.}\ }\textbf {\bibinfo {volume} {12}},\
  \bibinfo {pages} {026} (\bibinfo {year} {2022})},\ \Eprint
  {https://arxiv.org/abs/2107.06482} {arXiv:2107.06482 [hep-ph]} \BibitemShut
  {NoStop}%
\bibitem [{\citenamefont {Maris}\ \emph {et~al.}(1998)\citenamefont {Maris},
  \citenamefont {Roberts},\ and\ \citenamefont {Tandy}}]{Maris:1997hd}%
  \BibitemOpen
  \bibfield  {author} {\bibinfo {author} {\bibfnamefont {P.}~\bibnamefont
  {Maris}}, \bibinfo {author} {\bibfnamefont {C.~D.}\ \bibnamefont {Roberts}},\
  and\ \bibinfo {author} {\bibfnamefont {P.~C.}\ \bibnamefont {Tandy}},\
  }\bibfield  {title} {\bibinfo {title} {{Pion mass and decay constant}},\
  }\href {https://doi.org/10.1016/S0370-2693(97)01535-9} {\bibfield  {journal}
  {\bibinfo  {journal} {Phys. Lett. B}\ }\textbf {\bibinfo {volume} {420}},\
  \bibinfo {pages} {267} (\bibinfo {year} {1998})},\ \Eprint
  {https://arxiv.org/abs/nucl-th/9707003} {arXiv:nucl-th/9707003} \BibitemShut
  {NoStop}%
\bibitem [{\citenamefont {Gasser}\ and\ \citenamefont
  {Leutwyler}(1984)}]{Gasser:1983yg}%
  \BibitemOpen
  \bibfield  {author} {\bibinfo {author} {\bibfnamefont {J.}~\bibnamefont
  {Gasser}}\ and\ \bibinfo {author} {\bibfnamefont {H.}~\bibnamefont
  {Leutwyler}},\ }\bibfield  {title} {\bibinfo {title} {{Chiral Perturbation
  Theory to One Loop}},\ }\href {https://doi.org/10.1016/0003-4916(84)90242-2}
  {\bibfield  {journal} {\bibinfo  {journal} {Annals Phys.}\ }\textbf {\bibinfo
  {volume} {158}},\ \bibinfo {pages} {142} (\bibinfo {year}
  {1984})}\BibitemShut {NoStop}%
\bibitem [{\citenamefont {Guo}\ \emph {et~al.}(2015)\citenamefont {Guo},
  \citenamefont {Guo}, \citenamefont {Oller},\ and\ \citenamefont
  {Sanz-Cillero}}]{Guo:2015xva}%
  \BibitemOpen
  \bibfield  {author} {\bibinfo {author} {\bibfnamefont {X.-K.}\ \bibnamefont
  {Guo}}, \bibinfo {author} {\bibfnamefont {Z.-H.}\ \bibnamefont {Guo}},
  \bibinfo {author} {\bibfnamefont {J.~A.}\ \bibnamefont {Oller}},\ and\
  \bibinfo {author} {\bibfnamefont {J.~J.}\ \bibnamefont {Sanz-Cillero}},\
  }\bibfield  {title} {\bibinfo {title} {{Scrutinizing the $\eta$-$\eta'$
  mixing, masses and pseudoscalar decay constants in the framework of U(3)
  chiral effective field theory}},\ }\href
  {https://doi.org/10.1007/JHEP06(2015)175} {\bibfield  {journal} {\bibinfo
  {journal} {JHEP}\ }\textbf {\bibinfo {volume} {06}},\ \bibinfo {pages}
  {175}},\ \Eprint {https://arxiv.org/abs/1503.02248} {arXiv:1503.02248
  [hep-ph]} \BibitemShut {NoStop}%
\bibitem [{\citenamefont {Gao}\ \emph {et~al.}(2022)\citenamefont {Gao},
  \citenamefont {Guo}, \citenamefont {Xiao},\ and\ \citenamefont
  {Zhou}}]{Gao:2022dln}%
  \BibitemOpen
  \bibfield  {author} {\bibinfo {author} {\bibfnamefont {X.-L.}\ \bibnamefont
  {Gao}}, \bibinfo {author} {\bibfnamefont {Z.-H.}\ \bibnamefont {Guo}},
  \bibinfo {author} {\bibfnamefont {Z.}~\bibnamefont {Xiao}},\ and\ \bibinfo
  {author} {\bibfnamefont {Z.-Y.}\ \bibnamefont {Zhou}},\ }\bibfield  {title}
  {\bibinfo {title} {{Scrutinizing \ensuremath{\pi}\ensuremath{\pi} scattering
  in light of recent lattice phase shifts}},\ }\href
  {https://doi.org/10.1103/PhysRevD.105.094002} {\bibfield  {journal} {\bibinfo
   {journal} {Phys. Rev. D}\ }\textbf {\bibinfo {volume} {105}},\ \bibinfo
  {pages} {094002} (\bibinfo {year} {2022})},\ \Eprint
  {https://arxiv.org/abs/2202.03124} {arXiv:2202.03124 [hep-ph]} \BibitemShut
  {NoStop}%
\bibitem [{\citenamefont {Braun}\ \emph {et~al.}(2024)\citenamefont {Braun},
  \citenamefont {Chen}, \citenamefont {Fu}, \citenamefont {Gao}, \citenamefont
  {Geissel}, \citenamefont {Huang}, \citenamefont {Ihssen}, \citenamefont {Lu},
  \citenamefont {Pawlowski}, \citenamefont {Rennecke}, \citenamefont {Sattler},
  \citenamefont {Schallmo}, \citenamefont {Tan}, \citenamefont {T{\"o}pfel},
  \citenamefont {Wen}, \citenamefont {Wessely}, \citenamefont {Yin},
  \citenamefont {Wang},\ and\ \citenamefont {Zorbach}}]{fQCD}%
  \BibitemOpen
  \bibfield  {author} {\bibinfo {author} {\bibfnamefont {J.}~\bibnamefont
  {Braun}}, \bibinfo {author} {\bibfnamefont {Y.-r.}\ \bibnamefont {Chen}},
  \bibinfo {author} {\bibfnamefont {W.-j.}\ \bibnamefont {Fu}}, \bibinfo
  {author} {\bibfnamefont {F.}~\bibnamefont {Gao}}, \bibinfo {author}
  {\bibfnamefont {A.}~\bibnamefont {Geissel}}, \bibinfo {author} {\bibfnamefont
  {C.}~\bibnamefont {Huang}}, \bibinfo {author} {\bibfnamefont
  {F.}~\bibnamefont {Ihssen}}, \bibinfo {author} {\bibfnamefont
  {Y.}~\bibnamefont {Lu}}, \bibinfo {author} {\bibfnamefont {J.~M.}\
  \bibnamefont {Pawlowski}}, \bibinfo {author} {\bibfnamefont {F.}~\bibnamefont
  {Rennecke}}, \bibinfo {author} {\bibfnamefont {F.}~\bibnamefont {Sattler}},
  \bibinfo {author} {\bibfnamefont {B.}~\bibnamefont {Schallmo}}, \bibinfo
  {author} {\bibfnamefont {Y.-y.}\ \bibnamefont {Tan}}, \bibinfo {author}
  {\bibfnamefont {S.}~\bibnamefont {T{\"o}pfel}}, \bibinfo {author}
  {\bibfnamefont {R.}~\bibnamefont {Wen}}, \bibinfo {author} {\bibfnamefont
  {J.}~\bibnamefont {Wessely}}, \bibinfo {author} {\bibfnamefont
  {S.}~\bibnamefont {Yin}}, \bibinfo {author} {\bibfnamefont {Z.-n.}\
  \bibnamefont {Wang}},\ and\ \bibinfo {author} {\bibfnamefont
  {N.}~\bibnamefont {Zorbach}},\ }\href@noop {} {\bibinfo {title} {{fQCD
  collaboration}}} (\bibinfo {year} {2024})\BibitemShut {NoStop}%
\bibitem [{\citenamefont {Klevansky}(1992)}]{Klevansky:1992qe}%
  \BibitemOpen
  \bibfield  {author} {\bibinfo {author} {\bibfnamefont {S.~P.}\ \bibnamefont
  {Klevansky}},\ }\bibfield  {title} {\bibinfo {title} {{The Nambu-Jona-Lasinio
  model of quantum chromodynamics}},\ }\href
  {https://doi.org/10.1103/RevModPhys.64.649} {\bibfield  {journal} {\bibinfo
  {journal} {Rev. Mod. Phys.}\ }\textbf {\bibinfo {volume} {64}},\ \bibinfo
  {pages} {649} (\bibinfo {year} {1992})}\BibitemShut {NoStop}%
\bibitem [{\citenamefont {Braun}\ \emph {et~al.}(2020)\citenamefont {Braun},
  \citenamefont {Leonhardt}, \citenamefont {Pawlowski},\ and\ \citenamefont
  {Rosenbl\"uh}}]{Braun:2020mhk}%
  \BibitemOpen
  \bibfield  {author} {\bibinfo {author} {\bibfnamefont {J.}~\bibnamefont
  {Braun}}, \bibinfo {author} {\bibfnamefont {M.}~\bibnamefont {Leonhardt}},
  \bibinfo {author} {\bibfnamefont {J.~M.}\ \bibnamefont {Pawlowski}},\ and\
  \bibinfo {author} {\bibfnamefont {D.}~\bibnamefont {Rosenbl\"uh}},\
  }\bibfield  {title} {\bibinfo {title} {{Chiral and effective $U(1)_{\rm A}$
  symmetry restoration in QCD}},\ }\href@noop {} {\  (\bibinfo {year}
  {2020})},\ \Eprint {https://arxiv.org/abs/2012.06231} {arXiv:2012.06231
  [hep-ph]} \BibitemShut {NoStop}%
\bibitem [{\citenamefont {Braun}\ \emph {et~al.}(2025)\citenamefont {Braun},
  \citenamefont {Geißel}, \citenamefont {Pawlowski}, \citenamefont {Sattler},\
  and\ \citenamefont {Wink}}]{TensorBases2025}%
  \BibitemOpen
  \bibfield  {author} {\bibinfo {author} {\bibfnamefont {J.}~\bibnamefont
  {Braun}}, \bibinfo {author} {\bibfnamefont {A.}~\bibnamefont {Geißel}},
  \bibinfo {author} {\bibfnamefont {J.~M.}\ \bibnamefont {Pawlowski}}, \bibinfo
  {author} {\bibfnamefont {F.~R.}\ \bibnamefont {Sattler}},\ and\ \bibinfo
  {author} {\bibfnamefont {N.}~\bibnamefont {Wink}},\ }\href@noop {} {\bibfield
   {journal} {\bibinfo  {journal} {in preparation}\ } (\bibinfo {year}
  {2025})}\BibitemShut {NoStop}%
\bibitem [{\citenamefont {Braun}\ \emph {et~al.}(2019)\citenamefont {Braun},
  \citenamefont {Leonhardt},\ and\ \citenamefont {Pawlowski}}]{Braun:2018svj}%
  \BibitemOpen
  \bibfield  {author} {\bibinfo {author} {\bibfnamefont {J.}~\bibnamefont
  {Braun}}, \bibinfo {author} {\bibfnamefont {M.}~\bibnamefont {Leonhardt}},\
  and\ \bibinfo {author} {\bibfnamefont {J.~M.}\ \bibnamefont {Pawlowski}},\
  }\bibfield  {title} {\bibinfo {title} {{Renormalization group consistency and
  low-energy effective theories}},\ }\href
  {https://doi.org/10.21468/SciPostPhys.6.5.056} {\bibfield  {journal}
  {\bibinfo  {journal} {SciPost Phys.}\ }\textbf {\bibinfo {volume} {6}},\
  \bibinfo {pages} {056} (\bibinfo {year} {2019})},\ \Eprint
  {https://arxiv.org/abs/1806.04432} {arXiv:1806.04432 [hep-ph]} \BibitemShut
  {NoStop}%
\bibitem [{\citenamefont {Cyrol}\ \emph
  {et~al.}(2018{\natexlab{b}})\citenamefont {Cyrol}, \citenamefont {Mitter},
  \citenamefont {Pawlowski},\ and\ \citenamefont {Strodthoff}}]{Cyrol:2017qkl}%
  \BibitemOpen
  \bibfield  {author} {\bibinfo {author} {\bibfnamefont {A.~K.}\ \bibnamefont
  {Cyrol}}, \bibinfo {author} {\bibfnamefont {M.}~\bibnamefont {Mitter}},
  \bibinfo {author} {\bibfnamefont {J.~M.}\ \bibnamefont {Pawlowski}},\ and\
  \bibinfo {author} {\bibfnamefont {N.}~\bibnamefont {Strodthoff}},\ }\bibfield
   {title} {\bibinfo {title} {{Nonperturbative finite-temperature Yang-Mills
  theory}},\ }\href {https://doi.org/10.1103/PhysRevD.97.054015} {\bibfield
  {journal} {\bibinfo  {journal} {Phys. Rev.}\ }\textbf {\bibinfo {volume}
  {D97}},\ \bibinfo {pages} {054015} (\bibinfo {year} {2018}{\natexlab{b}})},\
  \Eprint {https://arxiv.org/abs/1708.03482} {arXiv:1708.03482 [hep-ph]}
  \BibitemShut {NoStop}%
\bibitem [{\citenamefont {Pawlowski}\ \emph {et~al.}(2017)\citenamefont
  {Pawlowski}, \citenamefont {Scherer}, \citenamefont {Schmidt},\ and\
  \citenamefont {Wetzel}}]{Pawlowski:2015mlf}%
  \BibitemOpen
  \bibfield  {author} {\bibinfo {author} {\bibfnamefont {J.~M.}\ \bibnamefont
  {Pawlowski}}, \bibinfo {author} {\bibfnamefont {M.~M.}\ \bibnamefont
  {Scherer}}, \bibinfo {author} {\bibfnamefont {R.}~\bibnamefont {Schmidt}},\
  and\ \bibinfo {author} {\bibfnamefont {S.~J.}\ \bibnamefont {Wetzel}},\
  }\bibfield  {title} {\bibinfo {title} {{Physics and the choice of regulators
  in functional renormalisation group flows}},\ }\href
  {https://doi.org/10.1016/j.aop.2017.06.017} {\bibfield  {journal} {\bibinfo
  {journal} {Annals Phys.}\ }\textbf {\bibinfo {volume} {384}},\ \bibinfo
  {pages} {165} (\bibinfo {year} {2017})},\ \Eprint
  {https://arxiv.org/abs/1512.03598} {arXiv:1512.03598 [hep-th]} \BibitemShut
  {NoStop}%
\bibitem [{\citenamefont {Zorbach}\ \emph {et~al.}(2024)\citenamefont
  {Zorbach}, \citenamefont {Stoll},\ and\ \citenamefont
  {Braun}}]{Zorbach:2024zjx}%
  \BibitemOpen
  \bibfield  {author} {\bibinfo {author} {\bibfnamefont {N.}~\bibnamefont
  {Zorbach}}, \bibinfo {author} {\bibfnamefont {J.}~\bibnamefont {Stoll}},\
  and\ \bibinfo {author} {\bibfnamefont {J.}~\bibnamefont {Braun}},\ }\bibfield
   {title} {\bibinfo {title} {{Optimization and Stabilization of Functional
  Renormalization Group Flows}},\ }\href@noop {} {\  (\bibinfo {year}
  {2024})},\ \Eprint {https://arxiv.org/abs/2401.12854} {arXiv:2401.12854
  [hep-ph]} \BibitemShut {NoStop}%
\bibitem [{\citenamefont {Huber}\ and\ \citenamefont
  {Braun}(2012)}]{Huber:2011qr}%
  \BibitemOpen
  \bibfield  {author} {\bibinfo {author} {\bibfnamefont {M.~Q.}\ \bibnamefont
  {Huber}}\ and\ \bibinfo {author} {\bibfnamefont {J.}~\bibnamefont {Braun}},\
  }\bibfield  {title} {\bibinfo {title} {{Algorithmic derivation of functional
  renormalization group equations and Dyson-Schwinger equations}},\ }\href
  {https://doi.org/10.1016/j.cpc.2012.01.014} {\bibfield  {journal} {\bibinfo
  {journal} {Comput. Phys. Commun.}\ }\textbf {\bibinfo {volume} {183}},\
  \bibinfo {pages} {1290} (\bibinfo {year} {2012})},\ \Eprint
  {https://arxiv.org/abs/1102.5307} {arXiv:1102.5307 [hep-th]} \BibitemShut
  {NoStop}%
\bibitem [{\citenamefont {Huber}\ \emph {et~al.}(2020)\citenamefont {Huber},
  \citenamefont {Cyrol},\ and\ \citenamefont {Pawlowski}}]{Huber:2019dkb}%
  \BibitemOpen
  \bibfield  {author} {\bibinfo {author} {\bibfnamefont {M.~Q.}\ \bibnamefont
  {Huber}}, \bibinfo {author} {\bibfnamefont {A.~K.}\ \bibnamefont {Cyrol}},\
  and\ \bibinfo {author} {\bibfnamefont {J.~M.}\ \bibnamefont {Pawlowski}},\
  }\bibfield  {title} {\bibinfo {title} {{DoFun 3.0: Functional equations in
  Mathematica}},\ }\href {https://doi.org/10.1016/j.cpc.2019.107058} {\bibfield
   {journal} {\bibinfo  {journal} {Comput. Phys. Commun.}\ }\textbf {\bibinfo
  {volume} {248}},\ \bibinfo {pages} {107058} (\bibinfo {year} {2020})},\
  \Eprint {https://arxiv.org/abs/1908.02760} {arXiv:1908.02760 [hep-ph]}
  \BibitemShut {NoStop}%
\bibitem [{\citenamefont {Sattler}\ and\ \citenamefont
  {Pawlowski}(2024)}]{Sattler:2024ozv}%
  \BibitemOpen
  \bibfield  {author} {\bibinfo {author} {\bibfnamefont {F.~R.}\ \bibnamefont
  {Sattler}}\ and\ \bibinfo {author} {\bibfnamefont {J.~M.}\ \bibnamefont
  {Pawlowski}},\ }\bibfield  {title} {\bibinfo {title} {{DiFfRG: A
  Discretisation Framework for functional Renormalisation Group flows}},\
  }\href@noop {} {\  (\bibinfo {year} {2024})},\ \Eprint
  {https://arxiv.org/abs/2412.13043} {arXiv:2412.13043 [hep-ph]} \BibitemShut
  {NoStop}%
\end{thebibliography}%

\end{document}